%% file: main.tex
\documentclass[11pt]{article}
\RequirePackage{etex}

\usepackage[T1]{fontenc}
\usepackage[utf8]{inputenc}

\usepackage{fancyhdr}

\usepackage{comment}
\usepackage{booktabs}
\usepackage{nicefrac}
\usepackage{microtype}
\usepackage{lipsum}
\usepackage{graphicx}
\usepackage{multirow}
\usepackage{mathrsfs}
\usepackage[title]{appendix}
\usepackage[x11names]{xcolor}
\usepackage{textcomp}
\usepackage{manyfoot}
\usepackage{algorithm}
\usepackage{algorithmicx}
\usepackage{algpseudocode}
\usepackage{listings}
\usepackage{float}

\usepackage{natbib}
\makeatletter
\renewcommand{\citet}[1]{%
  \citeauthor{#1}, \citeyear{#1}%
}
\makeatother

\usepackage{mathtools}
\mathtoolsset{showonlyrefs}

\DeclareUnicodeCharacter{2113}{\ensuremath{\ell}}   % ℓ
\DeclareUnicodeCharacter{221E}{\ensuremath{\infty}}

\usepackage{url}
\usepackage{subcaption}

\usepackage{tikz}
\usetikzlibrary{shapes,decorations,arrows,calc,arrows.meta,fit,positioning,decorations.pathreplacing}
\usepackage{tikz-network}
\usepackage{pgfplots}
\pgfplotsset{compat=1.18}
\usepackage{tikz}
\usepackage{physics}
\usepgfplotslibrary{groupplots}
\usepackage[inline]{enumitem}
\usepackage{multicol}
\usepackage{times}

\usepackage{amsmath,amsfonts,amsthm,amssymb,bm,dsfont}
\usepackage[a4paper,top=2.5cm,bottom=2.5cm,left=2cm,right=2cm]{geometry}
\usepackage{numprint}
\npthousandsep{,}
\npthousandthpartsep{}
\npdecimalsign{.}

\theoremstyle{plain}
\newtheorem{theorem}{Theorem}

\newtheorem{lemma}{Lemma}
\newtheorem{proposition}{Proposition}

\newtheorem{definition}{Definition}
\newtheorem{assumption}{Assumption}

\DeclareMathOperator*{\argmin}{argmin}

\newcommand{\titledoc}{Spectral embedding of inhomogeneous Poisson processes on multiplex networks}

\providecommand{\keywords}[1]{{\small{\textbf{\textit{Keywords ---}} #1}}}

\usepackage{authblk}

\author{Joshua Corneck}
\author{Edward A. K. Cohen}
\author{Francesco Sanna Passino}

\affil{Department of Mathematics, Imperial College London \\ 180 Queen's Gate, SW7 2AZ, London (United Kingdom)}

\date{}

\title{\Huge\textbf{\titledoc}}

\allowdisplaybreaks

\usepackage[colorlinks=true,linkcolor=blue,citecolor=blue,urlcolor=blue]{hyperref}
\numberwithin{equation}{section}



\newif\ifsupplementaryrefs
\supplementaryrefsfalse % or \supplementaryrefstrue

\ifsupplementaryrefs
  \usepackage{multibib}
  \newcites{SM}{Supplementary references}
\else
  \newcommand{\citeSM}{\cite}
  \newcommand{\citepSM}{\citep}
  \newcommand{\citetSM}{\citet}
\fi

\begin{document}

\maketitle

\begin{abstract}
\input{sections/abstract.tex}    
\end{abstract}


\input{sections/introduction}
\input{sections/notation_and_mathematical_background}
\input{sections/model}

\input{sections/theoretical_results}

\input{sections/simulations}

\input{sections/data}

\input{sections/conclusion}

\input{sections/acknowledgements}

\input{sections/code}

\bibliographystyle{rss}  
\bibliography{references} 

\newpage
\input{sections/appendix}

\ifsupplementaryrefs
\bibliographystyleSM{rss}
\bibliographySM{references}
\fi

\end{document}

%% file: sections/abstract.tex
In many real-world networks, data on the edges evolve in continuous time, naturally motivating representations based on point processes. Heterogeneity in edge types further gives rise to multiplex network point processes. In this work, we propose a model for multiplex network data observed in continuous-time. We establish two-to-infinity norm consistency and asymptotic normality for spectral-embedding-based estimation of the model parameters as both network size and time resolution increase. Drawing inspiration from random dot product graph models, each edge intensity is expressed as the inner product of two low-dimensional latent positions: one dynamic and layer-agnostic, the other static and layer-dependent. These latent positions constitute the primary objects of inference, which is conducted via spectral embedding methods. Our theoretical results are established under a histogram estimator of the network intensities and provide justification for applying a doubly unfolded adjacency spectral embedding method for estimation. Simulations and real-data analyses demonstrate the effectiveness of the proposed model and inference procedure.

%% file: sections/introduction.tex
\section{Introduction}
\label{sec:intro}

% \subsection{Summary of our contribution}

% \subsection{Problem motivation}

Network models provide a principled way to encode complex patterns of interaction among large collections of interdependent entities. In many modern applications, such data exhibit rich structure, encompassing multiple types of relationships and evolving continuously over time. To capture heterogeneity of connection types, \textit{multiplex} (also known as \textit{multilayer}) networks extend this framework by considering multiple \textit{layers}, each encoding a distinct form of relation among the same set of nodes \citep{domenico2015}. By further allowing temporal point processes to exist on the edges of multiplex networks, continuous-time interaction data can be more accurately understood. Such networks naturally arise in many domains \citep[for a survey of common applications, see][]{goldenberg2010}; for example, on a social media platform, layers may represent “follow’’ or “retweet’’ interactions with time-stamps defining a point process \citep{greene2013}. Similarly, on a trade network, times of trades could be considered as a point process with layers corresponding to different products \citep{corneck2025b}. %For a survey of common applications, see, for example, \cite{goldenberg2010}.

A common approach for modelling static networks is to associate each node with a low-dimensional latent position governing its connectivity, as in the latent position model of \cite{hoff2002}. The random dot-product graph (RDPG) is a widely studied special case, for which spectral embedding provides a natural and tractable method of inference \citep{athreya2018}. Spectral embedding has become a central tool in statistical network analysis, providing consistent estimators of latent positions under the stochastic block model \citep{sussman2012, rohe2010}, and enabling tasks such as clustering and community detection \citep{athreya2018} via limiting distributional results.

Despite significant advances in the literature on spectral embedding for static and multiplex graphs, few approaches have been developed to handle dynamic multiplex graphs observed in continuous time. In this work, we address this gap by introducing the \textit{multiplex inhomogeneous Poisson process dot-product graph (MIPP-DPG)}, a continuous-time multiplex dot-product model characterised via low-dimensional latent positions. The MIPP-DPG characterises marked point process data as a multiplex, edge-level network point process (NPP), with observations being quadruples describing the time of arrival, the source and recipient nodes, and the corresponding network layer of the arrival. The multiplex NPP is modelled as a collection of conditionally independent, inhomogeneous Poisson processes, with intensities given by an inner product between a dynamic, layer-agnostic position and a static, layer-dependent position. We estimate these positions by binning time to form a histogram estimator of edge intensities and then applying spectral embedding to the resulting matrix. Despite this discretisation step, our theory recovers the continuous-time nature of the latent positions asymptotically. Under non-degeneracy and smoothness conditions, and as the numbers of nodes increase and the time resolution is refined, we prove row-wise two-to-infinity norm consistency and asymptotic normality of the estimated positions. Simulations and a real-data application demonstrate that the method recovers interpretable dynamic and layer-dependent structure.

%To capture heterogeneity of connection types, multiplex (or multilayer) networks extend this framework by considering multiple \textit{layers}, each encoding a distinct form of relation among the same set of nodes \citep{domenico2015}. Such networks arise in many domains: neuroimaging, where brain regions are connected across patients \citep{paul2018}; social media, where layers may represent “follow’’ or “retweet’’ interactions \citep{greene2013}; and trade networks, where layers correspond to different products \citep{corneck2025b}. For a survey of applications, see \cite{goldenberg2010}.

The remainder of this manuscript is organised as follows. Section~\ref{sec:mathematical_backround} introduces the related literature and outlines the notation used. % and provides the necessary background on notions of probabilistic asymptotic equivalence and on random dot-product graphs and spectral embedding.
Section~\ref{sec:model} presents our proposed model and related estimation procedure, before Section~\ref{sec:theory} introduces the main theoretical results of this work, as well as key results for their proof. In conclusion, Section~\ref{sec:simulations} examines inference with simulated data, with Section~\ref{sec:real_data} then exploring an application of our model to a global air transportation network.

\subsection{Background and related literature} \label{sec:background}

A graph is mathematically represented as $\mathcal{G} = (\mathcal{V}, \mathcal{E})$, with a node set $\mathcal{V} := [N]$, where $[N] := \{1,\cdots,N\},\ N \in \mathbb{N}$, and an edge set $\mathcal{E} \subseteq \mathcal{V} \times \mathcal{V}$ that characterises interactions between these nodes. We write $(i,j) \in \mathcal{E}$ if there is a connection between nodes $i,j\in\mathcal{V}$. We say that the graph is undirected if and only if $(i,j) \in \mathcal{E}$ implies $(j,i) \in \mathcal{E}$, otherwise the graph is considered to be directed. A graph can be equivalently represented by its adjacency matrix $\boldsymbol{A}\in \{0,1\}^{N \times N}$, with $A_{ij} = \mathbb{I}_{\mathcal{E}}\{(i,j)\}$, where $\mathbb{I}_\cdot\{\cdot\}$ denotes the indicator function. 

A common model for such a directed graph in the single layer setting is the latent position model \citep[LPM; ][]{hoff2002}. In the undirected graph case, each node $i \in \mathcal{V}$ is associated with a latent position vector $X_i \in \mathcal{X} \subseteq \mathbb{R}^d$, where the latent space $\mathcal{X}$ has dimension $d \ll N$. Once equipped with a kernel function $\kappa: \mathcal{X} \times \mathcal{X} \to [0,1]$, the latent positions fully characterise the network by expressing the edge probabilities as $\mathbb{P}(A_{ij}=1)=\kappa(X_i,X_j)$. A popular choice of the kernel function is $\kappa(X_i,X_j) = X_i^\top X_j$, in which case the LPM is referred to as a random dot product graph \citep[RDPG; see, for example,][]{athreya2018}. Spectral embedding of the adjacency matrices can be used to estimate the latent positions in RDPGs, with the left and right singular vectors of the associated adjacency matrix sorted by the magnitude of the associated singular values \citep{athreya2018}. Spectral embedding estimators produce uniformly consistent estimates of up to orthogonal or general linear transformation in various RDPG-based models, with asymptotically Gaussian error  \citep{lyzinski2015, athreya2018, cape_2019b, cape_2019, jones_2021, gallagher2024, baum2025}.

Latent positions in latent space models can be assumed to be either deterministic and unknown, or to be random and unknown \citep[see, for example, the discussion in Section 4.1 of][]{athreya2018}. In the case of random positions, the latent representations can be sampled independently for each node from a pre-specified distribution on a suitable space. As discussed in \cite{wang2025}, deterministic latent positions are more suited to the case of labelled dynamic networks, meaning that each node has a deterministic identity across time. In unlabelled networks, the focus is on the overall population structure at each time, rather than individual trajectories, and inference is on recovering the latent positions. Contrastingly, in the labelled case, inference is anchored at the node level, aiming to track each node’s latent trajectory through time. We will focus on the labelled case and thus assume that our latent positions are \textit{deterministic but unknown}.

Networks that evolve over time are termed \textit{dynamic}, with temporal dependence manifesting in the edge set, the node set, or both. Much of the literature has focused on discrete-time formulations of dynamic networks, where networks are observed through aggregated snapshots, and most theoretical results are established in this setting \citep[see, for example,][]{sewell2015,Shlomovich22,billio2024,athreya2024}. In contrast, many real-world interaction datasets are inherently continuous-time, such as email exchanges within a company \citep{enron2004}, social media activity \citep{socialnetworks2017}, or bike-sharing networks \citep{corneck2025}. Models for such continuous-time data typically view interactions as edge-level NPPs, for example through semiparametric stochastic block models \citep{Matias2018-rx}, Cox process formulations \citep{Perry13}, or mutually exciting processes with intensities depending only on node-specific parameters \citep{sannapassino23}. An alternative formulation of a dynamic network is to view the network itself as static, but to consider a collection of point processes on the edges (an NPP), with the edge-level point process capturing the temporal dynamics of the data \citep[see, for example,][]{corneck2025, Matias2018-rx, modell2024}. It is this formulation on which the MIPP-DPG is modelled.

When the graph $\mathcal{G}$ consists of multiple connection types between its nodes, it is called a \textit{multiplex network}. Each of the $L$ connection types is described by a unique layer, with each layer $\ell \in [L]$ being a graph $\mathcal{G}_\ell = (\mathcal{V}, \mathcal{E}_\ell)$ on a node set common to all layers. A multiplex network $\mathcal{G}$ with $L$ types of connection is mathematically denoted by $\mathcal{G} = (\mathcal{V}, \{\mathcal{E}_1,\dots,\mathcal{E}_L\})$. %Each graph $\mathcal{G}_\ell$ could equivalently be represented by its adjacency matrix $A_\ell = \{A_{\ell ij}\}_{i,j=1}^N \in \{0,1\}^{N\times N}$.
Methodological development for multiplex networks is relatively recent \citep{sosa2022, huang2023, lei2023}. Within the context of RDPG models, a relevant extension is the multilayer RDPG \cite[MRDPG;][]{jones_2021}, where each node is assigned a layer-specific latent position $Y_{\ell j}$, so that the probability of connection between nodes $i$ and $j$ in layer $\ell$ is $X_i^\top Y_{\ell j}$. The MRDPG admits inference via unfolded adjacency spectral embedding \citep[UASE;][]{jones_2021}, wherein the adjacency matrices of the network are stacked and simultaneously embedded. Asymptotic theoretical properties for UASE have been established, with \cite{gallagher2021} showing that it yields stable and consistent latent position estimates across layers. More broadly, theoretical properties of the UASE estimator are supported by the extensive theory of spectral embedding for single-layer networks \citep{athreya2018}.
%Dynamic graphs can also be incorporated into this framework. In the discrete-time setting, when a time-indexed sequence of multiplex graphs are observed, \cite{baum2025} extend this model by assigning nodes a time-dependent latent position $X_{it}$, for $t \in \mathbb{N}$, so that the probability of connection between nodes $i$ and $j$ in layer $\ell$ at time point $t$ is $X_{it}^\top Y_{\ell j}$.

While multiplex and dynamic networks have each been studied, the setting where both aspects are present has received comparatively limited attention. Examples include the work of \cite{oselio2014, hoff2015, durante2017, loyal2023, wang2025}. Among the available approaches, the dynamic multiplex random dot-product graph (DMRDPG) of \cite{baum2025} is particularly relevant to this work. The DMRDPG extends the MRDPG to a sequence of discrete-time multiplex graphs. Nodes are assigned a time-dependent latent position $X_{it}$, for $t \in \mathbb{N}$, so that the probability of connection between nodes $i$ and $j$ in layer $\ell$ at time point $t$ is $X_{it}^\top Y_{\ell j}$. The authors develop an inference procedure called doubly unfolded adjacency spectral embedding \citep[DUASE;][]{baum2025}, establishing methodology and theoretical results for the discrete-time and binary edge setting. In this work, we adapt DUASE to the weighted, continuous-time setting by assuming a dot product structure for the edge-process intensity functions, proving that the estimators are asymptotically unbiased with normally distributed errors.

Recent work has introduced spectral methods for continuous-time networks, such as the Intensity Profile Projection \citep[IPP;][]{modell2024}, which provides representations of nodes in single-layer point process networks, or the Common Subspace Independent Process \citep[COSIP;][]{romero2025multiresolution}. However, the literature on continuous-time networks remain largely separate from the RDPG framework that underpins much of the theory for static and discrete-time random graph models. Furthermore, continuous-time approaches of the kind in \cite{modell2024} and \cite{romero2025multiresolution} have not yet been developed for multiplex networks. All available estimation procedures for continuous-time data on networks must involve discretising the observation window. Our work utilises this approach too; however, our theory will recover the continuous-time setting in the limit of a growing network. 

%% file: sections/notation_and_mathematical_background.tex
\label{sec:mathematical_backround}
\subsection{Matrix and norm notation}

We write matrices using boldface and vectors and scalars in standard font. For a matrix $\boldsymbol{M} \in \mathbb{R}^{m \times n}$, we denote its $i$-th row and $j$-th column by $M_{i,\ast}\in \mathbb{R}^n$ and $M_{\ast,j} \in \mathbb{R}^m$, treated as row and column vectors, respectively. We access the $(i,j)$-th element of $\boldsymbol{M}$ as $M_{ij}$, but when writing the $(i,j)$-th element of a product $\boldsymbol{M}_1\boldsymbol{M}_2$, we write $(\boldsymbol{M}_1\boldsymbol{M}_2)_{ij}$. When matrices carry subscripts, such as $\boldsymbol{M}_\ell$ or $\boldsymbol{M}_{m\ell}$, we access the $(i,j)$-th element as $M_{\ell, ij}$ or $M_{m\ell,ij}$. If we consider a block matrix $\boldsymbol{M} \in \mathbb{R}^{an \times bm}$, we write $\boldsymbol{M}^{(r,\ast)}$ to be the $n \times bm$ matrix consisting of the $r$-th set of $n$ rows, and $\boldsymbol{M}^{(\ast,s)}$ to be the $an \times m$ matrix consisting of the $s$-th set of $m$ columns, and $\boldsymbol{M}^{(r,s)}$ to be the $(r,s)$-th $n\times n$ submatrix. 

When $\boldsymbol{M}$ is square, the $k$-th largest eigenvalue of $\boldsymbol{M}$ is written as $\lambda_k(\boldsymbol{M})$, and when $m \neq n$, the $k$-th largest singular value is written as $\sigma_k(\boldsymbol{M})$. For $\alpha > 0$, we define the vector norm $\norm{\cdot}_{\alpha}$ on $\mathbb{R}^n$ by $\norm{x}_\alpha = \left(\sum_{i=1}^n \abs{x_i}^\alpha\right)^{1/\alpha}$. The matrix $\boldsymbol{M}$ induces a linear operator from $\mathbb{R}^{n}$ to $\mathbb{R}^{m}$, and we define the operator norm of $\boldsymbol{M}$ as a mapping from $\mathbb{R}^n$ to $\mathbb{R}^m$, equipped with $\norm{\cdot}_\alpha$ and $\norm{\cdot}_\beta$, respectively, as $\norm{\boldsymbol{M}}_{\alpha, \beta} := \sup_{\norm{x}_\alpha = 1} \norm{\boldsymbol{M}x}_\beta$. An important case is the two-to-infinty norm of $\boldsymbol{M}$, written $\norm{\boldsymbol{M}}_{2,\infty} := \sup_{\norm{x}_2 = 1}\norm{\boldsymbol{M}x}_\infty$, %. This norm is 
equivalent to $\norm{\boldsymbol{M}}_{2,\infty} = \max_{i \in [m]}\norm{M_{i,\cdot}}_2$, %which is 
the maximum of the Euclidean norm of the rows of $\boldsymbol{M}$. In the case that $\alpha = \beta$, the operator norm of the matrix $\boldsymbol{M}$ is simply written as $\norm{\boldsymbol{M}}_\alpha$. We regularly refer to the spectral norm $\norm{\boldsymbol{M}}_2$, which %can be shown to 
corresponds to $\lambda_1(\boldsymbol{M}^\top \boldsymbol{M})^{1/2} = \sigma_1(\boldsymbol{M})$. Also of importance is the Frobenius norm of $\boldsymbol{M}$, defined as $\norm{\boldsymbol{M}}_F := (\sum_{i=1}^m\sum_{j=1}^n \abs{M_{ij}}^2)^{1/2}$. We frequently make use of the following inequalites \citep[see, for example,][]{cape_2019}:
\begin{align}
	\norm{\boldsymbol{M}}_2 \leq \norm{\boldsymbol{M}}_F \leq \sqrt{\rank(\boldsymbol{M})}\norm{\boldsymbol{M}}_2, \qquad \norm{\boldsymbol{M}}_{2,\infty} \leq \norm{\boldsymbol{M}}_2 \leq \sqrt{m}\norm{\boldsymbol{M}}_{2,\infty}.
\end{align}
For an $n \times m$ matrix $\boldsymbol{M}$ of rank $d$, we write its skinny singular value decomposition (SVD) as $\boldsymbol{M} = \boldsymbol{U}_{\boldsymbol{M}}\boldsymbol{\Sigma}_{\boldsymbol{M}}\boldsymbol{V}_{\boldsymbol{M}}^\top$, and we define its incoherence parameter as
$\mu(\boldsymbol{M}) =
\max\{ \tfrac{n}{d}\,\lVert \boldsymbol{U}_{\boldsymbol{M}}\rVert^2_{2,\infty},\allowbreak
\tfrac{m}{d}\,\lVert \boldsymbol{V}_{\boldsymbol{M}}\rVert^2_{2,\infty} \allowbreak\}$
and its condition number as $\kappa(\boldsymbol{M}) = \sigma_1(\boldsymbol{M}) / \sigma_d(\boldsymbol{M})$.

We denote by $\mathrm{GL}(d)$ the general linear group of dimension $d$, which is the collection of all $d \times d$ invertible matrices. We denote by $\mathbb{O}(d)$ the orthogonal group of dimension $d$, which is the collection of all matrices $\boldsymbol{O} \in \mathbb{R}^{d \times d}$ for which $\boldsymbol{O}^\top \boldsymbol{O} = \boldsymbol{O}\boldsymbol{O}^\top = \boldsymbol{I}_d$, where we use $\boldsymbol{I}_d$ to denote the identity matrix of shape $d \times d$. For a collection of matrices $\boldsymbol{M}_1,\dots, \boldsymbol{M}_r \in \mathbb{R}^{m \times n}$, we denote their vertical stacking by $[\boldsymbol{M}_1\mid\cdots\mid\boldsymbol{M}_r] \in \mathbb{R}^{rm \times n}$ and their horizontal stacking by $[\boldsymbol{M}_1,\dots, \boldsymbol{M}_r] \in \mathbb{R}^{m \times rn}$.

\subsection{Probabilistic asymptotic notation}

For two real valued, non-stochastic functions $f$ and $g$, we write $f(n) = \mathcal{O}\{g(n)\}$ and $g(n) = \Omega\{f(n)\}$ as $n \to \infty$ if there exist $N^*\in\mathbb{N}$ and $C^* > 0$ such that $f(n) \leq C^* g(n)$ for all $n > n^*$. If the bound is tight, we write instead $f(n) = o\{g(n)\}$ and $g(n) = \omega\{f(n)\}$. Furthermore, we write $f(n) = \Theta\{g(n)\}$ if $f(n) = \mathcal{O}\{g(n)\}$ and $g(n) = \mathcal{O}\{f(n)\}$. The theoretical results in this work will be shown to hold probabilistically, and in what follows we define the necessary terminology. 

An event $E$ is said to hold \textit{almost surely} if $\mathbb{P}(E) = 1$. We say that an event $E_n$, depending on $n\in\mathbb{N}$,  occurs with \textit{overwhelming probability} if, for any $\gamma>0$, there exists a finite $C_\gamma > 0$ such that $\mathbb{P}(E_n) \geq 1 - C_\gamma n^{-\gamma}$, with $C_\gamma$ independent of $n$ \citep[see, for example,][]{tao2010}. Given a real-valued function $f$ and a family of random variables $\{Z_n\}_{n\in\mathbb{N}}$, each defined on some space $(\Omega_n, \mathcal{F}_n, \mathbb{P}_n)$, we write $|Z_n| = \mathcal{O}_{\mathbb{P}_n}\{f(n)\}$ if 
there exists $n^* \in \mathbb{N}$ and a finite $C>0$ such that the event $|Z_n| \leq C f(n)$ holds with overwhelming probability for all $n \geq n^*$. 

As noted in \cite{tao2010}, with overwhelming probability implies eventual almost sure bounds, provided only countably many unions are taken. If we consider the sequence $\{Z_n\}_{n\in\mathbb{N}}$ on the product space $(\Omega,\mathcal{F}, \mathbb{P}) = \prod_{n\in\mathbb{N}}(\Omega_n, \mathcal{F}_n, \mathbb{P}_n)$, then a sample $\omega = (\omega_1,\omega_2,\dots)\in \Omega$ encodes a trajectory $Z_1(\omega_1),Z_2(\omega_2),\dots$, representing the values of $\{Z_n\}_{n\in\mathbb{N}}$ for the sample $\omega\in\Omega$. Since the bound holds for any $\gamma > 0$, we may choose $\gamma > 1$ so that $\sum_n n^{-\gamma} < \infty$. In this setting, Borel-Cantelli implies that overwhelming-probability bounds, combined with polynomial cardinality of events, upgrades our statements to \textit{almost sure eventual bounds}: with probability one, there exists an $n^*(\omega)$ such that $|Z_n(\omega_n)| \leq Cf(n)$ for all $n \geq n^*(\omega)$, in which case we say $|Z_n| = \mathcal{O}\{f(n)\}$ \textit{eventually almost surely}.

%% file: sections/model.tex
\section{Multiplex inhomogeneous Poisson process random dot product graphs}
\label{sec:model}

In this work, we propose a model for a point process on a multiplex graph $\mathcal{G} = (\mathcal{V}, \{\mathcal{E}_1,\dots,\mathcal{E}_L\})$ with $L\in\mathbb{N}$ layers, where $\mathcal{V}$ is a node set with cardinality $|\mathcal{V}| = N$ shared across all layers, and $\mathcal{E}_\ell\subseteq\mathcal{V}\times\mathcal{V}$ denotes the edge set for layer $\ell \in [L]$. Arrivals of events on the network are viewed as observations of a marked point process consisting of a stream of quadruples $(i_k, j_k, \ell_k, t_k)\in %\mathcal{E}_{\ell_k} 
\mathcal{V} \times \mathcal{V} \times [L] \times\mathbb R_+,\ k=1,2,\dots$, denoting directed interactions from node $i_k$ to node $j_k$ in layer $\ell_k$ at time $t_k$, where $t_k\leq t_{k^\prime}$ for $k<k^\prime$. This formulation induces a counting process on each edge $(i,j)$ and layer $\ell$, defined as: %denoted by $N_{\ell ij}(\cdot)$, where, for all $\ell \in [L]$ and all $(i,j)\in\mathcal{E}_\ell$, 
\begin{equation}
	N_{\ell ij}(t) := \sum_{k=1}^\infty 
	\mathbb I_{\{(\ell,i,j)\}}\{(i_k,j_k,\ell_k)\}\mathbb I_{(0,t]}(t_k).
	%\mathbb{I}\{(i_\ell,j_\ell,t_\ell) \in \{i\}\times\{j\}\times [0,t)\},
    \label{eq:counting_process}
\end{equation} 
We write $\boldsymbol{N}(t) = [\boldsymbol{N}^{(\ast,1)}(t),\dots,\boldsymbol{N}^{(\ast,L)}(t)] \in \mathbb{N}^{N \times NL}$ to be the full matrix-valued network counting process, where $\boldsymbol{N}^{(\ast,\ell)}(\cdot) \in \mathbb{N}^{N\times N}$ is the matrix of counting processes for all possible edges in layer $\ell \in [L]$, with $(i,j)$-th entry given by $N_{\ell ij}(\cdot)$, the counting process for the edge $(i,j)$ on layer $\ell$. 
Furthermore, we define the intensity function associated to the counting process $N_{\ell ij}(\cdot)$ as 
\begin{equation}
    \lambda_{\ell ij}(t)=\lim_{h\downarrow0}\frac{\mathbb{E}\left\{ N_{\ell ij}(t+h) - N_{\ell ij}(t) \mid \mathcal{F}_t \right\}}{h},
\end{equation}
where $\mathcal{F}_t=\{(i_k,j_k,\ell_k,t_k): t_k<t\}$ denotes the history of the process in $[0,t)$.

The main contribution of this work is a model for intensity functions of a multi-layer network point process, that we call the \textit{multiplex inhomogeneous Poisson process dot product graph (MIPP-DPG)}, defined in Section~\ref{sec:def_MIPP}. Additionally, we propose an inference procedure based on spectral embedding methods in Section~\ref{sec:est_for_MIPP-RDPG}, with convenient asymptotic properties that will be illustrated in Section~\ref{sec:theory}.

\subsection{Defining the MIPP-DPG} \label{sec:def_MIPP}

We assume the point processes to be inhomogeneous Poisson, implying that:
\begin{equation}
N_{\ell ij}(t) - N_{\ell ij}(s) \sim\mathrm{Poisson}\left(\int_{s}^t \lambda_{\ell ij}(u)\mathrm{d}u\right).
\label{eq:inhomogeneous_poisson}
\end{equation}
for all $i,j\in\mathcal{V},\ \ell\in[L]$ and $0\leq s < t$, with independent increments across disjoint time intervals.
Under this framework, we assume that each node $i \in \mathcal{V}$ in layer $\ell \in [L]$ has two latent positions %in some low-dimensional space 
of dimension $d \ll N$, called $X_i(t) \in \mathbb{R}^d$ and $Y_{\ell i} \in \mathbb{R}^d$. The latent position $X_i(t)$ is a time-varying representation that is shared across layers, whereas $Y_{\ell j}$ a static position that is unique to the layer. The intensity of the Poisson process on %$\mathcal{E}_{\ell ij}$, $\lambda_{\ell ij}(t)$, 
the edge $(i,j)\in\mathcal{V}\times\mathcal{V}$ on layer $\ell\in[L]$ is assumed to arise from an inner product of these positions, so that 
\begin{equation}
\lambda_{\ell ij}(t) = X_i(t)^\top Y_{\ell j}.
\end{equation}
The latent positions are thus required to satisfy that $X_i(t)^\top Y_{\ell j} > 0$ for all %possible quadruples 
$(i, j, \ell, t) \in\mathcal{V} \times\mathcal{V}\times [L] \times \mathcal{T}$, where $\mathcal{T}=[0,T]$ is an observation interval. As noted in Section~\ref{sec:background}, we %work in a labeled setting and thus 
consider the latent positions to be unobserved but non-random quantities. We %store 
denote the %time-varying intensities 
intensity functions for the full network as $\boldsymbol{\Lambda}(t) = [\boldsymbol{\Lambda}^{(\ast,1)}(t),\dots,\boldsymbol{\Lambda}^{(\ast,L)}(t)] \in \mathbb{R}_+^{N \times NL}$, where $\boldsymbol{\Lambda}^{(\ast,\ell)}$ is the $N\times N$ matrix of intensities for layer $\ell \in [L]$, with $(i,j)$-th entry corresponding to $\Lambda^{(\ast,\ell)}_{ij}(t) = \lambda_{\ell ij}(t)$. We can compactly express $\boldsymbol{\Lambda}(t)$ at time $t\in\mathcal{T}$ as a matrix product $\boldsymbol{\Lambda}(t) = \boldsymbol{X}(t)\boldsymbol{Y}^\top$, where we define $\boldsymbol{X}(t) = [X_1(t)^\top \mid \cdots \mid X_n(t)^\top] \in \mathbb{R}^{N \times d}$ and $\boldsymbol{Y} = [Y^{\top}_{11} \mid \cdots \mid Y^{\top}_{1N}\mid Y^{\top}_{21} \mid \cdots \mid Y^{\top}_{LN}] \in \mathbb{R}^{NL \times d}$. %The latent positions are constrained to be such that $(\boldsymbol{X}(t)\boldsymbol{Y}^\top)_{ij} \geq 0$ for all $t \in \mathcal{T}$, $i \in [N]$ and $j \in [NL]$. 
A network point process %$\mathcal{F}_T=\left\{ (i_k, j_k, \ell_k, t_k) \right\}_{k=1}^Q$ 
that arises in this way is said to be a %\textit{multiplex inhomogeneous Poisson process dot product graph (MIPP-DPG)}.
MIPP-DPG, and we write:
\begin{equation}
%\mathcal{F}_T 
\boldsymbol{N}(t) \sim \text{MIPP-DPG}\left\{\boldsymbol{X}(t),\boldsymbol{Y}\right\}.
\end{equation}

%This is much like the model of \cite{baum2025}, except that here the time-varying latent position is continuous and their inner-product is not constrained to lie in $[0,1].

\subsection{A spectral estimation procedure for the MIPP-DPG}\label{sec:est_for_MIPP-RDPG}

Given quadruples $(i_k, j_k, \ell_k, t_k)\in %\mathcal{E}_{\ell_k} 
\mathcal{V} \times \mathcal{V} \times [L] \times\ \mathcal{T},\ k=1,2,\dots$, observed on a time interval $\mathcal{T}$, the inferential objective under the MIPP-DPG is to estimate the latent positions $X_i(t)$ and $Y_{i\ell}$ for all $i\in[N],\ \ell\in[L],\ t\in \mathcal{T}$. 
Without loss of generality, we set $\mathcal{T}=[0,1]$. 
For this inferential task, we resort to spectral methods for dynamic multiplex graphs observed at discrete time points, by leveraging doubly unfolded adjacency spectral embedding \citep[DUASE;][]{baum2025}, a dynamic extension of the UASE approach of \cite{jones_2021} for multiplex networks. 

\begin{definition}[Doubly unfolded adjacency spectral embedding, DUASE; \citet{baum2025}]
	Consider a set of adjacency matrices $\{\boldsymbol{A}_{m\ell}\}_{m\in[M],\ell \in [L]}$, for $M,L \in \mathbb{N}$, where $\boldsymbol{A}_{ml} \in \{0,1\}^{N \times N}$ for all pairs $(m,l) \in [M] \times [L]$, where $m$ and $\ell$ index time and layers, respectively. Define the doubly unfolded adjacency matrix as
	\begin{equation*}
		\boldsymbol{A} = \begin{bmatrix}
			\boldsymbol{A}_{11} & \cdots & \boldsymbol{A}_{1L} \\
			\vdots & \ddots & \vdots \\
			\boldsymbol{A}_{M1} & \cdots & \boldsymbol{A}_{ML}
		\end{bmatrix} \in \{0,1\}^{NM \times NL}.
	\end{equation*}
	Consider the singular value decomposition
	\begin{equation*}
		\boldsymbol{A} = \boldsymbol{U}\boldsymbol{\Sigma}\boldsymbol{V}^\top + \boldsymbol{U}_\perp \boldsymbol{\Sigma}_\perp \boldsymbol{V}_\perp^\top,
	\end{equation*}
	where $\boldsymbol{\Sigma} \in \mathbb{R}^{d \times d}$ is a diagonal matrix containing the $d$ largest singular values of $\boldsymbol{A}$, $\boldsymbol{U} \in \mathbb{R}^{NM \times d}$ and $\boldsymbol{V} \in \mathbb{R}^{NL \times d}$ contain the corresponding left and right singular vectors, respectively, and $\boldsymbol{\Sigma}_\perp, \boldsymbol{U}_\perp, \boldsymbol{V}_\perp$ contain the remaining singular values and left and right singular vectors, respectively. Then, the doubly unfolded adjacency spectral embedding of $\{\boldsymbol{A}_{m\ell}\}_{m \in [M], \ell \in [L]}$ into $\mathbb{R}^d$ is
	\begin{equation*}
		\hat{\boldsymbol{X}} = \boldsymbol{U}\boldsymbol{\Sigma}^{1/2} \in \mathbb{R}^{NM \times d} \quad \text{and} \quad \hat{\boldsymbol{Y}} = \boldsymbol{V}\boldsymbol{\Sigma}^{1/2} \in \mathbb{R}^{NL \times d}.
	\end{equation*}
	We refer to $\hat{\boldsymbol{X}}$ and $\hat{\boldsymbol{Y}}$ as the left and right doubly unfolded adjacency spectral embeddings, respectively, and we write $(\hat{\boldsymbol{X}}, \hat{\boldsymbol{Y}})=\mathrm{DUASE}(\boldsymbol{A})$.
\end{definition}

%DUASE is shown to consistently recover the latent positions of nodes (in the sense of the two-to-infinity norm) in the setting where a sequence of binary multiplex networks are observed. In this work, we consider how DUASE can be used to analyse a MIPP-RDPG. 

In the case that each $\boldsymbol{A}_{m\ell}$ is a random dot product graph with $\mathbb{E}\left\{\boldsymbol{A}_{m\ell}\right\} %\equiv \mathbb{E}\left\{\boldsymbol{A}^{(m,\ell)}\right\} 
= \boldsymbol{X}^{(m,\ast)}\boldsymbol{Y}^{(\ell,\ast) \ \top}$, for some $\boldsymbol{X} \in \mathbb{R}^{NM \times d}$ and $\boldsymbol{Y} \in \mathbb{R}^{NL \times d}$, DUASE can be shown to consistently recover these latent positions  at each discrete time point and layer with overwhelming probability in the sense of the two-to-infinity norm as $N\to\infty$. %In particular, $\hat{\boldsymbol{X}}^{(m,\ast)}$ and $\hat{\boldsymbol{Y}}^{(\ell,\ast)}$ of shapes $N \times d$, estimate the latent positions at each time point and layer and converge with overwhelming probability in two-to-infinity norm as $N \to \infty$. This result is formally stated in Theorem 1 of \cite{baum2025}.
In this work, we propose to use a procedure based on DUASE to obtain estimators $\hat{\boldsymbol{X}}$ and $\hat{\boldsymbol{Y}}$ for the latent positions $\boldsymbol{X}(t)$ and $\boldsymbol{Y}$ of the MIPP-DPG model, proving almost sure consistency in the sense of the two-to-infinity norm of the proposed estimator. 

The DUASE estimator is not immediately amenable to inference within the MIPP-DPG modelling framework due to the continuous-time nature of $\boldsymbol{X}(t)$ and that the graphs under consideration are weighted. As such, we propose to discretise the observation window $\mathcal{T}$ by dividing it into $M$ bins $\{B_m\}_{m=1}^M$, with $B_m := ((m-1)/M, m/M]$, where we let the number of bins $M$ potentially depend on the number of nodes $N$. 
We construct estimates $\hat{\boldsymbol{X}}$ and $\hat{\boldsymbol{Y}}$ for the MIPP-DPG latent positions by applying DUASE on a histogram estimate of $\boldsymbol{\Lambda}$, %written as a matrix $\hat{\boldsymbol{\Lambda}} \in \mathbb{R}^{NM \times NL}$, whose $m$-th set of $N$ rows is the histogram estimate to $\boldsymbol{\Lambda}$ on $B_m$. A histogram estimator is 
representing the simplest possible %realistic 
approach. Similarly to \cite{modell2024}, we argue that, by providing theoretical results for the case of this simple estimator, we can intuitively justify that our guarantees will be at least as strong in the case that more sophisticated estimation procedures are used.
%In the $m$-th bin, we 
In particular, we construct a histogram estimate of $\lambda_{\ell ij}(t)$ as 
\begin{equation}
\hat{\lambda}_{\ell ij}(t) = \sum_{m=1}^M \hat{\lambda}_{\ell ij}^m \mathbb{I}_{B_m}(t),
\end{equation}
where $\hat{\lambda}_{\ell ij}^m$ provides a point estimate for $\lambda_{\ell ij}(t)$ on $B_m$, defined as
$\hat{\lambda}_{\ell ij}^m := 
M [N_{\ell ij}(m/N) - N_{\ell ij}((m-1)/M)]$,
%M N_{\ell ij}(B_m)$, 
where $N_{\ell ij}(\cdot)$ is the counting process in \eqref{eq:counting_process}.
%with $\hat{\lambda}_{\ell ij}^m$ providing a point estimate for $\lambda_{\ell ij}(t)$ on $B_m$. 
All components of the histogram estimators for the edge intensities are then stacked into the block matrix $\hat{\boldsymbol{\Lambda}}$ defined as
\begin{equation}
	\hat{\boldsymbol{\Lambda}} = \begin{bmatrix}
		\hat{\boldsymbol{\Lambda}}^{(1,1)} & \cdots & \hat{\boldsymbol{\Lambda}}^{(1,L)} \\
		\vdots & \ddots & \vdots \\
		\hat{\boldsymbol{\Lambda}}^{(M,1)} & \cdots & \hat{\boldsymbol{\Lambda}}^{(M,L)}
	\end{bmatrix} \in \mathbb{R}^{NM \times NL},
\end{equation} 
with each block matrix $\hat{\boldsymbol{\Lambda}}^{(m, \ell)} \in \mathbb{R}^{N \times N}$ having $(i,j)$-th entry $\hat{\Lambda}^{(m,\ell)}_{ij} = \hat{\lambda}_{\ell ij}^m$.
The matrix $\hat{\boldsymbol{\Lambda}}^{(m,\ell)}$ is used as an estimate $\boldsymbol{\Lambda}^{(\ast,\ell)}(t)$ for all $t \in B_m$. 
Importantly, from the inhomogeneous Poisson process assumption in \eqref{eq:inhomogeneous_poisson}, each $\hat{\lambda}_{\ell ij}^m/M$ is a Poisson random variable with mean $\int_{B_m}\lambda_{\ell ij}(s) \dd s$. This fact allows us to construct the matrix %$\bar{\boldsymbol{\Lambda}}$, defined as
\begin{equation}
	\bar{\boldsymbol{\Lambda}} = \begin{bmatrix}
    		\bar{\boldsymbol{\Lambda}}^{(1,1)} & \cdots & \bar{\boldsymbol{\Lambda}}^{(1,L)} \\
		\vdots & \ddots & \vdots \\
		\bar{\boldsymbol{\Lambda}}^{(M,1)} & \cdots & \bar{\boldsymbol{\Lambda}}^{(M,L)}
	\end{bmatrix} \in \mathbb{R}^{NM \times NL},
\end{equation} 
with block matrices $\bar{\boldsymbol{\Lambda}}^{(m,\ell)}\in\mathbb{R}^{N\times N}$ having entries $\bar{\Lambda}^{(m,l)}_{ij} = \bar{\lambda}_{\ell ij}^m$, where we define $\bar{\lambda}^m_{\ell ij} =: M\int_{B_m}\lambda_{\ell ij}(s) \dd s$. 
Note that
$\bar{\boldsymbol{\Lambda}} = \mathbb{E}\{\hat{\boldsymbol{\Lambda}}\}$, 
and additionally %, writing 
%$\bar{\boldsymbol{\Lambda}}^{(m,\ast)} = [\bar{\boldsymbol{\Lambda}}^{(m,1)},\dots,\bar{\boldsymbol{\Lambda}}^{(m,L)}]\in\mathbb{R}^{N\times NL}$, 
this time-averaged matrix is seen to arise from a dot product:
\begin{equation}
	\bar{\boldsymbol{\Lambda}}^{(m,\ast)} = M\int_{B_m} \boldsymbol{X}(t)\boldsymbol{Y}^\top \dd t = \left(M\int_{B_m}\boldsymbol{X}(t)\dd t\right)\boldsymbol{Y}^\top := \tilde{\boldsymbol{X}}^{(m,\ast)} \boldsymbol{Y}^\top,
\end{equation}
where we write the average of $\boldsymbol{X}(t)$ over bin $B_m$ as
\begin{equation}
    \tilde{\boldsymbol{X}}^{(m,\ast)}:=M\int_{B_m}\boldsymbol{X}(t)\dd t \in \mathbb{R}^{N\times d}, 
\end{equation}
so that $\bar{\boldsymbol{\Lambda}} = \tilde{\boldsymbol{X}}\boldsymbol{Y}^\top$, with 
\begin{equation}
    \tilde{\boldsymbol{X}} = [\tilde{\boldsymbol{X}}^{(1,\ast)}\mid\cdots\mid\tilde{\boldsymbol{X}}^{(M,\ast)}] \in \mathbb{R}^{NM \times d}. 
\end{equation}
Building off of the theoretical properties of DUASE in the binary case presented in \cite{baum2025}, we will show that doubly unfolded left and right embeddings of $\hat{\boldsymbol{\Lambda}}$ to provide asymptotically good estimates of $\tilde{\boldsymbol{X}}$ and $\boldsymbol{Y}$ in the sense of the two-to-infinity norm. Therefore, we propose to use the following estimates for the latent positions under the MIPP-DPG model: 
\begin{equation}
    (\hat{\boldsymbol{X}}, \hat{\boldsymbol{Y}})=\mathrm{DUASE}(\hat{\boldsymbol{\Lambda}}),\ \qquad \hat{\boldsymbol{X}}\in\mathbb{R}^{NM\times d}, \quad \hat{\boldsymbol{Y}}\in\mathbb{R}^{NL\times d}.
    \label{eq:duase_estimates}
\end{equation}
%% In the next section, we derive the theoretical properties of our proposed estimator.

%% file: sections/theoretical_results.tex
\section{Theoretical results}
\label{sec:theory}

In this section, we present key theoretical results about recovery of the MIPP-DPG latent positions via the proposed DUASE estimator. In particular, we extend existing theoretical results for DUASE from the binary to the weighted setting, and demonstrate that the estimators recover the continuous dynamic when an appropriate limit is taken. These results are natural extensions of the results of \cite{jones_2021, gallagher2024, baum2025} to weighted, multiplex graphs observed in continuous time.

Consistency is derived in the sense of the two-to-infinity norm, which is a particularly meaningful metric in the context of our model as it is the natural norm to capture the maximum error between the set of estimates of the latent positions and the truth under the MIPP-DPG. Consider DUASE estimates $\hat{\boldsymbol{Y}}$ for $\boldsymbol{Y}$, and $\hat{\boldsymbol{X}}$ for $\boldsymbol{X}(t)$ across the set of bins, such that $\hat{\boldsymbol{X}}^{(m,\ast)}$ estimates $\boldsymbol{X}(t)$ on $B_m$, \textit{cf.} \eqref{eq:duase_estimates}. We can look to bound the following norms as $N \to \infty$:
\begin{equation}
	\max_{m \in [M]}\sup_{t \in B_m}\lVert\hat{\boldsymbol{X}}^{(m,\ast)}\boldsymbol{W}^N_{X} - \boldsymbol{X}(t)\rVert_{2,\infty} \quad \text{and} \quad \lVert \hat{\boldsymbol{Y}}\boldsymbol{W}^N_{Y} - \boldsymbol{Y}\rVert_{2,\infty},
    \label{eq:target}
\end{equation} 
for sequences of matrices $\boldsymbol{W}^N_{X}, \boldsymbol{W}^N_{Y} \in \mathrm{GL}(d)$. Here, we talk of sequences as $\boldsymbol{W}^N_X$ and $\boldsymbol{W}^N_Y$ will change with $N$. $\boldsymbol{W}^N_X$ and $\boldsymbol{W}^N_Y$ are defined as the product of an orthogonal and a general linear transformation, 
\textit{cf.} Section \ref{app_sec:proof_thm_1}. They arise because the true latent positions are fundamentally unidentifiable in both scale and orientation. This is clear by noting that, for any $\boldsymbol{Q} \in \mathbb{O}(d)$, $\boldsymbol{X}(t)\boldsymbol{Q}$ and $\boldsymbol{Y}\boldsymbol{Q}$ give rise to the same inner product as the untransformed latent positions. Furthermore, scaling $\boldsymbol{X}(t)$ by $s$ and $\boldsymbol{Y}$ by $1/s$ and will result in the same inner product and thus the same intensity matrix. 

The second norm in \eqref{eq:target} is a standard target in related RDPG literature  \citep[see, for example,][]{athreya2018, jones_2021, gallagher2024, baum2025}, whereas the first is a continuous-time adaptation, similar in nature to that of \cite{modell2024}. In particular, a bound on the first norm would bound the ``worst case scenario'', and will account for errors introduced through discretisation. It should be emphasised that despite binning the observation windows, our theory will recover the continuous setting as we can allow $M\to \infty$, provided the rate is sufficiently slow in comparison with the growth of $N$. These statements will be made more precise shortly.

Before stating the main theoretical results, %around the %asymptotic consistency and limiting distribution
%asymptotic behaviour of the DUASE estimator for the MIPP-DGP,  
we %extensively 
discuss the necessary assumptions on the latent positions required to obtain these properties. 
We divide the assumptions into two classes, %stated in Assumptions~\ref{ass:thm1_variance} and~\ref{ass:thm1_bias} respectively, 
related to variance and bias of the proposed DUASE estimator for the latent positions. These assumption naturally occur from the following bias-variance decomposition of the bound on the first of the two norms in \eqref{eq:target}:
\begin{multline}
	\max_{m \in [M]}\sup_{t \in B_m} \lVert\hat{\boldsymbol{X}}^{(m,\ast)}\boldsymbol{W}^N_{X} - \boldsymbol{X}(t)\rVert_{2,\infty} \leq \\ \lVert\hat{\boldsymbol{X}}\boldsymbol{W}^N_{X,2} - \tilde{\boldsymbol{X}}\rVert_{2,\infty} + \max_{m \in [M]}\sup_{t \in B_m} \lVert\tilde{\boldsymbol{X}}^{(m,\ast)}\boldsymbol{W}^N_{X,1}- \boldsymbol{X}(t)\rVert_{2,\infty},
    \label{eq:bias_variance_decomposition}
\end{multline}
with $\boldsymbol{W}^N_X = \boldsymbol{W}^N_{X,2}\boldsymbol{W}_{X,1}^{N,\top}$, $\boldsymbol{W}^N_{X,1},\boldsymbol{W}^N_{X,2}\in\mathrm{GL}(d)$. Similarly, variance assumptions are needed to establish properties for the decay of the term $\lVert \hat{\boldsymbol{Y}}\boldsymbol{W}^N_2 - \boldsymbol{Y} \rVert_{2,\infty}$ in \eqref{eq:target}. Assumptions on the latent positions needed for the variance components of the bounds to decay in $N$ are summarised in Assumption~\ref{ass:thm1_variance}. 

%The assumptions needed to derive guarantees on the decay of the first of these terms are stated in Section \ref{sec:variance_assumptions}, and those for the second are in Section \ref{sec:bias_assumptions}.

% For the two-to-infinity norm on $\hat{\boldsymbol{Y}}$ as an estimator of $\boldsymbol{Y}$:
% \begin{equation}
% 	\lVert \hat{\boldsymbol{Y}}\boldsymbol{W}_2 - \boldsymbol{Y} \rVert_{2,\infty},
% \end{equation}
% the assumptions needed are stated in Section \ref{sec:variance_assumptions}. For the bound describing $\hat{\boldsymbol{X}}$ as an estimator of $\boldsymbol{X}(t)$, we will decompose our norm into bias and variance terms as:

%In the following, we state the necessary assumptions required to obtain our main theoretical results. We then detail the theoretical guarantees for inference using DUASE and a histogram estimator.

% \subsection{Main results}\label{sec:variance_assumptions}

\begin{assumption}[Variance assumptions]\label{ass:thm1_variance}
The latent positions $\boldsymbol{X}(t)$ and $\boldsymbol{Y}$ of the MIPP-DPG %model
satisfy %: %each of 
all the following assumptions:
\begin{enumerate}[label=\roman*.,ref=\theassumption.\roman*,leftmargin=*,nosep]
  \item\label{ass:bound_and_int}
    (Bounded and integrable) $\lVert \boldsymbol{X}(t)\rVert_{2,\infty}, \lVert \boldsymbol{Y}\rVert_{2,\infty}=\mathcal{O}(1)$ for $N\to\infty$, and $\boldsymbol{X}(t)$ is integrable on $\mathcal{T}$,
  \item\label{ass:moment_stab}
    (Stable moments) There exist $\boldsymbol{Q}_X, \boldsymbol{Q}_Y \succ 0$ such that, for $N\to\infty$, we have entrywise convergence:
    \[
      \frac{1}{NM_N}\tilde{\boldsymbol{X}}^\top \tilde{\boldsymbol{X}} \to \boldsymbol{Q}_X,\qquad
      \frac{1}{NL_N}\boldsymbol{Y}^\top \boldsymbol{Y} \to \boldsymbol{Q}_Y,
    \]
  \item\label{ass:M_L_growth}
    (Growth of $M_N$ and $L_N$) $M_N=o(N^{1/2}/\log^4 N)$ and $L_N=o(N^{1/2})$.
\end{enumerate}
\end{assumption}

Here we have changed notation to introduce $N$ subscripts onto $M$ and $L$ to emphasise their dependence on the number of nodes, $N$. Assumption~\ref{ass:bound_and_int} is on the behaviour of the integrability of the time-dependent latent positions and the boundedness of the positions in general. It is analogous to Assumption ii in \cite{gallagher2024} and Assumption 1 in \cite{modell2024}. Assumption~\ref{ass:moment_stab} is on the limiting behaviour of the collection of latent positions. Assumptions of this form in the case of random positions is standard in the literature \citep[see, for example][]{athreya2018,jones_2021, baum2025}. Assumption~\ref{ass:M_L_growth} is on the growth rate of the number of bins and the number of layers. These are analogous to the requirements of the growth of $T_n$ and $K_n$ in \cite{baum2025}. The required behaviour of $M_N$ and $L_N$ %in Assumption \ref{ass:M_L_growth} 
are such that the derived two-to-infinity norm bound decays in $N$. The number of bins $M_N$ should be interpreted as controlling the sparsity of $\hat{\boldsymbol{\Lambda}}$. % and it is thus interesting to note that the
The maximum growth rate of $M_N$ is analogous to the maximum decay rate of the sparsity parameter in the work of \cite{baum2025}. The number of layers $L_N$ need not to grow to obtain asymptotic consistency, whereas $M_N$ is only required to grow for the bias term to vanish, and is not required to grow for the recovery of each $Y_{\ell j}$.

We note that Assumptions~\ref{ass:bound_and_int}~and~\ref{ass:moment_stab} provide the following proposition (proved in Appendix~\ref{proof:prop1}) that characterises the behaviour of the matrix of latent positions in the limit. 

\begin{proposition}[Behaviour of latent position matrices]\label{prop:behaviour_latent_position_matrices}
    Under Assumptions \ref{ass:bound_and_int} and \ref{ass:moment_stab}, as $N \to \infty$, we have
	\begin{equation}
		\sigma_i(\tilde{\boldsymbol{X}}) = \Theta(M_N^{1/2}N^{1/2}),\quad  \sigma_i(\boldsymbol{Y}) = \Theta(L_N^{1/2}N^{1/2}), \quad \sigma_i(\bar{\boldsymbol{\Lambda}}) = \Theta(M_N^{1/2}L_N^{1/2}N),
	\end{equation}
	for all $i \in [d]$, and $\sigma_{d+1}(\bar{\boldsymbol{\Lambda}}) = 0$. Furthermore, we have that $\kappa(\tilde{\boldsymbol{X}}),\ \kappa(\boldsymbol{Y}),\ \mu(\tilde{\boldsymbol{X}}),\ \mu(\boldsymbol{Y}) = \mathcal{O}(1)$, where $\kappa$ and $\mu$ denote the condition number and incoherence parameter, respectively, as defined in Section \ref{sec:mathematical_backround}.
\end{proposition}

Under Assumption~\ref{ass:thm1_variance} (combined with additional technical assumptions detailed in Sections~\ref{sec:asymptotic_consistency}~and~\ref{sec:clt}), we derive theoretical results that demonstrate the asymptotic consistency of DUASE applied to the MIPP-DPG and a subsequent central limit theorem.

\subsection{Asymptotic consistency} \label{sec:asymptotic_consistency}

The first main theoretical result involves asymptotic consistency of the DUASE estimator for the MIPP-DPG latent positions. 
In order to prove the result, an additional assumption on the behaviour of the latent positions $\boldsymbol{X}(t),\ t\in\mathcal{T}$ is needed on the bias term arising from the decomposition in \eqref{eq:bias_variance_decomposition}. Two options are described in Assumption~\ref{ass:thm1_bias}.

\begin{assumption}[Bias assumptions]\label{ass:thm1_bias}
The latent positions $\boldsymbol{X}(t)$ of the MIPP-DPG model
satisfy \underline{one} of the following assumptions:
\begin{enumerate}[label=\roman*.,ref=\theassumption.\roman*,leftmargin=*,nosep]
  \item\label{ass:no_rotation_lipschitz_continuity}
    (Rotation-free Lipschitz continuity) For each $s,t\in \mathcal{T}$, and for each $N \in \mathbb{N}$, there exists a $K_N = o(M_N)$ such that:
	\begin{equation}
		\lVert \boldsymbol{X}(t) - \boldsymbol{X}(s) \rVert_{2,\infty} \leq K_N |t - s|,
	\end{equation}	

\item\label{ass:subspace_lipschitz_continuity}
  (Subspace Lipschitz continuity) Define the projector $\boldsymbol{P}(\cdot)$ as 
    \begin{equation}
		\boldsymbol{P}(t) = \boldsymbol{X}(t)\{\boldsymbol{X}(t)^\top \boldsymbol{X}(t)\}^{-1}\boldsymbol{X}(t)^\top.
    \end{equation}
  For each $s,t \in \mathcal{T}$, and for each $N \in \mathbb{N}$, there exists $K_{N,1}, K_{N,2} = o(M_N)$ such that
  \begin{itemize}
  \item (Subspace smoothness)
	%\begin{equation}
    $
		\lVert \boldsymbol{P}(t) - \boldsymbol{P}(s) \rVert_{2} \leq K_{N,1} |t - s| 
    $ 
    %\quad \text{(subspace smoothness)},
	%\end{equation}
	and 
	\item (Coordinate smoothness)
    %\begin{equation}
	$	
    \lVert \boldsymbol{P}(t)\{\boldsymbol{X}(t) - \boldsymbol{X}(s)\} \rVert_{2,\infty} \leq K_{N,2}|t-s|, 
    $
    \end{itemize}
    %\quad \text{(coordinate smoothness)},
	%\end{equation}
	
	%\end{equation}
\end{enumerate}
\end{assumption}

Assumption~\ref{ass:no_rotation_lipschitz_continuity} is a strong assumption that asserts that there is a global coordinate system in which the latent positions evolve smoothly. This is a statement on the orientation and scale up to an overall fixed right-orthogonal matrix. On the other hand, Assumption \ref{ass:subspace_lipschitz_continuity} is a statement in two parts on the dynamics of the latent positions $\boldsymbol{X}(t)$. The first part, involving subspace smoothness, says that the subspace spanned by the columns of $\boldsymbol{X}(t)$ moves Lipschitz-continuously in time, measured on the Grassmann manifold via projector distance. Intuitively, it corresponds to a bound on the angular velocity of the latent position directional structure. On the other hand, the second part, involving coordinate smoothness, is a bound on how far the coordinates at time $s$ are from the best representation in the subspace at time $t$. It gives a bound on how fast the coordinates move in the space at time $t$. Combined, the two conditions in Assumption~\ref{ass:subspace_lipschitz_continuity} give a rotation-invariant and interpretable notion of temporal smoothness for dynamic latent positions. 
Note that Assumption~\ref{ass:subspace_lipschitz_continuity} is strictly weaker than Assumption~\ref{ass:no_rotation_lipschitz_continuity}.

An important observation to make here is that the Lipschitz constants in Assumption \ref{ass:thm1_bias} as allowed to grow with $N$, provide they grow no faster than $M_N$. This weakens the assumption, as we are not demanding the existence of a universal constant to govern the smoothness, but instead allow the dynamics to become less smooth as the nextwork grows.

Under Assumptions~\ref{ass:thm1_variance}~and~\ref{ass:thm1_bias}, we can then establish asymptotic consistency. The result is stated in Theorem~\ref{thm:big_O}.

\begin{theorem}[Two-to-infinity norm bound]\label{thm:big_O}
	Suppose that $\boldsymbol{N}(t)$ is a counting process arising from a MIPP-DPG on $\mathcal{T}=(0,1]$ with $N$ nodes, $L_N$ layers, intensity matrix $\boldsymbol{\Lambda}(t)$ and latent position matrices $\boldsymbol{X}(t)$ and $\boldsymbol{Y}$, as defined in Section \ref{sec:def_MIPP}. Let $\hat{\boldsymbol{\Lambda}}$ be the histogram estimator of $\boldsymbol{\Lambda}$ using $M_N$ bins, and $\hat{\boldsymbol{X}}$ and $\hat{\boldsymbol{Y}}$ be estimators obtained as defined in Section \ref{sec:est_for_MIPP-RDPG}. Then, if Assumption \ref{ass:thm1_variance}, is met, there exists sequences of matrices $\boldsymbol{W}^N_{X},\boldsymbol{W}^N_{Y}\in \mathrm{GL}(d)$, such that
	\begin{align}
		\max_{m \in [M_N]} \sup_{t \in B_m}\lVert \hat{\boldsymbol{X}}^{(m,\ast)} \boldsymbol{W}^N_{X} - \boldsymbol{X}(t) \rVert_{2,\infty} &= \mathcal{O}\left(M_NL_N^{-1/2}N^{-1/2}\log^{3/2}(N)\right) + B_{N},\\ 
		\lVert \hat{\boldsymbol{Y}}\boldsymbol{W}^N_{Y} - \boldsymbol{Y}\rVert_{2,\infty} &= \mathcal{O}\left(M_N^{1/2}N^{-1/2}\log^{3/2}(N)\right),
	\end{align}
	eventually almost surely, where $B_N$ is:
	\begin{enumerate}
		\item $\mathcal{O}(K_NM_N^{-1})$ under Assumption \ref{ass:no_rotation_lipschitz_continuity},
		\item $\mathcal{O}((K_{N,1} + K_{N,2})M_N^{-1})$ under Assumption \ref{ass:subspace_lipschitz_continuity}.
	\end{enumerate}
\end{theorem}

The proof of Theorem \ref{thm:big_O} is provided in Appendix \ref{app_sec:proof_thm_1}. It should be noted that Theorem \ref{thm:big_O} quantifies that a growing $M_N$ increases the variance term of our estimator, but decreases the bias. This is an intuitive result as larger $M_N$ corresponds to a smaller bin width, thereby reproducing the classical bias–variance tradeoff in histogram density estimation. A similar result is observed in the non-asymptotic result of \cite{modell2024}.

\subsection{Central limit theorem} \label{sec:clt}

The second result, stated in Theorem \ref{thm:asymptotic_normality}, is a statement on the asymptotic normality of a studentised vector of differences between the estimated and true latent positions. In addition to Assumption~\ref{ass:thm1_variance}, a further weak technical assumption, stated in Assumption~\ref{ass:rank_prev_suff_dens}, is required.

\begin{assumption}[Rank preservation and sufficient density]\label{ass:rank_prev_suff_dens}
    There exists $N_1$ such that, whenever $N > N_1$, there exists two sets of $d$ indices $i_1,\dots,i_d \in [NM_N]$ and $j_1,\dots,j_d \in [NL_N]$ such that: 
    \begin{enumerate}
        \item the set $\{Y_{j_k,\ast}\}_{k=1}^d$ are linearly independent and $\tilde{X}_{i,\ast}^{(m,\ast)}Y_{j_k,\ast}^\top > 0$ for all $k=1,\dots,d$,
        \item the set $\{\tilde{X}_{i_k,\ast}\}_{k=1}^d$ are linearly independent and $\tilde{X}_{i_k,\ast} Y_{i,\ast}^\top > 0$ for all $k=1,\dots,d$.
    \end{enumerate} 
\end{assumption}
Assumption~\ref{ass:rank_prev_suff_dens} ensures that, for large enough $N$, for any node in the network we can always find two sets of $d$ indices (not necessarily distinct) such that the the intensities on the edges from that node to these indices are positive, and the corresponding latent positions are linearly independent. The existence of such a finite set of indices in the limit of growing $N$ is intuitively reasonable, and guarantees that matrices arising in our central limit theorem are invertible. Under this additional assumption, then the following studentised CLT holds:

\begin{theorem}[Studentised asymptotic normality]\label{thm:asymptotic_normality}
	Suppose that $\boldsymbol{N}(t)$ is a counting process arising from a MIPP-RDPG on $(0,1]$ with $N$ nodes, $L_N$ layers, intensity matrix $\boldsymbol{\Lambda}(t)$ and latent position matrices $\boldsymbol{X}(t)$ and $\boldsymbol{Y}$, as defined in Section \ref{sec:def_MIPP}. Let $\hat{\boldsymbol{\Lambda}}$ be the histogram estimator of $\boldsymbol{\Lambda}$ using $M_N$ bins, and $\hat{\boldsymbol{X}}$ and $\hat{\boldsymbol{Y}}$ be estimators obtained as defined in Section \ref{sec:est_for_MIPP-RDPG}. Then, if Assumptions~\ref{ass:thm1_variance} and \ref{ass:rank_prev_suff_dens} are met, there exists sequences of matrices $\boldsymbol{W}^N_{X},\boldsymbol{W}^N_{Y}\in \mathrm{GL}(d)$, such that, for $N \to \infty$, 
    \begin{gather}
        N^{1/2}L_N^{1/2}\left(\boldsymbol{Q}_{\boldsymbol{X}}^{-1} \boldsymbol{C}_{i,m}^{N}\boldsymbol{Q}_{\boldsymbol{X}}^{-1}\right)^{-1/2}(\hat{\boldsymbol{X}}^{(m,\ast)} \boldsymbol{W}^N_{X} - \tilde{\boldsymbol{X}}^{(m,\ast)})_{i,\ast}^\top \rightarrow \mathcal{N}(0, \boldsymbol{I}_d),\\
        N^{1/2}M_N^{1/2}\left(\boldsymbol{Q}_{\boldsymbol{Y}}^{-1} \boldsymbol{D}_{i,\ell}^{N}\boldsymbol{Q}_{\boldsymbol{Y}}^{-1}\right)^{-1/2}(\hat{\boldsymbol{Y}}^{(m,\ast)} \boldsymbol{W}^N_{Y} - \tilde{\boldsymbol{Y}}^{(m,\ast)})_{i,\ast}^\top \rightarrow \mathcal{N}(0, \boldsymbol{I}_d),
    \end{gather}
    in distribution, where the matrices $\boldsymbol{C}_{i,m}^{N}$ and $\boldsymbol{D}_{i,\ell}^{N}$ take the form
    \begin{align}
        \boldsymbol{C}_{i,m}^{N} = \frac{1}{NL_N}\sum_{j=1}^{NL_N}(\tilde{X}^{(m,\ast)}_{i,\ast}Y_{j,\ast}^\top) Y_{j,\ast}^\top Y_{j,\ast}, &&
        \boldsymbol{D}_{i,\ell}^N = \frac{1}{NM_N}\sum_{j=1}^{NM_N}(\tilde{X}_{j,\ast} Y_{i,\ast}^{(\ell, \ast)\ \top})\tilde{X}_{j,\ast}^\top\tilde{X}_{j,\ast},
    \end{align}
    and are invertible provided Assumption \ref{ass:rank_prev_suff_dens} holds.
\end{theorem}

The result is proved in Appendix~\ref{sec:proof_thm2}.
It should be noted that $\lim_{N \to \infty}\boldsymbol{C}_{i,m}^N$ and $\lim_{N \to \infty}\boldsymbol{D}_{i,m}^N$
need not necessarily exist without further strong assumptions on the third moments of our latent positions. Assumptions~\ref{ass:bound_and_int}~and~\ref{ass:moment_stab} are sufficient to ensure that these limits are finite, but Assumption~\ref{ass:rank_prev_suff_dens} is required to ensure that they are invertible. As the limit need not exist, Theorem~\ref{thm:asymptotic_normality} is stated in a studentised form.

%% file: sections/simulations.tex
\section{Simulation studies}
\label{sec:simulations}

In order to demonstrate the performance of the proposed estimator for the latent position of the MIPP-DPG, we construct two simulation studies. Both studies use a block structure intensity matrix to examine recovery of an underlying group structure. Section \ref{sec:sim1} examines a smoothly evolving dynamic latent position and the interplay between network size and granularity in the recovery of the latent positions. The study of Section \ref{sec:sim2} examines a similar setting, but where we introduce discontinuities into the latent positions to qualitatively assess how a violation of Assumption \ref{ass:thm1_bias} affects recovery.

\subsection{Block structure MIPP-DPG}
\label{sec:sim1}

We consider a simulation study with a smooth intensity matrix $\boldsymbol{\Lambda}(t)$ taking a block structure, based on an extension of the dynamic multiplex stochastic block of \cite{baum2025}. We assume that the nodes are assigned to one of $G_1\in\mathbb{N}$ groups when functioning as a source node, and to one of $G_2\in\mathbb{N}$ groups when acting as a destination. We define a collection of functions $\mu_g(t): \mathcal{T} \to \mathbb{R}^d$ for each $g \in [G_1]$, and constant vectors $\gamma^\ell_q \in \mathbb{R}^d$ for each $\ell \in [L]$ and $q \in [G_2]$. Each node $i \in \mathcal{V}$ is assigned to a group for its temporal dynamics $z_i \in [G_1]$ and to a group for its layer-specific behaviour $v^{\ell}_i \in [G_2]$. We then define the node-specific latent positions as $X_i(t) = \mu_{z_i}(t)$ and $Y_{\ell j} = \gamma^\ell_{v_j^\ell}$. In this example, we consider $d=2$ and functions $\mu_g(t)$ of the form:
\begin{equation}
    \mu_g(t) = \begin{bmatrix} c_{1,g} + R_g \sin(2\pi t + \theta_g) & c_{2,g} + R_g \cos(2\pi t + \theta_g)\end{bmatrix}^\top,
    \label{eqn:mu_g}
\end{equation}
where $c_{1,g}, c_{2,g}, R_g, \theta_g\in\mathbb{R}$ for $g\in[G_1]$ are group-dependent constants. 
%and projections $\gamma_q^\ell$ of the form:
For the layer-specific latent positions, we use: 
\begin{equation}
    \gamma_q^\ell = \begin{bmatrix} d^\ell_q + \cos(\phi^\ell_q) &  d^\ell_q + \sin(\phi^\ell_q)\end{bmatrix}^\top,
    \label{eqn:gamma_lq}
\end{equation}
where $d_q^\ell, \phi_q^\ell$ for $\ell \in [L]$ and $q\in[G_2]$ are again constants depending on the group.

\begin{figure}[p]
\centering

\begin{subfigure}{0.975\textwidth}
\centering
\includegraphics[height=0.425\textheight, keepaspectratio]{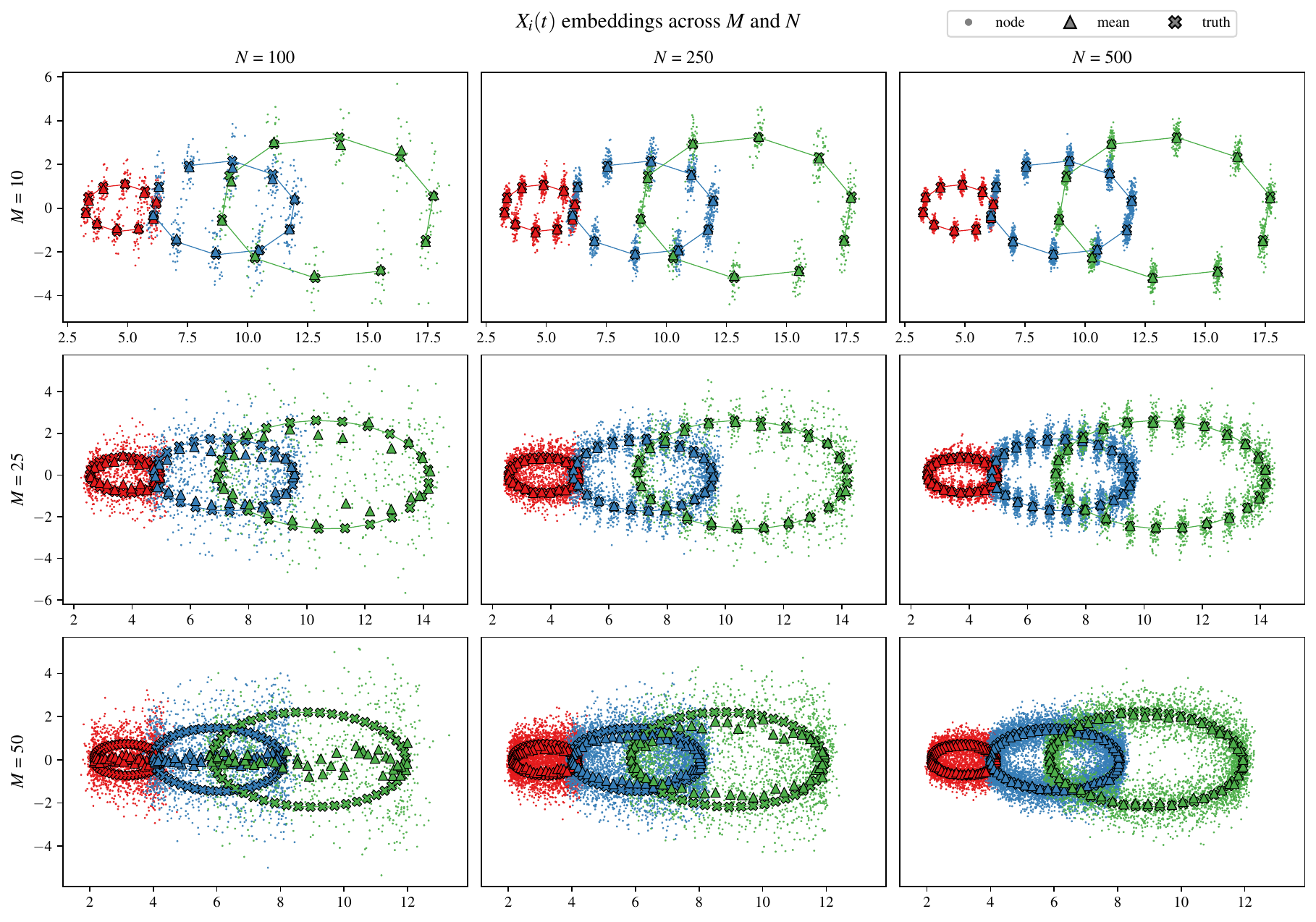}
\caption{$\hat{X}_i(t)$ for networks of different sizes and with different numbers of bins.}
\label{fig:X_estimates_N_M}
\end{subfigure}

\begin{subfigure}{0.975\textwidth}
\centering
\includegraphics[height=0.425\textheight, keepaspectratio]{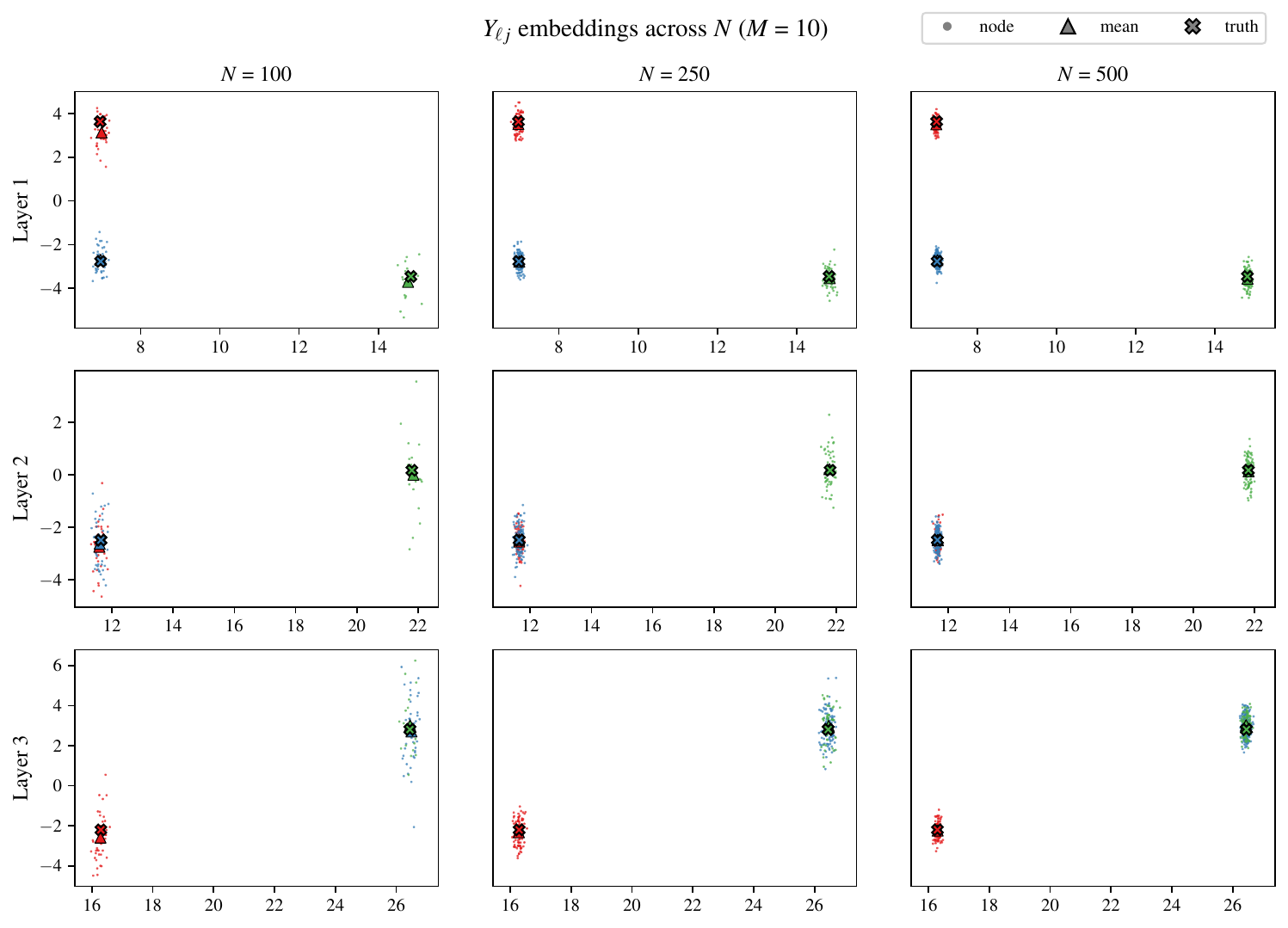}
\caption{$\hat{Y}_{\ell j}(t)$ for each layer for networks of different sizes, $M=10$.}
\label{fig:Y_estimates_N_M}
\end{subfigure}

\caption{Estimated and true latent positions for the simulation study described in Section~\ref{sec:sim1}, for different values of the number of nodes $N$ and number of bins $M$.}
\end{figure}

We examine the recovery of the latent positions $\{X_i(t)\}_{i=1}^N$ via the DUASE-based estimator in a graph with $L = 3$ layers generated using the group-specific positions in Equations~\eqref{eqn:mu_g}~and~\eqref{eqn:gamma_lq}. The nodes are distributed among $G_1 = 3$ groups, such that the first two groups each receive 40\% of the nodes, and the third group receives the remaining 20\%. For the layer groups, we take $G_2=3$, but merge groups 1 and 2 in the second layer and groups 2 and 3 in the third. For the group-specific time-varying latent positions $\mu_g(t)$, we take the following parameters for $g\in\{1,2,3\}$: $ R_g = 5g,\ c_{1,g} = 2R_g + 1,\ c_{2,g} = 2R_g + 1,\ \theta_g = g\pi$. On the other hand, for the layer-specific latent positions $\gamma_q^\ell$, in the first layer we take $d_q^1 = q, \phi_q^1 = \pi/q$, for the second we take $d_1^2 = d_2^2 = 2, d_3^2 = 4$ and $\phi_1^2 = \phi_2^2 = \pi / 2, \phi_3^2 = \pi / 3$, and for the third $d_1^3 = 3, d_2^3 = d_3^3 = 5$ and $\phi_1^3 = \pi / 2, \phi_2^3 = \phi_3^3 = \pi/6$. We simulate networks with $N\in\{100,200,500\}$ and run the DUASE inference procedure described in Section~\ref{sec:est_for_MIPP-RDPG} for $M\in\{10, 25, 50\}$ to examine the effect of decreasing bin width and increasing the number of nodes on the recovery of the latent positions. Results are reported in Figures~\ref{fig:X_estimates_N_M} and~\ref{fig:Y_estimates_N_M}.

% \begin{figure}[p]
% \centering
% \begin{subfigure}{0.975\textwidth}
% \centering
%     \caption{$\hat{X}_i(t)$ for networks of different sizes and with different numbers of bins.}
%     \label{fig:X_estimates_N_M}
%     \includegraphics[width=0.925\textwidth]{figures/simulations/overlapping/X_estimates_N_M.pdf}
% \end{subfigure}
% \begin{subfigure}{0.975\textwidth}
% \centering
%     \caption{$\hat{Y}_{\ell j}(t)$ for each layer for networks of different sizes, $M=10$.}
%     \label{fig:Y_estimates_N_M}
%     \includegraphics[width=0.925\textwidth]{figures/simulations/overlapping/Y_estimates_N_M.pdf}
% \end{subfigure}
% \caption{Estimated and true latent positions for the simulation study described in Section~\ref{sec:sim1}, for different values of the number of nodes $N$ and number of bins $M$.}
% \end{figure}

In each panel of Figure~\ref{fig:X_estimates_N_M}, the different coloured crosses show the true simulation values of $\mu_1(t), \mu_2(t)$ and $\mu_3(t)$ at the centres of each bin $B_m$. The small points show the estimates of each $X_i(t)$ for the bin $B_m$ and the triangles the group-specific means of these estimates. We have aligned the estimated latent positions with the truth using orthogonal Procrustes to aid visualisation. We see that increasing $M$ in general leads to worse estimates of the latent positions, but that the decrease in the estimation quality is smaller for larger $N$. This aligns with the statement of Theorem~\ref{thm:big_O}. In Figure~\ref{fig:Y_estimates_N_M}, we produce an analogous plot for the recovery of $\{\gamma_q^\ell\}_{q,\ell = 1}^3$, for the case when $M=10$. We see that the merging of groups in layers 2 and 3 is correctly detected. 

\subsection{Discontinuous MIPP-DPG}

\label{sec:sim2}

\begin{figure}[t]
\centering
\begin{subfigure}{0.975\textwidth}
\centering
    \caption{True generating functions $\mu_{g}(t)$ %for $X_i(t)$: plots of $\mu_{g,1}(t)$ against $\mu_{g,2}(t))$ 
    for $t \in (0,1]$}
    \label{fig:true_mu}
    \includegraphics[width=0.925\textwidth]{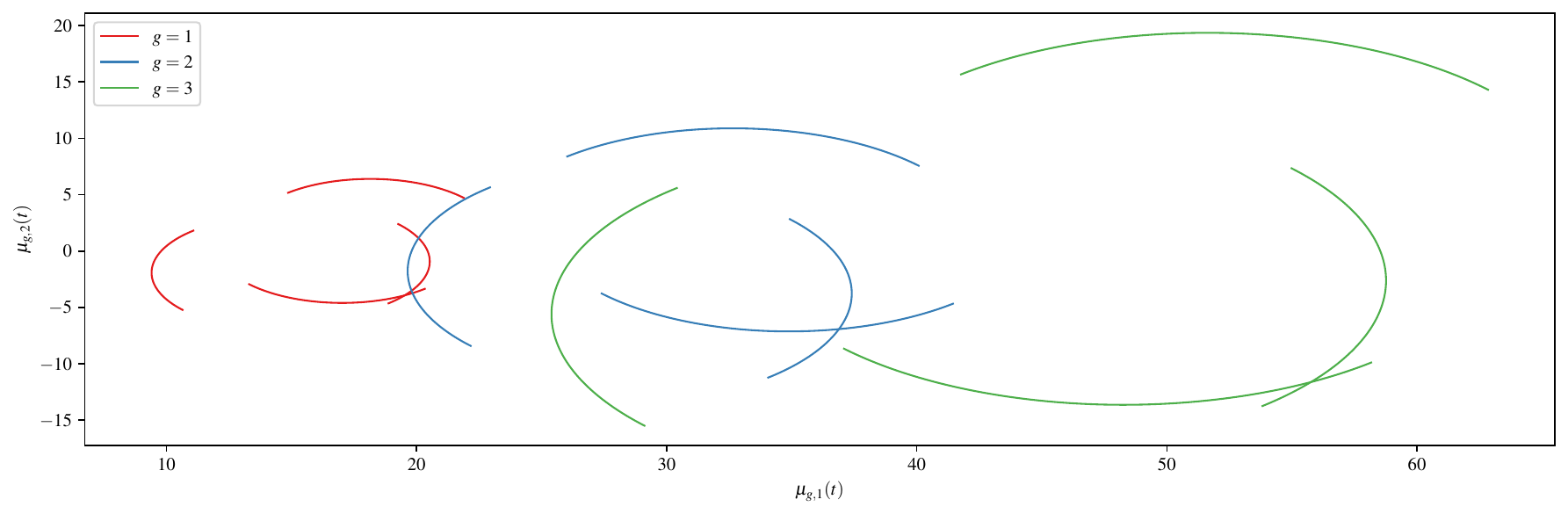}
\end{subfigure}
\begin{subfigure}{0.975\textwidth}
\centering
    \caption{$\hat{X}_i(t)$ for networks of different sizes for $M=50$}
    \label{fig:X_estimates_N_50}
    \includegraphics[width=0.925\textwidth]{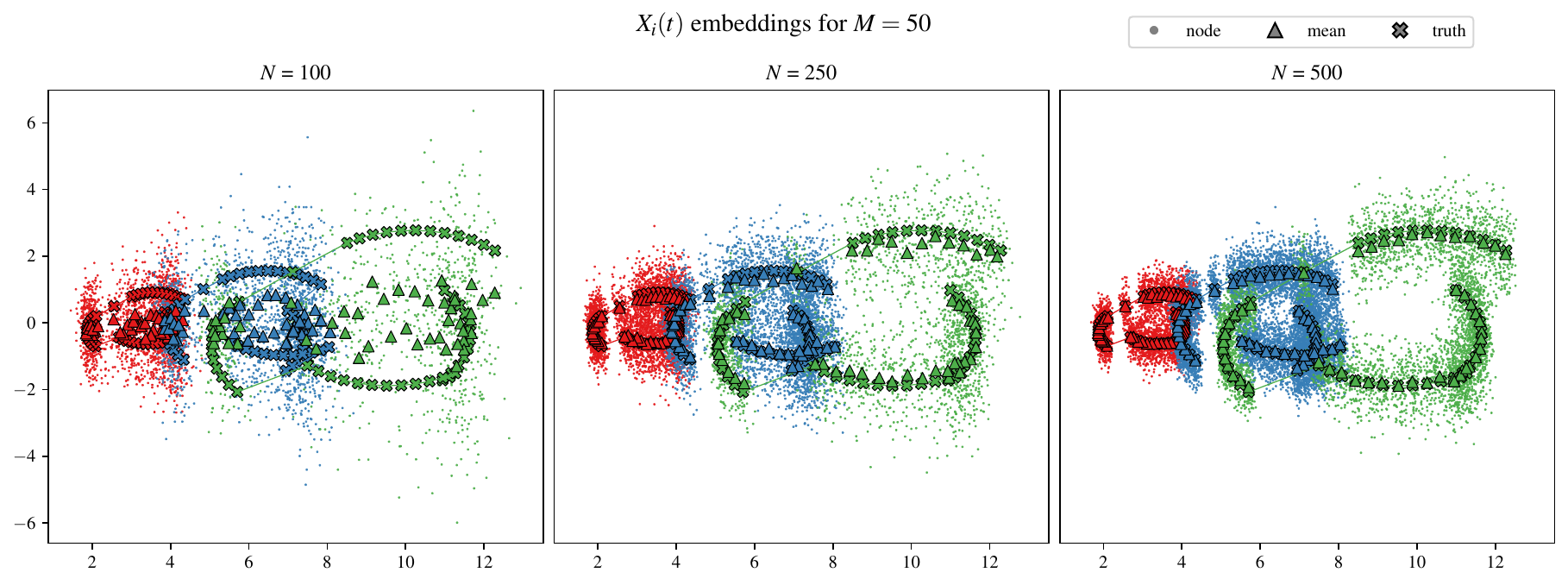}
\end{subfigure}
\caption{Estimated and true latent positions for the simulation study described in Section~\ref{sec:sim2}, for different values of the number of nodes $N$ and number of bins $M=50$.}
\end{figure}

We now consider discontinuities in our latent positions that violate Assumption \ref{ass:thm1_bias}. We examine a simulation study similar to Section~\ref{sec:sim1}, but with with an intensity matrix $\boldsymbol{\Lambda}(t)$ that is discontinuous. In this example, we again consider $d=2$, and functions $\mu_g(t)$ of the form
\begin{equation}
    \mu_g(t)
    =
    \begin{bmatrix}
        c_{1,g} + R_g \sin(2\pi t + \theta_g)
        &
        c_{2,g} + R_g \cos(2\pi t + \theta_g) + \delta(t)
    \end{bmatrix}^\top,
\end{equation}
where $c_{1,g}, c_{2,g}, R_g, \theta_g\in\mathbb{R}$ for $g\in[G_1]$ are the same as in Section \ref{sec:sim1}, and $\delta_g:\mathbb{R}_+\to\mathbb{R}_+$ is the following step function:
\begin{equation}
    \delta_g(t)
    =
    \begin{cases}
        0, & 0 < t \le \tau_1,\\
        0.4R_g, & \tau_1 < t \le \tau_2,\\
        -0.3R_g, & \tau_2 < t \le \tau_3,\\
        0.7R_g, & \tau_3 < t \le 1,
    \end{cases}
\end{equation}
for step locations $\tau_1,\tau_2,\tau_3\in\mathbb{R}_+$ with $\tau_1<\tau_2<\tau_3$. We select $\tau_1 = 0.25, \tau_2 = 0.5$ and $\tau_3 = 0.75$. This function is discontinuous and thus violates Assumption \ref{ass:thm1_bias}. A plot of this function (after alignment) is seen in Figure~\ref{fig:true_mu}. For the layer-specific latent positions, we use the same functions as in Section~\ref{sec:sim1}.

In each panel of Figure~\ref{fig:X_estimates_N_50}, the different colours and symbols are as described in Section~\ref{sec:sim1}, %. The plots displayed are 
for a choice of $M=50$. Again, we observe that increasing $N$ leads to improved recovery of the underlying latent positions. The discontinuities do not affect recovery for $N = 250$ and $N = 500$; however, for $N = 100$ nodes, recovery is poor for such a large value of $M$, as in Section~\ref{sec:sim1}. 

% \begin{align}
%     R_g &= 5g, & c_{1,g} &= 2R_g + 1, & c_{2,g} &= 2R_g + 1, & \theta_g &= g\pi, 
% \end{align}
% and 
% \begin{align}
%     d_q^\ell &= (\ell - 1) + q, & \phi_q^\ell &= 2^{1 - \ell}\pi/q.
% \end{align}

%% file: sections/data.tex
\section{Application to a global air transportation network}
\label{sec:real_data}

We provide a real-data application of the proposed method on a global air transportation network. We analyse the crowdsourced OpenSky Network for the month of April in 2021, which details every recorded flights between airports across the globe \citep{strohmeier2021}. Each flight consists of an origin and destination airport, the aircraft model for the flight and time stamps of when the first and last message was received from the flight by the OpenSky Network. We process the network by restricting to the 10 most used aircraft models across the dataset and use these to construct $L=10$ layers. We consider a set of $N=567$ airports having 5 or more in or out flights over the month. Quadruples $(i_k,j_k,\ell_k,t_k)$ consist of directed flights between airports $i_k$ and $j_k$ on an aircraft of type $\ell_k$, landed at time $t_k$. We run the DUASE inference procedure described in Section~\ref{sec:est_for_MIPP-RDPG} with a choice of $d=10$ and binning into $M=28$ bins. This choice corresponds to the days of the month, and we pick $d$ by inspecting the scree plot of singular values of the resulting matrix $\hat{\boldsymbol{\Lambda}}$. 

We consider each element of the resulting sequence $\{\hat{\boldsymbol{X}}^{(m,\ast)}\}_{m=1}^{M}$ as the realisation of $N$ multivariate $d$-dimensional time series at the $m$-th time point, and examine clustering the nodes using these. To cluster the embedding time series, we implement a simple approach. We normalise each dimension by centring and dividing by its standard deviation, and subsequently smooth it using a box kernel with a window of width 5, obtaining estimates $\{\breve{\boldsymbol{X}}^{(m,\ast)}\}_{m=1}^M$. To cluster the time series, we define the node-specific matrices
$\breve{\boldsymbol{X}}_i = [\breve{X}^{(1,\ast)\top}_{i}\mid\cdots\mid \breve{X}^{(M,\ast)\top}_{i}] \in \mathbb{R}^{M \times d}$, summarising the time-varying latent positions of node $i$ within all bins.  
%$\hat{\boldsymbol{X}}_i$ as having $(p,q)$-th element $\hat{X}_{i,pq} = \hat{X}^{(p,\ast)}_{i,q}$, the estimate to dimension $q$ of $X_i(t)$ in bin $p$. 
We then construct a pairwise distance matrix %$\boldsymbol{D} \in \mathbb{R}_{\geq 0}^{N \times N}$ with $D_{ij} = \lVert \hat{\boldsymbol{X}}_i - \hat{\boldsymbol{X}}_j\rVert_F$.
$\boldsymbol{D} \in \mathbb{R}_{\geq 0}^{N \times N}$ with entries $D_{ij} = \lVert \breve{\boldsymbol{X}}_i - \breve{\boldsymbol{X}}_j\rVert_F$, and then implement an agglomerative, hierarchical clustering algorithm using average linkage and $K=6$ clusters, with dissimilarity $\boldsymbol{D}$. This choice of $K$ is made to compare our clusters to the continents that the nodes %(airports) 
are situated in. 

\begin{figure}[p]
\centering

\begin{subfigure}{0.975\textwidth}
\centering
\includegraphics[
  height=0.31\textheight,
  keepaspectratio
]{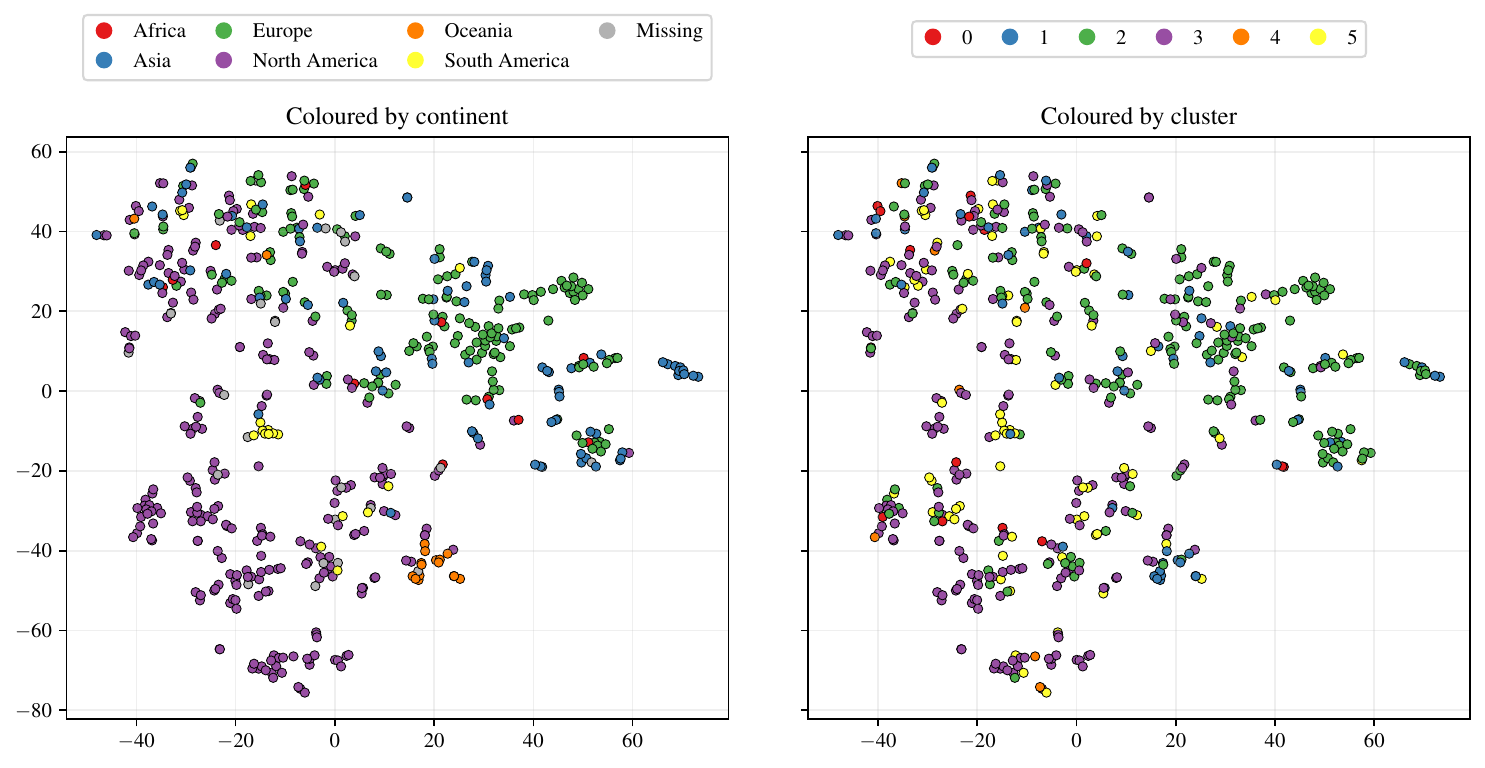}
\caption{t-SNE embedding of $[\hat{\boldsymbol{X}}^{(1,\ast)}, \dots, \hat{\boldsymbol{X}}^{(M,\ast)}]\in\mathbb{R}^{N\times Md}$.}
\label{fig:X_tsne_all_time}
\end{subfigure}

\begin{subfigure}{0.975\textwidth}
\centering
\includegraphics[
  height=0.475\textheight,
  keepaspectratio
]{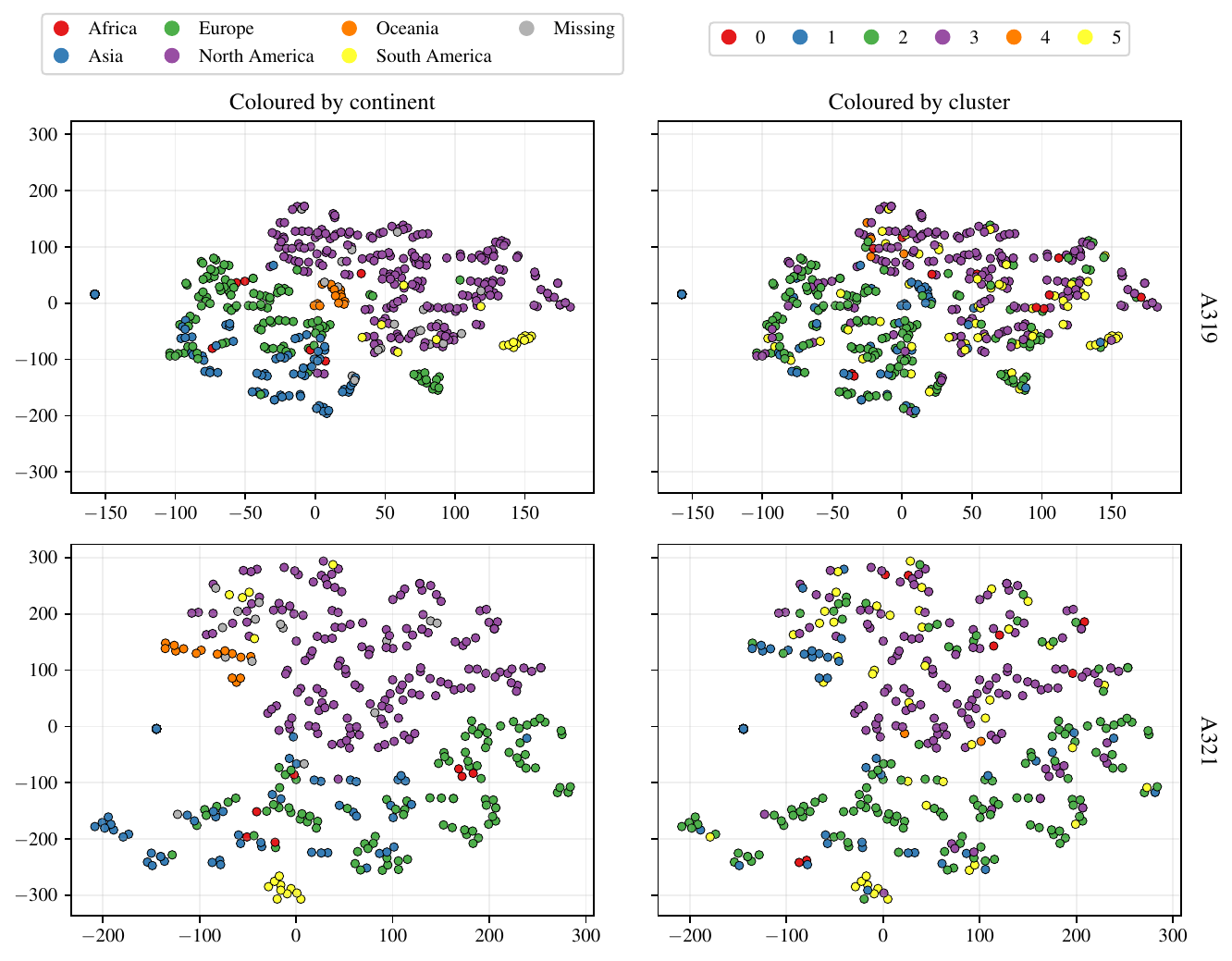}
\caption{t-SNE embedding of $\hat{\boldsymbol{Y}}^{(\ell,\ast)}$ for two layers, corresponding to aircraft types A319 and A321.}
\label{fig:Y_tsne}
\end{subfigure}

\caption{t-SNE embeddings of the inferred airport latent vectors, shown (a) after stacking the dynamic position across time and (b) for two representative layers. Each point represents an airport in the 2D t-SNE space. Points are coloured by continent (left column) and by inferred cluster membership (right column).}
\label{fig:results_air}
\end{figure}

In Figures~\ref{fig:X_tsne_all_time}~and~\ref{fig:Y_tsne}, we use t-SNE, a popular non-linear dimension reduction tool, to visualise the resulting low-dimensional embedding vectors. Figure~\ref{fig:X_tsne_all_time} shows the two-dimensional t-SNE embedding of the matrix $[\hat{\boldsymbol{X}}^{(1,\ast)}, \dots, \hat{\boldsymbol{X}}^{(M,\ast)}]\in\mathbb{R}^{N\times Md}$, coloured in the left panel by the continent of the node, and in the right panel by the cluster assigned to the node by the hierarchical clustering procedure described above. %This matrix corresponds to embedding the entire trajectory at once. 
Note that this visualisation embeds the entire trajectories all at once, and independent t-SNE representations of each of the estimates for each bin would not be interpretable due to the randomness in the t-SNE procedure. Figure~\ref{fig:Y_tsne} shows an analogous plot for $\hat{\boldsymbol{Y}}^{(\ell,\ast)}$ for two choice of aircrafts, identified by the typecodes A319 and A321. %In the first row is the layer that corresponds to typecode (aircraft type) A319 and the second to A321.  
When coloured by continent, the plots exhibit clear geographic structure: airports from the same continent form largely contiguous regions. When coloured by the output of our clustering algorithm, several clusters align closely with these continental groups (notably those corresponding to the main European and North-American regions), whereas others cut across continents and instead group airports with similar functional roles in the network (e.g. intercontinental hubs versus regional airports). This indicates that, while geographic location is strongly reflected in the flow-induced geometry, the clustering captures additional structure beyond continents, organising airports according to their traffic patterns rather than purely their physical position.

% \begin{figure}[p]
% \centering
% \begin{subfigure}{0.975\textwidth}
% \centering
%     \caption{t-SNE embedding of $[\hat{\boldsymbol{X}}^{(1,\ast)}, \dots, \hat{\boldsymbol{X}}^{(M,\ast)}]\in\mathbb{R}^{N\times Md}$.}
%     \label{fig:X_tsne_all_time}
%     \includegraphics[width=\textwidth]{figures/data/tSNE_X_all_time.pdf}
% \end{subfigure}
% \begin{subfigure}{0.975\textwidth}
% \centering
%     \caption{t-SNE embedding of $\hat{\boldsymbol{Y}}^{(\ell,\ast)}$ for two layers, corresponding to aircraft types A319 and A321.}
%     \label{fig:Y_tsne}
%     \includegraphics[width=\textwidth]{figures/data/tSNE_Y_0_2.pdf}
% \end{subfigure}
% \caption{t-SNE embeddings of the inferred airport latent vectors, shown (a) after stacking the dynamic position across time and (b) for two representative layers. Each point represents an airport in the 2D t-SNE space. Points are coloured by continent (left column) and by inferred cluster membership (right column).}
% \label{fig:results_air}
% \end{figure}

%% file: sections/conclusion.tex
\section{Conclusion and discussion}

This paper introduced the multiplex inhomogeneous Poisson process dot product graph (MIPP-DPG), a model for multiplex networks with continuous-time interaction data. Theorems \ref{thm:big_O} and \ref{thm:asymptotic_normality} show that, under mild conditions, doubly unfolded adjacency spectral embedding \citep[DUASE;][]{baum2025} yields latent position estimators that are asymptotically consistent and asymptotically normal. Our theory allows the number of discretization bins to grow with the number of nodes, thereby recovering the continuous setting. This is demonstrated in our two simulation studies. Furthermore, the asymptotic normality result is particularly relevant for downstream community detection, an important topic in network literature, as it provides theoretical support for fitting Gaussian mixture models to the embedded positions when clustering nodes. Despite requiring Lispchitz continuity in the latent dynamics, our second simulation established that DUASE is able to recover the latent positions even when the assumptions required for our theoretical guarantees were violated. In an application to a real-world air traffic network, we demonstrated that inference under our model was able to recover latent structure beyond simple geography. 

Some paths for future research naturally arise as a result of this work. An interesting avenue would be to explore when the network structure is not assumed to be known, and how spectral methods can be adapted to infer both the latent positions and the network structure itself. Another route would be to investigate more complex point process models, such as a self-exciting process. 

%% file: sections/acknowledgements.tex
\section*{Acknowledgements}

Joshua Corneck acknowledges funding from the Engineering and Physical Sciences Research Council (EPSRC), grant number EP/S023151/1. Ed Cohen acknowledges funding from the EPSRC NeST Programme, grant number EP/X002195/1. Francesco Sanna Passino acknowledges funding from the EPSRC, grant number EP/Y002113/1.

%% file: sections/code.tex
\section*{Code}

The Python code for implementing the methods proposed in this article and reproducing the results is available in the GitHub repository \href{https://github.com/joshcorneck/MIPP-DPG.git}{joshcorneck/MIPP-DPG}.

%% file: sections/appendix.tex
% \begin{appendices}
\appendix

\section{Prerequisites for Theorems \ref{thm:big_O} and \ref{thm:asymptotic_normality}}

To minimise notation, for the duration of this Appendix, we denote the spectral norm by $\lVert \cdot \rVert$, dropping the subscript.

\subsection{Background on Poisson processes and subexponential random variables}
\label{app_sec:poisson_sub_exp}

A (simple) temporal point process on a time domain $\mathcal{T} \subseteq \mathbb{R}$ is a random process whose realisation is a list of discrete times $\mathcal{H} = \{t_k\} \subset \mathcal{T}$. Associated with a temporal point process is a counting process $N(\cdot) \in \mathbb{N}$. For $a < b$, the counting process is valued $N((a,b]) =  \sum_{t_k \in \mathcal{H}} \mathbb{I}_{(a,b]}\{t_k\}$ so that it counts the number of events in $(a,b]$. Here we write $\mathbb{I}_\cdot\{\cdot\}$ for the indicator function. We will write $N(t)$ to mean $N((0,t])$. In this work, we will refer to $N(\cdot)$ as the point process. When a point process comes with additional information, that value is referred to as a mark.

The intensity of $N(\cdot)$, denoted by $\lambda(\cdot)$, is the expected number of arrivals per unit time, defined as $\lambda(t) = \lim_{h \to 0} \mathbb{P}(N((t, t+h]) > 0) / h$. If $N((a,b]) \sim \text{Poisson}(\int_a^b\lambda(t) \dd t)$ and $N((a,b])$ and $N((c,d])$ are independent random variables whenever $(a,b]\cap (c,d] = \emptyset$, we refer to $N(\cdot)$ as an inhomogeneous Poisson process (IPP). If $\lambda(t) \equiv \lambda$ for all $t$, then it is an homogeneous Poisson process (HPP). If a collection of IPPs $N_1(\cdot),\dots,N_m(\cdot)$ are independent with intensities $\lambda_1,\dots,\lambda_m$, then $\sum_r N_r(\cdot)$ is an IPP with intensity $\sum_r\lambda(\cdot)$. 

A real-valued random variable $X$ is subexponential if its Orlicz/subexponential norm $\lVert \cdot \rVert_{\psi_1}$ is finite, where we define:
\begin{equation}
	\lVert X \rVert_{\psi_1} := \inf \left\{K>0 : \mathbb{E}\{|X|/K\} \leq 2\right\}. \label{eqn:orlicz_norm}
\end{equation}
An important equivalent definition is to have that here exists $K_1 > 0$ such that:
\begin{equation}
	\mathbb{P}(|X| \geq t) \leq 2 \exp\left\{-t/K_1\right\} \quad \text{ for all $t \geq 0$}.
\end{equation}
This is a statement on the decay rate of the tail probabilities of $X$. We have the following lemma that shows that a centred Poisson random variable is subexponential. The fact that a centred Poisson random variable is subexponential will be used frequently to prove the theorems in this work.

\begin{lemma}[Poisson tail bound]
	\label{lem:poisson_tail_bound}
	Let $X \sim \mathrm{Poisson}(\lambda)$. Then
	\begin{equation}
		\mathbb{P}\left(|X - \lambda| \geq t\right) \leq 2 \exp\left(-\frac{t^2}{2(\lambda + t/3)}\right).
	\end{equation}
\end{lemma}

\subsection{Network point processes}

To allow for continuous-time interactions in a network, a point process can be placed on the network edges. The point processes are often modelled as Poisson \citepSM{matias-2017, fang2020-ix, corneck2025, modell2024}, but examples of self-exciting processes have also been considered \citepSM{Perry13, sannapassino23}. Under such models, a marked point process is observed, consisting of, in the case of a multiplex network, a stream of quadruples $(\ell_k, i_k, j_k, t_k)\in\mathcal [L_N] \times \mathcal{E}_{\ell_k}\times\mathbb R_+,\ k=1,2,\dots$, denoting directed interactions from node $i_k$ to node $j_k$ in layer $\ell_k$ at time $t_k$, where $t_k\leq t_{k^\prime}$ for $k<k^\prime$. The associated edge-specific counting process is denoted by $N_{\ell ij}(\cdot)$, where, for all $\ell \in [L_N]$ and all $(i,j)\in\mathcal{E}_\ell$, 
\begin{equation}
	N_{\ell ij}(t) := \sum_{k} 
	\mathbb I_{\{(\ell,i,j)\}}\{(\ell_k, i_k,j_k)\}\mathbb I_{(0,t]}(t_k).
	%\mathbb{I}\{(i_\ell,j_\ell,t_\ell) \in \{i\}\times\{j\}\times [0,t)\},
\end{equation}

\subsection{Wedin's \texorpdfstring{$\mathbf{\sin}\Theta$}{sintheta} theorem and subspace distances}\label{app_sec:wedin}

%\begin{lemma}[Weyl's inequality]\label{lemma:weyl}
	%Let $\boldsymbol{A}$ and $\boldsymbol{B}$ be Hermitian matrices on an inner product space $V$ with dimension $n$. Let the spectrums of $\boldsymbol{A}$ and $\boldsymbol{B}$ be ordered in descending order:  $\lambda_1(\cdot)\geq \dots \lambda_n(\cdot)$. Then
	%\[
	%\lambda_{i + j - 1}(\boldsymbol{A} + \boldsymbol{B}) \leq \lambda_i (\boldsymbol{A}) + \lambda_j(\boldsymbol{B}) \leq \lambda_{i + j - n}(\boldsymbol{A} + \boldsymbol{B}).
	%\]
	%Equivalently,
	%\[
	%\lambda_j(\boldsymbol{A}) + \lambda_n(\boldsymbol{B}) \leq \lambda_j(\boldsymbol{A} + \boldsymbol{B}) \leq \lambda_j(\boldsymbol{A}) + \lambda_1(\boldsymbol{B}).
	%\]
%\end{lemma}

The following theorem is classical in matrix perturbation theory. The exposition here will follow the description of \citeSM{chen_2021}. Given two $d$-dimensional subspaces $\mathcal{U}$ and $\mathcal{U}^*$ in $\mathbb{R}^n$, one can represent by matrices $\boldsymbol{U}$ and $\boldsymbol{U}^*$ whose columns form orthonormal bases of $\mathcal{U}$ and $\mathcal{U}^*$, respectively. There are a number of ways of measuring distance between two subspaces, any valid choice of which will account for a global rotation ambiguity: that $\boldsymbol{U}\boldsymbol{R}$ for any $\boldsymbol{R} \in \mathbb{O}(d)$ will also form an orthonormal bases for $\mathcal{U}$, and so even when the two subspaces coincide, a rotation could make their distance non-zero unless accounted for. The following notions of distance will be used in the proofs that follow:
\begin{itemize}
	\item Distance with optimal rotation:
	\begin{equation}
		\min_{\boldsymbol{R}\in\mathbb{O}(r)}\lVert  \boldsymbol{U}\boldsymbol{R} - \boldsymbol{U}^* \rVert,
	\end{equation}
	\item Distance with projection matrices:
	\begin{equation*}
		\norm{\boldsymbol{U}\boldsymbol{U}^\top - \boldsymbol{U}^*\boldsymbol{U}^{* \top}},
	\end{equation*}
	\item Distance with principal angles: let $\sigma_1\geq \dots \geq \sigma_r \geq 0$ be the singular values of $\boldsymbol{U}^\top \boldsymbol{U}^*$. All singular angles fall within the interval $[0,1]$. One can thus define the principal angles:
	\begin{equation*}
		\theta_i := \arccos(\sigma_i), \quad \text{for all } 1 \leq i \leq r,
	\end{equation*}
	which will all lie in $[0,\pi/2]$. The distance between the two subspaces can then be defined as $\lVert \sin \boldsymbol{\Theta}\rVert$, where $\boldsymbol{\Theta} = \text{diag}(\sin(\sigma_1),\dots,\sin(\sigma_r))$.
\end{itemize}
Despite the difference in appearance of these distance measure, it can be shown that they are all equivalent up to constant multiplicative factors. As such, in the context of asymptotic bounds, they will all be used interchangeably. 

\citeSM{wedin1972} proved the following theorem that controls the perturbation of the left and right subspaces when examining noisy observations of a matrix.

\begin{theorem}[Wedin's $\sin\Theta$ Theorem; e.g. Theorem 2.9 in \citetSM{chen_2021}]
	\label{thm:wedin}
	Let $\boldsymbol{M}^*$ and $\boldsymbol{M} = \boldsymbol{M}^* + \boldsymbol{E}$ be two matrices in $\mathbb{R}^{n_1 \times n_2}$ and denote by $\boldsymbol{U}^*, \boldsymbol{U}$ (respectively, $\boldsymbol{V}^*, \boldsymbol{V}$) the matrices whose columns contain the top $d$ left (respectively, right) singular vectors of $\boldsymbol{M}^*, \boldsymbol{M}$, respectively. If $\lVert\boldsymbol{E}\rVert < \sigma_d(\boldsymbol{M}^*) - \sigma_{d+1}(\boldsymbol{M}^*)$, then one has
	\begin{equation}
		\max\left\{\lVert\boldsymbol{U}\boldsymbol{U}^\top - \boldsymbol{U}^*\boldsymbol{U}^{*\top}\rVert, \lVert\boldsymbol{V}\boldsymbol{V}^\top - \boldsymbol{V}^*\boldsymbol{V}^{*\top}\rVert\right\} \leq \frac{2^{1/2}\max\left\{\norm{\boldsymbol{E}^\top \boldsymbol{U}^{*\top}}, \lVert\boldsymbol{E}\boldsymbol{V}^*\rVert\right\}}{\sigma_d(\boldsymbol{M}^*) - \sigma_{d+1}(\boldsymbol{M}^*) - \lVert\boldsymbol{E}\rVert}.
	\end{equation}
\end{theorem}

\subsection{Bernstein theorems}\label{app_sec:bernstein}

\begin{theorem}[Matrix Bernstein; e.g. Theorem 1.6.2 in \citetSM{tropp_2015}]\label{thm:matrix-Bernstein}
	Let $\boldsymbol{S}_1,\dots,\boldsymbol{S}_n$ be independent, centered random matrices with common dimension $d_1 \times d_2$, and assume that each one is uniformly bounded, that is:
	\begin{equation}
		\mathbb{E}\{\boldsymbol{S}_k\} = 0 \quad \text{and} \quad \norm{\boldsymbol{S}_k} \leq L_N \quad \text{for each $k=1,\dots,n$}.
	\end{equation}
	Introduce the sum $\boldsymbol{S} = \sum_k \boldsymbol{S}_k$ and let $v(\boldsymbol{S})$ denote the variance statistic of the sum:
	\begin{equation}
		v(\boldsymbol{S}) = \max\left\{\lVert\mathbb{E}\{\boldsymbol{S}\boldsymbol{S}^\top\}\rVert, \lVert\mathbb{E}\{\boldsymbol{S}^\top \boldsymbol{S}\}\rVert\right\} = \max\left\{\norm{\sum_{k=1}^n \mathbb{E}\{\boldsymbol{S}_k\boldsymbol{S}_k^\top\}}, \norm{\sum_{k=1}^n \mathbb{E}\{\boldsymbol{S}_k^\top \boldsymbol{S}_k\}}\right\}.
	\end{equation}
	Then 
	\begin{equation}
		\mathbb{P}(\norm{\boldsymbol{Z}} \geq t) \leq (d_1 + d_2) \exp\left(-\frac{t^2 / 2}{v(\boldsymbol{Z}) + Lt/3}\right) \quad \text{for all $t \geq 0$}.
	\end{equation}
\end{theorem}

\begin{theorem}[Subexponential Bernstein; e.g. Corollary 2.9.2 in \citetSM{vershynin_2018}]
	\label{thm:subexp_bern}
	Let $X_1,\dots, X_n$ be subexponential, mean zero random variables, and $a = (a_1,\dots,a_n)^\top \in \mathbb{R}^n$. Then, for every $t \geq 0$, we have
	\begin{equation}
		\mathbb{P}\left\{\abs{\sum_{i=1}^n a_iX_i} \geq t\right\} \leq 2\exp\left(-c\min\left\{\frac{t^2}{\left(\max_i\norm{X_i}_{\psi_1}\right)^2\norm{a}_2^2}, \frac{t}{\left(\max_i\norm{X_i}_{\psi_1}\right)\norm{a}_\infty}\right\}\right).
	\end{equation}
\end{theorem}

\subsection{Leave-one-out analysis}\label{app:LOO}

Leave-one-out (LOO) analysis is a proof technique used to decouple dependencies between data points. One constructs a LOO counterpart to an estimator and then compares them via a small perturbation. 

As an illustrative example, we examine our setting where $\hat{\boldsymbol{\Lambda}}$ is a matrix of Poisson random variables with expectation $\bar{\boldsymbol{\Lambda}}$. Write the SVDs of each of these matrices as
\begin{align}
	\hat{\boldsymbol{\Lambda}} &= \hat{\boldsymbol{U}}\hat{\boldsymbol{\Sigma}}\hat{\boldsymbol{V}}^\top + \hat{\boldsymbol{U}}_\perp\hat{\boldsymbol{\Sigma}}_\perp\hat{\boldsymbol{V}}^\top_\perp, \\
	\bar{\boldsymbol{\Lambda}} &= \bar{\boldsymbol{U}}\bar{\boldsymbol{\Sigma}}\bar{\boldsymbol{V}}^\top + \bar{\boldsymbol{U}}_\perp\bar{\boldsymbol{\Sigma}}_\perp\bar{\boldsymbol{V}}^\top_\perp,
\end{align}
and suppose that we wish to bound the object $\lVert(\hat{\boldsymbol{\Lambda}} - \bar{\boldsymbol{\Lambda}})(\boldsymbol{I} - \bar{\boldsymbol{V}}\bar{\boldsymbol{V}}^\top)\hat{\boldsymbol{V}}\hat{\boldsymbol{V}}^\top\rVert_{2,\infty}$. We cannot use concentration inequalities here in a standard way as there is a dependency between $\hat{\boldsymbol{\Lambda}}$ and $\hat{\boldsymbol{V}}$. This problem can be circumvented through LOO analysis. 

For each $i \in [NM]$, define a matrix
$\hat{\boldsymbol{\Lambda}}^{(i)}$ as
\begin{equation}
	\hat{\Lambda}^{(i)}_{jk} = \begin{cases}
		\hat{\Lambda}_{jk}, & j \neq i, \\
		\bar{\Lambda}_{jk} & j = i
	\end{cases}.
\end{equation}
Then $\hat{\boldsymbol{\Lambda}}^{(i)}$ is $\hat{\boldsymbol{\Lambda}}$ with its $i$-th row replaced with its expectation. That is to say, $\hat{\boldsymbol{\Lambda}}^{(i)}$ is $\hat{\boldsymbol{\Lambda}}$ but with the randomness removed from the $i$-th row. In this way, we can write $\hat{\boldsymbol{\Lambda}} = \hat{\boldsymbol{\Lambda}}^{(i)} + \boldsymbol{E}^{(i)}$, where 
\begin{equation}
	E^{(i)}_{pq} = 
	\begin{cases}
		\hat{\Lambda}_{pq} - \bar{\Lambda}_{pq}, & p \neq i, \\
		0, & p = i.
	\end{cases}
\end{equation}
We compute the singular value decomposition of this de-noised matrix $\hat{\boldsymbol{\Lambda}}^{(i)}$ as
\begin{equation}
	\hat{\boldsymbol{\Lambda}}^{(i)} = \hat{\boldsymbol{U}}^{(i)}\hat{\boldsymbol{\Sigma}}^{(i)}\hat{\boldsymbol{V}}^{(i) \top} + \hat{\boldsymbol{U}}^{(i)}_\perp\hat{\boldsymbol{\Sigma}}^{(i)}_\perp\hat{ \boldsymbol{V}}^{(i)\top}_\perp.
\end{equation}
The point of this construction is that $\hat{\boldsymbol{\Lambda}}^{(i)}$, and therefore $\hat{\boldsymbol{V}}^{(i)}\hat{\boldsymbol{V}}^{(i)\top}$, are independent of the $i$-th row of $\hat{\boldsymbol{\Lambda}}$. To see the benefit of this, consider the following decomposition of our norm
\begin{align}
	\lVert e_i^\top (\hat{\boldsymbol{\Lambda}} - \bar{\boldsymbol{\Lambda}})(\boldsymbol{I} - \bar{\boldsymbol{V}}\bar{\boldsymbol{V}}^\top)\hat{\boldsymbol{V}}\hat{\boldsymbol{V}}^\top\rVert &\leq \lVert e_i^\top (\hat{\boldsymbol{\Lambda}} - \bar{\boldsymbol{\Lambda}}) (\boldsymbol{I} - \bar{\boldsymbol{P}})\hat{\boldsymbol{P}}^{(i)}\rVert + \lVert e_i^\top (\hat{\boldsymbol{\Lambda}} - \bar{\boldsymbol{\Lambda}})(\boldsymbol{I} - \bar{\boldsymbol{P}})(\hat{\boldsymbol{P}} - \hat{\boldsymbol{P}}^{(i)})\rVert,\\
	&= \lVert (\hat{\Lambda}_{i,\ast} - \bar{\Lambda}_{i,\ast}) (\boldsymbol{I} - \bar{\boldsymbol{P}})\hat{\boldsymbol{P}}^{(i)}\rVert + \lVert (\hat{\Lambda}_{i,\ast} - \bar{\Lambda}_{i,\ast})(\boldsymbol{I} - \bar{\boldsymbol{P}})(\hat{\boldsymbol{P}} - \hat{\boldsymbol{P}}^{(i)})\rVert,
\end{align}
where we define $\bar{\boldsymbol{P}} = \bar{\boldsymbol{V}}\bar{\boldsymbol{V}}^\top$, and have analogous definitions for $\hat{\boldsymbol{P}}$ and $\hat{\boldsymbol{P}}^{(i)}$. It is here that we see the power of LOO analysis as we now have the luxury of applying concentration inequalities for sums of independent random variables to the first term as we have decoupled the dependent terms. For the second, we can applying Wedin's Theorem to the operator norm $\lVert \hat{\boldsymbol{P}} - \hat{\boldsymbol{P}}^{(i)} \rVert$ which are projections onto very ``close''subspaces as they only differ by one row. This argument will be presented in depth when applied in later propositions.

%Proposition \ref{prop:poisson_tail_bound} tells us that the event $X > \beta \log(n)$ occurs with high probability as we can always pick a $\rho > 0$.

\section{Key propositions}
\label{app_sec:key_props}

In this section, we provide several propositions that will be used to prove the main theorems of this work. The proofs in this Appendix combine and extend the standard machinery used to prove results in \citeSM{jones_2021, gallagher2024, baum2025}. Throughout this section, as in the main text, we will write $\hat{\boldsymbol{\Lambda}} \in \mathbb{R}_+^{NM \times NL}$ to be a matrix of independent Poisson random variables with expectation $\bar{\boldsymbol{\Lambda}} \in \mathbb{R}_+^{NM \times NL}$. We will write the SVDs of these matrices as:
\begin{align}
	\hat{\boldsymbol{\Lambda}} &= \hat{\boldsymbol{U}}\hat{\boldsymbol{\Sigma}}\hat{\boldsymbol{V}}^\top + \hat{\boldsymbol{U}}_\perp\hat{\boldsymbol{\Sigma}}_\perp\hat{\boldsymbol{V}}_\perp^\top,\\
	\bar{\boldsymbol{\Lambda}} &= \bar{\boldsymbol{U}}\bar{\boldsymbol{\Sigma}}\bar{\boldsymbol{V}}^\top + \bar{\boldsymbol{U}}_\perp\bar{\boldsymbol{\Sigma}}_\perp\bar{\boldsymbol{V}}_\perp^\top.
\end{align}

We will write $\hat{\boldsymbol{\Lambda}}^{(i)} \in \mathbb{R}^{NM \times NL}_+$ to be the $i$-th  leave-one-out counterpart of $\hat{\boldsymbol{\Lambda}}$, as detailed in Section \ref{app:LOO}, with $\hat{\boldsymbol{\Lambda}}^{(i)}$ having its $(p,q)$-th entry defined as
\begin{equation}
	\hat{\Lambda}^{(i)}_{pq} = \begin{cases}
		\hat{\Lambda}_{pq}, & p \neq i \\
		\bar{\Lambda}_{pq}, & p = q.
	\end{cases}
\end{equation}
We write the SVD of $\hat{\boldsymbol{\Lambda}}^{(i)}$ in an analogous way to that of $\hat{\boldsymbol{\Lambda}}$ and $\bar{\boldsymbol{\Lambda}}$.

Proposition \ref{prop:poisson_tail_bound} provides a high probability decay bound on Poisson random variables that will be used throughout our proofs. 

\begin{proposition}[Poisson tail bound]\label{prop:poisson_tail_bound}
	Let $X \sim \mathrm{Poisson}(\lambda)$. Then, for $N \in \mathbb{N}$, for each $\rho > 0$ there exists some $\beta_\rho$ such that
	\begin{equation}
		\mathbb{P}(X < \beta_\rho \log (N)) > 1 - N^{-\rho}.
	\end{equation}
\end{proposition}

We prove Proposition \ref{prop:poisson_tail_bound} using a simple application of Markov's inequality and the moment generating function of a Poisson random variable. The next proposition controls the spectral norm between the Poisson random matrix $\hat{\boldsymbol{\Lambda}}$ and its expectation $\bar{\boldsymbol{\Lambda}}$. The bound is seen to grow the $N,M_N$ and $L_N$.

\begin{proposition}\label{prop:hat_lambda_bar_lambda}
    $\lVert\hat{\boldsymbol{\Lambda}} - \bar{\boldsymbol{\Lambda}}\rVert = \mathcal{O}\left(\max\{M_N,L_N\}^{1/2}M_NN^{1/2}\log^{3/2}(N)\right)$ almost surely.
\end{proposition}

Proposition \ref{prop:hat_lambda_bar_lambda} is proved by combining Propostion \ref{prop:poisson_tail_bound} with Bernstein's theorem and Popoviciu’s inequality. An appliation of Borel-Cantelli extends the argument to hold eventually almost surely. Through Proposition \ref{prop:sing_vals_lam_hat}, we provide control on the growth of the singular values of $\hat{\boldsymbol{\Lambda}}$. In particular, the growth of the eigengap is guaranteed with growing $M_N,N$ and $L_N$. These bounds hold under Assumption \ref{ass:thm1_variance} and a simple application of Corollary 7.3.5 from \citeSM{horn2012}.

\begin{proposition}[Control of $\sigma_i(\hat{\boldsymbol{\Lambda}})$]\label{prop:sing_vals_lam_hat}
    Under Assumption \ref{ass:thm1_variance}, we have 
    \begin{equation}
        \sigma_1\left(\hat{\boldsymbol{\Lambda}}\right),\dots, \sigma_d\left(\hat{\boldsymbol{\Lambda}}\right) = \Theta\left(M_N^{1/2}L_N^{1/2}N\right),
    \end{equation}
    and 
    \begin{equation}
        \sigma_{d+1}(\hat{\boldsymbol{\Lambda}}) = \mathcal{O}\left(\max\{M_N,L_N\}^{1/2}M_NN^{1/2}\log^{3/2}(N)\right),
    \end{equation}
    almost surely.
\end{proposition}

We require control on the growth of the term  $\lVert\bar{\boldsymbol{U}}^\top(\hat{\boldsymbol{\Lambda}} - \bar{\boldsymbol{\Lambda}})\bar{\boldsymbol{V}}\rVert_F = \mathcal{O}\left(M_N\log^{3/2}(N)\right)$ as $N \to \infty$. This is provided by Proposition \ref{prop:U_lambda_hat_lambda_bar_V}.

\begin{proposition}\label{prop:U_lambda_hat_lambda_bar_V}
	We have $		\lVert\bar{\boldsymbol{U}}^\top(\hat{\boldsymbol{\Lambda}} - \bar{\boldsymbol{\Lambda}})\bar{\boldsymbol{V}}\rVert_F = \mathcal{O}\left(M_N\log^{3/2}(N)\right)$, almost surely.
\end{proposition}

Proposition \ref{prop:U_lambda_hat_lambda_bar_V} is proved by conidering the term inside the norm as a sum of scaled, independent random variables and applying Hoeffding's inequality and a union bound.

In Proposition \ref{prop:lots_of_bounds}, we provide a number of Frobenius and spectral norm bounds that are employed in the proofs of the main theorems. 

\begin{proposition}\label{prop:lots_of_bounds}
    The following bounds hold almost surely:
    \begin{enumerate}
        \item 
        \begin{enumerate}
            \item $\lVert\hat{\boldsymbol{U}}\hat{\boldsymbol{U}}^\top - \bar{\boldsymbol{U}}\bar{\boldsymbol{U}}^\top\rVert = \mathcal{O}\left(\max\{M_N,L_N\}^{1/2}M_N^{1/2}L_N^{-1/2}N^{-1/2}\log^{3/2}(N)\right)$,
            \item $\lVert\hat{\boldsymbol{V}}\hat{\boldsymbol{V}}^\top - \bar{\boldsymbol{V}}\bar{\boldsymbol{V}}^\top\rVert = \mathcal{O}\left(\max\{M_N,L_N\}^{1/2}M_N^{1/2}L_N^{-1/2}N^{-1/2}\log^{3/2}(N)\right),$
        \end{enumerate}
        \item 
        \begin{enumerate}
            \item $\lVert\hat{\boldsymbol{U}} - \bar{\boldsymbol{U}}\bar{\boldsymbol{U}}^\top \hat{\boldsymbol{U}}\rVert_F = \mathcal{O}\left(\max\{M_N,L_N\}^{1/2}M_N^{1/2}L_N^{-1/2}N^{-1/2}\log^{3/2}(N)\right)$,
            \item $\lVert\hat{\boldsymbol{V}} - \bar{\boldsymbol{V}}\bar{\boldsymbol{V}}^\top \hat{\boldsymbol{V}}\rVert_F = \mathcal{O}\left(\max\{M_N,L_N\}^{1/2}M_N^{1/2}L_N^{-1/2}N^{-1/2}\log^{3/2}(N)\right)$,
        \end{enumerate}
        \item 
        \begin{enumerate}
            \item $\lVert\bar{\boldsymbol{U}}^\top \hat{\boldsymbol{U}}\hat{\boldsymbol{\Sigma}} - \bar{\boldsymbol{\Sigma}}\bar{\boldsymbol{V}}^\top\hat{\boldsymbol{V}}\rVert_F =\mathcal{O}\left(\max\{M_N,L_N\}M_N^{3/2}L_N^{-1/2}\log^3(N)\right)$,
            \item $\lVert\bar{\boldsymbol{\Sigma}} \bar{\boldsymbol{U}}^\top\hat{\boldsymbol{U}} - \bar{\boldsymbol{V}}^\top\hat{\boldsymbol{V}}\hat{\boldsymbol{\Sigma}}\rVert_F =\mathcal{O}\left(\max\{M_N,L_N\}M_N^{3/2}L_N^{-1/2}\log^3(N)\right)$,
        \end{enumerate}
        \item $\lVert\bar{\boldsymbol{U}}^\top \hat{\boldsymbol{U}} - \bar{\boldsymbol{V}}^\top \hat{\boldsymbol{V}}\rVert_F =  \mathcal{O}\left(\max\{M_N,L_N\}M_NL_N^{-1}N^{-1}\log^3(N)\right)$.
    \end{enumerate}
\end{proposition}

Proposition \ref{prop:def_W} is important as it defines the orthogonal matrix $\boldsymbol{W}$, which is key to defining the general linear transformation up to which we recover the latent positions. The result is proved by applying a result from \citeSM{schonemann_1966} and parts of Proposition \ref{prop:lots_of_bounds}.

\begin{proposition}\label{prop:def_W}
    Let $\bar{\boldsymbol{U}}^\top\hat{\boldsymbol{U}} + \bar{\boldsymbol{V}}^\top\hat{\boldsymbol{V}}$ admit the singular value decomposition
    \begin{equation}
        \bar{\boldsymbol{U}}^\top\hat{\boldsymbol{U}} + \bar{\boldsymbol{V}}^\top\hat{\boldsymbol{V}} = \boldsymbol{W}_1\boldsymbol{\Sigma} \boldsymbol{W}_2,
    \end{equation}
    and let $\boldsymbol{W} = \boldsymbol{W}_1\boldsymbol{W}_2^\top$. Then
    \begin{equation}
        \max\left\{\lVert\bar{\boldsymbol{U}}^\top\hat{\boldsymbol{U}} - \boldsymbol{W}\rVert_F, \lVert\bar{\boldsymbol{V}}^\top\hat{\boldsymbol{V}} - \boldsymbol{W}\rVert_F\right\} = \mathcal{O}\left(\max\{M_N,L_N\}M_NL_N^{-1}N^{-1}\log^{3}(N)\right),
    \end{equation}
    almost surely.
\end{proposition}

With $\boldsymbol{W}$ defined, Proposition \ref{prop:W_bounds} provides Frobenius norm bounds on terms involving $\boldsymbol{W}, \hat{\boldsymbol{\Sigma}}$ and $\bar{\boldsymbol{\Sigma}}$. These bounds are employed in proofs of subsequent propositions, in particular Proposition \ref{prop:residuals}. 

\begin{proposition}\label{prop:W_bounds}
    With $\boldsymbol{W}$ defined as in Proposition \ref{prop:def_W}, we have the following bounds holding almost surely:
    \begin{enumerate}
        \item $\lVert\boldsymbol{W}\hat{\boldsymbol{\Sigma}} - \bar{\boldsymbol{\Sigma}}\boldsymbol{W}\rVert_F = \mathcal{O}\left(\max\{M_N,L_N\}M_N^{3/2}L_N^{-1/2}\log^3(N)\right),$ \\
        \item $\lVert\boldsymbol{W}\hat{\boldsymbol{\Sigma}}^{1/2} - \bar{\boldsymbol{\Sigma}}^{1/2}\boldsymbol{W}\rVert_F = \mathcal{O}\left(\max\{M_N,L_N\}M_N^{5/4}L_N^{-3/4}N^{-1/2}\log^3(N)\right)$, \\
        \item $\lVert\boldsymbol{W}\hat{\boldsymbol{\Sigma}}^{-1/2} - \bar{\boldsymbol{\Sigma}}^{-1/2}\boldsymbol{W}\rVert_F = \mathcal{O}\left(\max\{M_N,L_N\}M_N^{3/4}L_N^{-5/4}N^{-3/2}\log^3(N)\right)$.
    \end{enumerate}
\end{proposition}

Proposition \ref{prop:KR} defines the general linear transformations $\breve{\boldsymbol{K}}$ and $\breve{\boldsymbol{R}}$ that relate the latent position matrices estimated from $\bar{\boldsymbol{\Lambda}}$, namely $\bar{\boldsymbol{X}}$ and $\bar{\boldsymbol{Y}}$, to the true positions $\tilde{\boldsymbol{X}}$ and $\boldsymbol{Y}$. These are key transformations, as they combine with $\boldsymbol{W}$ to define $\boldsymbol{W}_X$ and $\boldsymbol{W}_Y$ in Theorem \ref{thm:big_O}.

\begin{proposition}\label{prop:KR}
		Define $\tilde{\boldsymbol{X}}$ to be $\tilde{\boldsymbol{X}} = \left(\tilde{\boldsymbol{X}}_1\mid\cdots\mid\tilde{\boldsymbol{X}}^{(m,\ast)}\right) \in \mathbb{R}^{NM_N \times d}$ where $\tilde{\boldsymbol{X}}^{(m,\ast)} := M_N \int_{B_m}\boldsymbol{X}(t)\dd t \in \mathbb{R}^{N\times d}$. Let the singular value decomposition of $\bar{\boldsymbol{\Lambda}} = \tilde{\boldsymbol{X}}\boldsymbol{Y}^\top$ be $\bar{\boldsymbol{U}}\bar{\boldsymbol{\Sigma}}\bar{\boldsymbol{V}}^\top + \bar{\boldsymbol{U}}_\perp \bar{\boldsymbol{\Sigma}}_\perp\bar{\boldsymbol{V}}_\perp^\top$. Define $\bar{\boldsymbol{X}} = \bar{\boldsymbol{U}}\bar{\boldsymbol{\Sigma}}^{1/2}$ and $\bar{\boldsymbol{Y}} = \bar{\boldsymbol{V}}\bar{\boldsymbol{\Sigma}}^{1/2}$. Then, there exists matrices $\breve{\boldsymbol{K}} \in \mathrm{GL}(d)$ and $\breve{\boldsymbol{R}} \in \mathrm{GL}(d)$ such that
        $\bar{\boldsymbol{X}} = \tilde{\boldsymbol{X}}\breve{\boldsymbol{K}}$ and $\bar{\boldsymbol{Y}} = \boldsymbol{Y}\breve{\boldsymbol{R}}$. Furthermore, we have $\breve{\boldsymbol{K}}\breve{\boldsymbol{R}}^\top = I_d$.
\end{proposition}

Proposition \ref{prop:LOO_incoherence} provides a bound on the growth of the left and right singular vectors that arise from the leave-one-out analysis. This is needed to provide the decay rates of the remainder terms in Proposition \ref{prop:residuals}.

\begin{proposition}\label{prop:LOO_incoherence}
	Suppose that $\hat{\boldsymbol{\Lambda}}^{(i)}$ is the matrix formed by setting the $i$-th row of $\hat{\boldsymbol{\Lambda}}$ equal to $\bar{\Lambda}_{i,\ast}$. Write the SVD of $\hat{\boldsymbol{\Lambda}}^{(i)}$ as
	\begin{equation} \hat{\boldsymbol{\Lambda}}^{(i)} = \hat{\boldsymbol{U}}^{(i)}\hat{\boldsymbol{\Sigma}}^{(i)}\hat{\boldsymbol{V}}^{(i) \top} + \hat{\boldsymbol{U}}^{(i)}_\perp\hat{\boldsymbol{\Sigma}}^{(i)}_\perp\hat{\boldsymbol{V}}^{(i) \top}_\perp,
	\end{equation}
	and the SVDs of $\hat{\boldsymbol{\Lambda}}$ and $\bar{\boldsymbol{\Lambda}}$ using analogous notation. Then, under Assumption \ref{ass:bound_and_int}, we have
	\begin{equation}
		\lVert\hat{\boldsymbol{U}}\rVert_{2,\infty},\lVert\hat{\boldsymbol{U}}^{(i)}\rVert_{2,\infty},\lVert\hat{\boldsymbol{V}}\rVert_{2,\infty},\lVert\hat{\boldsymbol{V}}^{(i)}\rVert_{2,\infty} = \mathcal{O}\left(\max\{M_N,L_N\}^{1/2}M_N^{1/2}L_N^{-1/2}N^{-1/2}\log^{3/2}(N)\right).
	\end{equation}
\end{proposition}

The next proposition bounds the spectral norms of the general linear transformations $\tilde{\boldsymbol{K}}$ and $\tilde{\boldsymbol{R}}$ and also of their inverses. These are needed in the proof of Theorem \ref{thm:big_O} when we transform from the theoretical latent positions to their estimates.

\begin{proposition}\label{prop:KR_spec_norm}
    The matrices $\breve{\boldsymbol{K}}$ and $\breve{\boldsymbol{R}}$ as defined in Proposition \ref{prop:KR} satisfy:
    \begin{itemize}
        \item $\breve{\norm{\boldsymbol{K}}} = \mathcal{O}\left(M_N^{-1/4}L_N^{1/4}\right)$, \qquad $\lVert\breve{\boldsymbol{K}}^{-1}\rVert = \mathcal{O}\left(M_N^{1/4}L_N^{-1/4}\right)$,
        \item $\breve{\norm{\boldsymbol{R}}} = \mathcal{O}\left(M_N^{1/4}L_N^{-1/4}\right)$, \qquad $\lVert\breve{\boldsymbol{R}}^{-1}\rVert = \mathcal{O}\left(M_N^{-1/4}L_N^{1/4}\right)$.
    \end{itemize}
\end{proposition}

Proposition \ref{prop:residuals} is key for the proof of Theorem \ref{thm:big_O}. In the theorem, we decompose the two-to-infinity norm into multiple remainders. The following proposition shows that these remainders decay with $N$ and quantifies the rate at which they decay. 

\begin{proposition}\label{prop:residuals}
    Define the following:
    \begin{alignat}{4}
  \boldsymbol{R}_{1,1} \,&=\, \bar{\boldsymbol{U}}\bigl(\bar{\boldsymbol{U}}^\top\hat{\boldsymbol{U}}\,\hat{\boldsymbol{\Sigma}}^{1/2}
                     - \bar{\boldsymbol{\Sigma}}^{1/2}\boldsymbol{W}\bigr),
  &\quad
  \boldsymbol{R}_{2,1} \,&=\, \bar{\boldsymbol{V}}\bigl(\bar{\boldsymbol{V}}^\top \hat{\boldsymbol{V}}\,\hat{\boldsymbol{\Sigma}}^{1/2}
                     - \bar{\boldsymbol{\Sigma}}^{1/2}\boldsymbol{W}\bigr),\\[1ex]
  \boldsymbol{R}_{1,2} \,&=\, (\boldsymbol{I} - \bar{\boldsymbol{U}}\bar{\boldsymbol{U}}^\top)\,(\hat{\boldsymbol{\Lambda}} - \bar{\boldsymbol{\Lambda}})\,
                     (\hat{\boldsymbol{V}} - \bar{\boldsymbol{V}}\boldsymbol{W})\,\hat{\boldsymbol{\Sigma}}^{-1/2},
  &\quad
  \boldsymbol{R}_{2,2} \,&=\, (\boldsymbol{I} - \bar{\boldsymbol{V}}\bar{\boldsymbol{V}}^\top)\,(\hat{\boldsymbol{\Lambda}} - \bar{\boldsymbol{\Lambda}})\,
                     (\hat{\boldsymbol{U}} - \bar{\boldsymbol{U}}\boldsymbol{W})\,\hat{\boldsymbol{\Sigma}}^{-1/2},\\[1ex]
  \boldsymbol{R}_{1,3} \,&=\, -\,\bar{\boldsymbol{U}}\bar{\boldsymbol{U}}^\top(\hat{\boldsymbol{\Lambda}} - \bar{\boldsymbol{\Lambda}})\,
                     \bar{\boldsymbol{V}}\boldsymbol{W}\,\hat{\boldsymbol{\Sigma}}^{-1/2},
  &\quad
  \boldsymbol{R}_{2,3} \,&=\, -\,\bar{\boldsymbol{V}}\bar{\boldsymbol{V}}^\top(\hat{\boldsymbol{\Lambda}} - \bar{\boldsymbol{\Lambda}})\,
                     \bar{\boldsymbol{U}}\boldsymbol{W}\,\hat{\boldsymbol{\Sigma}}^{-1/2},\\[1ex]
  \boldsymbol{R}_{1,4} \,&=\, (\hat{\boldsymbol{\Lambda}} - \bar{\boldsymbol{\Lambda}})\,\bar{\boldsymbol{V}}\,
                     (\boldsymbol{W}\hat{\boldsymbol{\Sigma}}^{-1/2} - \bar{\boldsymbol{\Sigma}}^{-1/2}\boldsymbol{W}),
  &\quad
  \boldsymbol{R}_{2,4} \,&=\, (\hat{\boldsymbol{\Lambda}} - \bar{\boldsymbol{\Lambda}})\,\bar{\boldsymbol{U}}\,
                     (\boldsymbol{W}\hat{\boldsymbol{\Sigma}}^{-1/2} - \bar{\boldsymbol{\Sigma}}^{-1/2}\boldsymbol{W}).
\end{alignat}
    With these definitions, the following bounds hold almost surely:
    \begin{enumerate}
        \item $\norm{\boldsymbol{R}_{1,1}}_{2, \infty}, \norm{\boldsymbol{R}_{2,1}}_{2, \infty} = \mathcal{O}\left(\max\{M_N,L_N\}M_N^{3/4}L_N^{-3/4}N^{-1}\log^3(N)\right)$
        \item $\norm{\boldsymbol{R}_{1,2}}_{2, \infty}, \norm{\boldsymbol{R}_{2,2}}_{2, \infty} = \mathcal{O}\left(\max\{M_N,L_N\}^{3/2}M_N^{7/4}L_N^{-5/4}N^{-1}\log^{9/2}(N)\right)$,
        \item $\norm{\boldsymbol{R}_{1,3}}_{2, \infty}, \norm{\boldsymbol{R}_{2,3}}_{2, \infty} = \mathcal{O}\left(M_N^{1/4}L_N^{-1/4}N^{-1}\log^{3/2}(N)\right)$,
        \item $\norm{\boldsymbol{R}_{1,4}}_{2, \infty}, \norm{\boldsymbol{R}_{2,4}}_{2, \infty} = \mathcal{O}\left(\max\{M_N,L_N\}^{3/2}M_N^{7/4}L_N^{-5/4}N^{-1}\log^{9/2}(N)\right).$
    \end{enumerate}
\end{proposition}

The next two propositions are on the decay of the bias term in Theorem \ref{thm:big_O} under Assumptions \ref{ass:no_rotation_lipschitz_continuity} and \ref{ass:subspace_lipschitz_continuity}.

\begin{proposition}[Bias decay with Assumption \ref{ass:no_rotation_lipschitz_continuity}]\label{prop:bias_decay_no_rotation_lipschitz}
	Under Assumption \ref{ass:no_rotation_lipschitz_continuity}, we have the following decay:
	\begin{equation}
		\lVert\tilde{\boldsymbol{X}}^{(m,\ast)} - \boldsymbol{X}(t)\rVert_{2,\infty} = \mathcal{O}(K_NM_N^{-1}).
	\end{equation}
\end{proposition}

\begin{proposition}[Bias decay with Assumption \ref{ass:subspace_lipschitz_continuity}]\label{prop:bias_decay_subspace_lipschitz}
	Under Assumption \ref{ass:no_rotation_lipschitz_continuity}, we have the following decay:
    \begin{equation}
        \lVert \tilde{\boldsymbol{X}}^{(m,\ast)} \boldsymbol{W}_m - \boldsymbol{X}(t) \rVert_{2,\infty} = \mathcal{O}\left((K_1 + K_2)M_N^{-1}\right).
    \end{equation}
\end{proposition}

\section{Proof of Theorem \ref{thm:big_O}}
\label{app_sec:proof_thm_1}

The proof of Theorem \ref{thm:big_O} is constructed using the series of propositions and supplementary results stated in Section \ref{app_sec:key_props}.

The task at hand is to obtain bounds on the following norms:
\begin{equation}
	\max_{m \in [M_N]}\sup_{t \in B_m} \lVert\hat{\boldsymbol{X}}^{(m,\ast)}\boldsymbol{W}_{\boldsymbol{X}} - \boldsymbol{X}(t)\rVert_{2,\infty} \qquad \text{and} \qquad
	\lVert \hat{\boldsymbol{Y}}\boldsymbol{W}_{\boldsymbol{Y}} - \boldsymbol{Y}\rVert.
\end{equation}
To tackle the first of these norms, our strategy is decompose this norm into the sum of two terms as:
\begin{align}
	\max_{m \in [M_N]}\sup_{t \in B_m} \lVert\hat{\boldsymbol{X}}^{(m,\ast)}\boldsymbol{W}_X - \boldsymbol{X}(t)\rVert_{2,\infty} \leq \lVert\hat{\boldsymbol{X}}\boldsymbol{W}_2 - \tilde{\boldsymbol{X}}\rVert_{2,\infty} + \max_{m \in [M_N]}\sup_{t \in B_m} \lVert\tilde{\boldsymbol{X}}^{(m,\ast)}\boldsymbol{W}_1- \boldsymbol{X}(t)\rVert_{2,\infty},
\end{align}
with $\boldsymbol{W}_X = \boldsymbol{W}_2\boldsymbol{W}_1$ and $\boldsymbol{W}_1, \boldsymbol{W}_2 \in \mathbb{O}(d)$. Note this holds because the two-to-infinity norm of the $m$-th block is always upper bounded by the two-to-infinity norm of the full matrix. The second term is $B_N$, and the decay rates of this is proved in Proposition \ref{prop:bias_decay_no_rotation_lipschitz}.

We begin with the following decomposition:
\begin{align}
    \hat{\boldsymbol{X}} - \bar{\boldsymbol{X}}\boldsymbol{W} &= \hat{\boldsymbol{U}}\hat{\boldsymbol{\Sigma}}^{1/2} - \bar{\boldsymbol{U}}\bar{\boldsymbol{\Sigma}}^{1/2}W \\
    &= \hat{\boldsymbol{U}}\hat{\boldsymbol{\Sigma}}^{1/2} - \bar{\boldsymbol{U}}\bar{\boldsymbol{U}}^\top \hat{\boldsymbol{U}}\hat{\boldsymbol{\Sigma}}^{1/2} + \bar{\boldsymbol{U}}\left(\bar{\boldsymbol{U}}^\top\bar{\boldsymbol{U}} \hat{\boldsymbol{\Sigma}}^{1/2} - \bar{\boldsymbol{\Sigma}}^{1/2}W\right) \\
    &= \hat{\boldsymbol{U}}\hat{\boldsymbol{\Sigma}}^{1/2} - \bar{\boldsymbol{U}}\bar{\boldsymbol{U}}^\top\hat{\boldsymbol{U}}\hat{\boldsymbol{\Sigma}}^{1/2} + \boldsymbol{R}_{1,1},
\end{align}
where $\boldsymbol{R}_{1,1}$ is defined as in Proposition \ref{prop:residuals}. Continuing, we see
\begin{align}
    \hat{\boldsymbol{X}} - \bar{\boldsymbol{X}}\boldsymbol{W} &= (\hat{\boldsymbol{\Lambda}} - \bar{\boldsymbol{\Lambda}})\hat{\boldsymbol{V}}\hat{\boldsymbol{\Sigma}}^{-1/2} - (\bar{\boldsymbol{U}}\bar{\boldsymbol{U}}^\top \hat{\boldsymbol{\Lambda}} - \bar{\boldsymbol{\Lambda}})\hat{\boldsymbol{V}}\hat{\boldsymbol{\Sigma}}^{-1/2} + \boldsymbol{R}_{1,1} \\ 
    &= (\hat{\boldsymbol{\Lambda}} - \bar{\boldsymbol{\Lambda}})\hat{\boldsymbol{V}}\hat{\boldsymbol{\Sigma}}^{-1/2} - \bar{\boldsymbol{U}}\bar{\boldsymbol{U}}^\top (\hat{\boldsymbol{\Lambda}} - \bar{\boldsymbol{\Lambda}})\hat{\boldsymbol{V}}\hat{\boldsymbol{\Sigma}}^{-1/2} + \boldsymbol{R}_{1,1} \\
    &= (\boldsymbol{I} - \bar{\boldsymbol{U}}\bar{\boldsymbol{U}}^\top)(\hat{\boldsymbol{\Lambda}} - \bar{\boldsymbol{\Lambda}})\hat{\boldsymbol{V}}\hat{\boldsymbol{\Sigma}}^{-1/2} + \boldsymbol{R}_{1,1} \\
    &= (\boldsymbol{I} - \bar{\boldsymbol{U}}\bar{\boldsymbol{U}}^\top)(\hat{\boldsymbol{\Lambda}} - \bar{\boldsymbol{\Lambda}})[\bar{\boldsymbol{V}}\boldsymbol{W} + (\hat{\boldsymbol{V}} - \bar{\boldsymbol{V}}\boldsymbol{W})]\hat{\boldsymbol{\Sigma}}^{-1/2} + \boldsymbol{R}_{1,1}\\
    &= (\hat{\boldsymbol{\Lambda}} - \bar{\boldsymbol{\Lambda}})\bar{\boldsymbol{V}}\boldsymbol{W}\hat{\boldsymbol{\Sigma}}^{-1/2} + \boldsymbol{R}_{1,2} + \boldsymbol{R}_{1,3} + \boldsymbol{R}_{1,1} \\
    &= (\hat{\boldsymbol{\Lambda}} - \bar{\boldsymbol{\Lambda}})\bar{\boldsymbol{V}}\bar{\boldsymbol{\Sigma}}^{-1/2}\boldsymbol{W} + \boldsymbol{R}_{\boldsymbol{X}},
\end{align}
with $\boldsymbol{R}_{\boldsymbol{X}}:= \boldsymbol{R}_{1,4} + \boldsymbol{R}_{1,2} + \boldsymbol{R}_{1,3} + \boldsymbol{R}_{1,1}$, and each $\boldsymbol{R}_{\cdot,\cdot}$ defined as in Proposition \ref{prop:residuals}. By Proposition \ref{prop:residuals}, we know $\lVert \boldsymbol{R}_{\boldsymbol{X}}\rVert_{2,\infty} = \mathcal{O}\left(\max\{M_N,L_N\}^{3/2}M_N^{7/4}L_N^{-5/4}N^{-1}\log^{9/2}(N)\right)$ almost surely, which says
\begin{equation}
    \lVert\hat{\boldsymbol{X}} - \bar{\boldsymbol{X}}\boldsymbol{W}\rVert_{2,\infty} \leq \sigma_d(\bar{\boldsymbol{\Lambda}})^{-1/2}\lVert(\hat{\boldsymbol{\Lambda}} - \bar{\boldsymbol{\Lambda}})\bar{\boldsymbol{V}}\rVert_{2,\infty} + \mathcal{O}\left(\max\{M_N,L_N\}^{3/2}M_N^{7/4}L_N^{-5/4}N^{-1}\log^{9/2}(N)\right).
\end{equation}
 Consider the $(i,j)$-th element of $(\hat{\boldsymbol{\Lambda}} - \bar{\boldsymbol{\Lambda}})\bar{\boldsymbol{V}}$, which takes the form
\begin{equation}
    \left((\hat{\boldsymbol{\Lambda}} - \bar{\boldsymbol{\Lambda}})\bar{\boldsymbol{V}}\right)_{ij} = \sum_{k=1}^{NL_N}(\hat{\Lambda}_{ik} - \bar{\Lambda}_{ik})\bar{V}_{kj}.    
\end{equation}
Note that $\lVert\bar{V}_{\ast,j}\rVert_{\infty} \leq 1$ almost surely, and, as argued in the proof of Proposition \ref{prop:hat_lambda_bar_lambda}, for large enough $N$, for all $(i,j) \in \mathcal{X}^{M_N,L_N}$, the event $|\hat{\Lambda}_{ij} - \bar{\Lambda}_{ij}| < M_N \log(N)$ occurs almost surely. We can thus apply Hoeffding's inequality, by which we know
\begin{equation}
    \mathbb{P}\left(\abs{\sum_{k=1}^{NL_N} (\hat{\Lambda}_{ik} - \bar{\Lambda}_{ik})\bar{V}_{kj}} \geq t\right) \leq 2\exp\left(-\frac{2t^2}{\sum_{k=1}^{NL_N}\abs{\bar{V}_{kj}}^2M_N^2\log^2(N)}\right) \leq 2\exp\left(-\frac{2t^2}{M_N^2\log^2(N)}\right),
\end{equation}
from which it follows that 
\begin{equation}
    \abs{\left((\hat{\boldsymbol{\Lambda}} - \bar{\boldsymbol{\Lambda}})\bar{\boldsymbol{V}}\right)_{ij}} = \mathcal{O}\left(M_N\log^{3/2}(N)\right),
\end{equation}
almost surely. By taking a union over its $d$ elements, we deduce that $\lVert((\hat{\boldsymbol{\Lambda}} - \bar{\boldsymbol{\Lambda}})\bar{\boldsymbol{V}})_{i,\ast}\rVert = \mathcal{O}\left(M_N\log^{3/2}(N)\right)$, almost surely. By Proposition \ref{prop:behaviour_latent_position_matrices}, we know $\sigma_d(\bar{\boldsymbol{\Lambda}})^{-1/2} = \mathcal{O}\left(M_N^{-1/4}L_N^{-1/4}N^{-1/2}\right)$. Combining, it follows that
\begin{equation}
\sigma_d(\bar{\boldsymbol{\Lambda}})^{-1/2}\lVert(\hat{\boldsymbol{\Lambda}} - \bar{\boldsymbol{\Lambda}})\bar{\boldsymbol{V}}\rVert_{2,\infty} = \mathcal{O}\left(M_N^{3/4}L_N^{-1/4}N^{-1/2}\log^{3/2}(N)\right),
\end{equation}
almost surely. It follows that
%\textcolor{red}{Why do we split into $M$ pieces for no apparent reason?} Define $\hat{\boldsymbol{X}}^m$ and $\bar{\boldsymbol{X}}^m$ as the $N\times d$ matrices consisting of the set of rows from $N(m-1) + 1$ up to $Nm$, inclusive, of $\hat{\boldsymbol{X}}$ and $\bar{\boldsymbol{X}}$, respectively. Define analogously the $(N \times NL)$ matrices $\hat{\boldsymbol{\Lambda}}^m$ and $\bar{\boldsymbol{\Lambda}}^m$ and the $N \times d$ matrix $\boldsymbol{R}_{\boldsymbol{X}}^m$. We then clearly have
\begin{equation}
    \max_{m\in [M]}\lVert\hat{\boldsymbol{X}}^{(m,\ast)} - \bar{\boldsymbol{X}}^{(m,\ast)}\boldsymbol{W}\rVert_{2,\infty} = \lVert\hat{\boldsymbol{X}} - \bar{\boldsymbol{X}}\boldsymbol{W}\rVert_{2,\infty} = \mathcal{O}\left(M_N^{3/4}L_N^{-1/4}N^{-1/2}\log^{3/2}(N)\right).
\end{equation}
To complete this part of the proof, define $\boldsymbol{W}_{\boldsymbol{X}} = \boldsymbol{W}^{-1}\breve{\boldsymbol{K}}^{-1}$ and note
\begin{align}
    \lVert\hat{\boldsymbol{X}}\boldsymbol{W}_{\boldsymbol{X}} - \tilde{\boldsymbol{X}}\rVert_{2,\infty} \leq \lVert\hat{\boldsymbol{X}} - \tilde{\boldsymbol{X}}\boldsymbol{W}_{\boldsymbol{X}}^{-1}\rVert_{2,\infty} \lVert \boldsymbol{W}_{\boldsymbol{X}}\rVert = \lVert\hat{\boldsymbol{X}} - \bar{\boldsymbol{X}}\boldsymbol{W}\rVert_{2,\infty} \lVert\boldsymbol{W}^{-1}\rVert\lVert\breve{\boldsymbol{K}}^{-1}\rVert. 
\end{align}
Recalling that $\boldsymbol{W}$ is orthogonal, upon combining with Proposition \ref{prop:KR_spec_norm}, we obtain the final result
\begin{equation}
    \lVert\hat{\boldsymbol{X}}\boldsymbol{W}_{\boldsymbol{X}} - \tilde{\boldsymbol{X}}\rVert_{2,\infty} = \mathcal{O}\left(M_NL_N^{-1/2}N^{-1/2}\log^{3/2}(N)\right),
\end{equation}
almost surely. \\

An analogous set of algebraic steps gives
\begin{equation}
    \hat{\boldsymbol{Y}} - \bar{\boldsymbol{Y}}\boldsymbol{W} = (\hat{\boldsymbol{\Lambda}} - \bar{\boldsymbol{\Lambda}})^\top \bar{\boldsymbol{U}}\bar{\boldsymbol{\Sigma}}^{-1/2}\boldsymbol{W} + \boldsymbol{R}_{\boldsymbol{Y}},
\end{equation}
where $\boldsymbol{R}_{\boldsymbol{Y}} = \boldsymbol{R}_{2,1} + \boldsymbol{R}_{2,2} + \boldsymbol{R}_{2,3} + \boldsymbol{R}_{2,4}$, with the $\boldsymbol{R}_{2,\ast}$ defined as in Proposition \ref{prop:residuals}. The same proposition tells us $\lVert\boldsymbol{R}_{\boldsymbol{Y}}\rVert_{2,\infty} = \mathcal{O}\left(\max\{M_N,L_N\}^{3/2}M_N^{7/4}L_N^{-5/4}N^{-1}\log^{9/2}(N)\right)$ almost surely. Following the same approach, we consider the $(i,j)$-th element of $(\hat{\boldsymbol{\Lambda}} - \bar{\boldsymbol{\Lambda}})^\top\bar{\boldsymbol{U}}$, which takes the form
\begin{equation}
    \left((\hat{\boldsymbol{\Lambda}} - \bar{\boldsymbol{\Lambda}})^\top\bar{\boldsymbol{U}}\right)_{ij} = \sum_{k=1}^{NM}(\hat{\Lambda}_{ki} - \bar{\Lambda}_{ki})\bar{U}_{kj}.
\end{equation}
Again, we know $\norm{\bar{U}_{\ast, j}} \leq 1$ almost surely, and, for large enough $N$, for all $(i,j)\in\mathcal{X}^{N,M_N,L_N}$ the event $|\Lambda_{ij} - \bar{\Lambda}_{ij}| < M_N\log(N)$ occurs almost surely. We can apply Hoeffding's inequality in an analogous way to before to obtain
\begin{equation}
    \mathbb{P}\left(\abs{\sum_{k=1}^{NM_N} (\hat{\Lambda}_{ki} - \bar{\Lambda}_{ki})\bar{U}_{kj}} \geq t\right) \leq 2 \exp\left(-\frac{2t^2}{\sum_{k=1}^{NM_N}\abs{U_{kj}}^2M_N^2\log^2(N)}\right) \leq 2\exp\left(-\frac{2t^2}{M_N^2\log^2(N)}\right),
\end{equation}
from which it follows as before that 
\begin{equation}
    \abs{\left(\hat{\boldsymbol{\Lambda}} - \bar{\boldsymbol{\Lambda}}\bar{\boldsymbol{U}}\right)_{i,\ast}} = \mathcal{O}\left(M_N\log^{3/2}(N)\right),
\end{equation}
almost surely. We deduce
\begin{equation}
    \sigma_d(\bar{\boldsymbol{\Lambda}})^{-1/2}\lVert(\hat{\boldsymbol{\Lambda}} - \bar{\boldsymbol{\Lambda}})\bar{\boldsymbol{U}}\rVert_{2,\infty} = \mathcal{O}\left(M_N^{3/4}L_N^{-1/4}N^{-1/2}\log^{3/2}(N)\right),
\end{equation}
almost surely. Defining $\boldsymbol{W}_{\boldsymbol{Y}} = \boldsymbol{W}^{-1}\breve{\boldsymbol{R}}^{-1}$ and using Proposition \ref{prop:KR_spec_norm}, we obtain the result
\begin{equation}
    \lVert\hat{\boldsymbol{Y}}\boldsymbol{W}_{\boldsymbol{Y}} - \boldsymbol{Y}\rVert_{2,\infty} = \mathcal{O}\left(M_N^{1/2}N^{-1/2}\log^{3/2}(N)\right),
\end{equation}
almost surely. 

\section{Proof of Theorem \ref{thm:asymptotic_normality}} \label{sec:proof_thm2}

We start with the left embedding, and recall the identities $\boldsymbol{W}_{\boldsymbol{X}} = \boldsymbol{W}^{-1}\breve{\boldsymbol{K}}^{-1}$ and $\tilde{\boldsymbol{X}} = \bar{\boldsymbol{X}}\boldsymbol{K}^{-1}$. We can then write
\begin{align}
    N^{1/2}L_N^{1/2}(\hat{\boldsymbol{X}}\boldsymbol{W}_{\boldsymbol{X}} - \tilde{\boldsymbol{X}}) &= (\hat{\boldsymbol{X}} - \bar{\boldsymbol{X}}\breve{\boldsymbol{K}}^{-1}\boldsymbol{W}_{\boldsymbol{X}}^{-1})\boldsymbol{W}_{\boldsymbol{X}} \\
    &= N^{1/2}L_N^{1/2}(\hat{\boldsymbol{X}} - \bar{\boldsymbol{X}}\boldsymbol{W})\boldsymbol{W}_{\boldsymbol{X}} \\
    &=N^{1/2}L_N^{1/2}(\hat{\boldsymbol{\Lambda}} - \bar{\boldsymbol{\Lambda}})\bar{\boldsymbol{V}}\bar{\boldsymbol{\Sigma}}^{-1/2}\boldsymbol{W}\boldsymbol{W}_{\boldsymbol{X}} + N^{1/2}L_N^{1/2}\boldsymbol{R}_{\boldsymbol{X}}\boldsymbol{W}_{\boldsymbol{X}} \\
    &= N^{1/2}L_N^{1/2}(\hat{\boldsymbol{\Lambda}} - \bar{\boldsymbol{\Lambda}})\bar{\boldsymbol{V}}\bar{\boldsymbol{\Sigma}}^{-1/2}\breve{\boldsymbol{K}}^{-1} + N^{1/2}L_{N}^{1/2}\boldsymbol{R}_{\boldsymbol{X}}\boldsymbol{W}_{\boldsymbol{X}},
\end{align}
where $\boldsymbol{R}_{\boldsymbol{X}}$ is defined as in Section \ref{app_sec:proof_thm_1}. From our earlier work, we know that the second term on the right-hand side decays vanishes in two-to-infinity norm almost surely. 

Note that the definition $\bar{\boldsymbol{Y}} = \bar{\boldsymbol{V}}\bar{\boldsymbol{\Sigma}}^{1/2}$ implies that $\bar{\boldsymbol{V}}\bar{\boldsymbol{\Sigma}}^{1/2} = \bar{\boldsymbol{Y}}\bar{\boldsymbol{\Sigma}}^{-1} = \bar{\boldsymbol{Y}}\breve{\boldsymbol{R}}\bar{\boldsymbol{\Sigma}}^{-1}$, and so
\begin{equation}
    N^{1/2}L_N^{1/2}(\hat{\boldsymbol{\Lambda}} - \bar{\boldsymbol{\Lambda}})\bar{\boldsymbol{V}}\bar{\boldsymbol{\Sigma}}^{-1/2}\breve{\boldsymbol{K}}^{-1} = (\hat{\boldsymbol{\Lambda}} - \bar{\boldsymbol{\Lambda}})\bar{\boldsymbol{Y}}\breve{\boldsymbol{R}}\bar{\boldsymbol{\Sigma}}^{-1}\breve{\boldsymbol{K}}^{-1}.
\end{equation}
We consider the $m$-th block:
\begin{align}
    N^{1/2}L_N^{1/2}(\hat{\boldsymbol{X}}^{(m,\ast)} \boldsymbol{W}_{\boldsymbol{X}} - \tilde{\boldsymbol{X}}^{(m,\ast)}) = N^{1/2}L_N^{1/2}(\hat{\boldsymbol{\Lambda}}^{(m,\ast)} - \bar{\boldsymbol{\Lambda}}^{(m,\ast)})\bar{\boldsymbol{Y}}\breve{\boldsymbol{R}}\bar{\boldsymbol{\Sigma}}^{-1}\breve{\boldsymbol{K}}^{-1}.
\end{align}
Recall that $\breve{\boldsymbol{K}}\breve{\boldsymbol{R}}^\top = \boldsymbol{I}_d$ and so we can write
\begin{equation}
    \tilde{\boldsymbol{X}}\breve{\boldsymbol{K}}\bar{\boldsymbol{\Sigma}}\breve{\boldsymbol{K}}^\top \tilde{\boldsymbol{X}}^\top = \bar{\boldsymbol{X}}\bar{\boldsymbol{\Sigma}}\bar{\boldsymbol{X}}^\top = \bar{\boldsymbol{\Lambda}}\bar{\boldsymbol{\Lambda}}^\top = \tilde{\boldsymbol{X}}\boldsymbol{Y}^\top \boldsymbol{Y}\tilde{\boldsymbol{X}}^\top.
\end{equation}
It follows that $\breve{\boldsymbol{K}}\bar{\boldsymbol{\Sigma}}\breve{\boldsymbol{K}}^\top = \boldsymbol{Y}^\top \boldsymbol{Y}$ and
\begin{align}
    (\breve{\boldsymbol{R}}\bar{\boldsymbol{\Sigma}}^{-1}\breve{\boldsymbol{K}}^{-1})^\top = \breve{\boldsymbol{K}}^{-1\ \top} \bar{\boldsymbol{\Sigma}}^{-1} \breve{\boldsymbol{K}}^{-1}\breve{\boldsymbol{K}}\breve{\boldsymbol{R}}^\top = (\boldsymbol{Y}^\top \boldsymbol{Y})^{-1} \breve{\boldsymbol{K}}\breve{\boldsymbol{R}}^\top = (\boldsymbol{Y}^\top \boldsymbol{Y})^{-1}.
\end{align}
Using this result, we can decompose further and consider the transpose of the $i$-th row:
\begin{align}
    N^{1/2}L_N^{1/2}(\hat{\boldsymbol{X}}^{(m,\ast)} \boldsymbol{W}_{\boldsymbol{X}} - \tilde{\boldsymbol{X}}^{(m,\ast)})_{i,\ast}^\top &= N^{1/2}L_N^{1/2} (\breve{\boldsymbol{R}}\bar{\boldsymbol{\Sigma}}^{-1}\breve{\boldsymbol{K}}^{-1})^\top [(\hat{\boldsymbol{\Lambda}}^{(m,\ast)} - \bar{\boldsymbol{\Lambda}}^{(m,\ast)})\boldsymbol{Y}]_i^\top \\
    &= N^{1/2}L_N^{1/2}(\boldsymbol{Y}^\top \boldsymbol{Y})^{-1} \sum_{j=1}^{NL_N}  (\hat{\Lambda}^{(m,\ast)}_{ij} - \bar{\Lambda}^{(m,\ast)}_{ij})Y_{j,\ast}^\top \\
    &= NL_N(\boldsymbol{Y}^\top \boldsymbol{Y})^{-1} \frac{1}{N^{1/2}L_N^{1/2}}\sum_{j=1}^{NL_N}  (\hat{\Lambda}^{(m,\ast)}_{ij} - \bar{\Lambda}^{(m,\ast)}_{ij})Y_{j,\ast}^\top.
\end{align}
We first note that as $N\to \infty$, we have $NL_N(\boldsymbol{Y}^\top \boldsymbol{Y})^{-1} \to \boldsymbol{Q}_{\boldsymbol{Y}}^{-1}$ by Assumption \ref{ass:moment_stab}. Define $A^{m}_{ij}$ to be the random vector
\begin{equation}
    A_{ij}^{m} = (\hat{\Lambda}^{(m,\ast)}_{ij} - \bar{\Lambda}^{(m,\ast)}_{ij})Y_{j,\ast}^\top,
\end{equation}
which has covariance given by
\begin{equation}
   \text{Cov}(A_{ij}^{m}) = (\tilde{X}^{(m,\ast)}_{i,\ast}Y_{j,\ast}^\top) Y_{j,\ast}^\top Y_{j,\ast}.
\end{equation}
Now we define the matrix $\boldsymbol{C}_{i,m}^{N}$ to be
\begin{equation}
    \boldsymbol{C}_{i,m}^{N} = \frac{1}{NL_N}\sum_{j=1}^{NL_N}\text{Cov}(A_{ij}^{m}),
\end{equation}
which is invertible by Assumption \ref{ass:rank_prev_suff_dens}. It follows that
\begin{equation}
    \text{Cov}\left(N^{1/2}L_N^{1/2}(\hat{\boldsymbol{X}}^{(m,\ast)} \boldsymbol{W}_{\boldsymbol{X}} - \tilde{\boldsymbol{X}}^{(m,\ast)})_{i,\ast}^\top\right) = \boldsymbol{Q}_{\boldsymbol{X}}^{-1} \boldsymbol{C}_{i,m}^{N}(\tilde{\boldsymbol{X}})\boldsymbol{Q}_{\boldsymbol{X}}^{-1}.
\end{equation}
Because we work with Poisson random variables with intensities that are $\mathcal{O}(1)$, we have for all $u \in \mathbb{R}^d$ that
\begin{equation}
   \text{Cov}\left((\hat{\Lambda}^{(m,\ast)}_{ij} - \bar{\Lambda}^{(m,\ast)}_{ij})u^\top Y_{j,\ast}^\top\right) = \mathcal{O}(1) \quad \text{and} \quad \mathbb{E}\left\{|(\hat{\Lambda}^{(m,\ast)}_{ij} - \bar{\Lambda}^{(m,\ast)}_{ij})u^\top Y_{j,\ast}^\top|^3\right\} \leq C,
\end{equation}
uniformly in $j$ and so the Lyapnunov condition holds with $\delta = 1$. Consequently, by the multivariate Lindenberg-Feller central limit theorem, we can state that
\begin{equation}
    N^{1/2}L_N^{1/2}\left(\boldsymbol{Q}_{\boldsymbol{X}}^{-1} \boldsymbol{C}_{i,m}^{N}\boldsymbol{Q}_{\boldsymbol{X}}^{-1}\right)^{-1/2}(\hat{\boldsymbol{X}}^{(m,\ast)} \boldsymbol{W}_{\boldsymbol{X}} - \tilde{\boldsymbol{X}}^{(m,\ast)})_{i,\ast}^\top \rightarrow \mathcal{N}(0, \boldsymbol{I}_d),
\end{equation}
in distribution as $N \to \infty$. Note that this doesn't require that the sequence of $\boldsymbol{C}_i^{m,N}$ converges, just that they are eventually always invertible.

Similarly, since $\breve{\boldsymbol{K}} = (\tilde{\boldsymbol{X}}^\top \tilde{\boldsymbol{X}})^{-1}\tilde{\boldsymbol{X}}^\top \bar{\boldsymbol{X}}$ and $\bar{\boldsymbol{X}}^\top \bar{\boldsymbol{X}} = \bar{\boldsymbol{\Sigma}}$, we have
\begin{align}
(\breve{\boldsymbol{K}}\bar{\boldsymbol{\Sigma}}^{-1}\breve{\boldsymbol{R}}^{-1})^\top &= \breve{\boldsymbol{R}}^{-1 \ \top}\bar{\boldsymbol{\Sigma}}^{-1}\bar{\boldsymbol{X}}^\top \tilde{\boldsymbol{X}}(\tilde{\boldsymbol{X}}^\top \tilde{\boldsymbol{X}})^{-1} \\
    &= \breve{\boldsymbol{R}}^{-1\ \top}\bar{\boldsymbol{\Sigma}}^{-1}\bar{\boldsymbol{X}}^\top \bar{\boldsymbol{X}}\breve{\boldsymbol{K}}^{-1} (\tilde{\boldsymbol{X}}^\top \tilde{\boldsymbol{X}})^{-1} \\
    &= \breve{\boldsymbol{R}}^{-1\ \top}\bar{\boldsymbol{\Sigma}}^{-1}\bar{\boldsymbol{\Sigma}}\breve{\boldsymbol{K}}^{-1}(\tilde{\boldsymbol{X}}^\top \tilde{\boldsymbol{
    X}})^{-1} \\
    &= \breve{\boldsymbol{R}}^{-1\ \top} \breve{\boldsymbol{K}}^{-1}(\tilde{\boldsymbol{X}}^\top \tilde{\boldsymbol{X}})^{-1} \\
    &= (\tilde{\boldsymbol{X}}^\top \tilde{\boldsymbol{X}})^{-1}.
\end{align}
In a completely analogous way, we form the following decomposition:
\begin{equation}
    N^{1/2}M_N^{1/2}(\hat{\boldsymbol{Y}} \boldsymbol{W}_{\boldsymbol{Y}} - \boldsymbol{Y})_{i,\ast}^\top = NM_N(\tilde{\boldsymbol{X}}^\top \tilde{\boldsymbol{X}})^{-1}\frac{1}{N^{1/2}M_N^{1/2}}\sum_{j=1}^{NM_N}(\hat{\Lambda}_{ji}^{(\ast,\ell)}- \bar{\Lambda}_{ji}^{(\ast,\ell)})\tilde{X}_{j,\ast}^\top.
\end{equation}
Define the summand 
\begin{equation}
    B_{ji}^m = (\hat{\Lambda}_{ji}^{(\ast,\ell)} - \bar{\Lambda}_{ji}^{(\ast,\ell)})\tilde{X}_{j,\ast}^\top,
\end{equation}
so that we have
\begin{equation}
    \text{Cov}(B_{ji}^\ell) = (\tilde{X}_{j,\ast} Y_{i,\ast}^{(\ell, \ast)\ \top})\tilde{X}_{j,\ast}^\top\tilde{X}_{j,\ast}.
\end{equation}
If we define the matrices $\boldsymbol{D}_{i,\ell}^N$ to be
\begin{equation}
    \boldsymbol{D}_{i,\ell}^N = \frac{1}{NM_N}\sum_{j=1}^{NM_N} \text{Cov}(B_{ji}^\ell),
\end{equation}
which is again eventually always invertible by Assumption \ref{ass:rank_prev_suff_dens}, we can use an analogous argument to the above to obtain that
\begin{equation}
    N^{1/2}M_N^{1/2}\left(\boldsymbol{Q}_{\boldsymbol{Y}}^{-1} \boldsymbol{D}_{i,\ell}^{N}\boldsymbol{Q}_{\boldsymbol{Y}}^{-1}\right)^{-1/2}(\hat{\boldsymbol{Y}}^{(m,\ast)} \boldsymbol{W}_{\boldsymbol{Y}} - \tilde{\boldsymbol{Y}}^{(m,\ast)})_{i,\ast}^\top \rightarrow \mathcal{N}(0, \boldsymbol{I}_d),
\end{equation}
in distribution as $N \to \infty$.

\section{Proofs of key propositions}

\subsection{Proof of Proposition~\ref{prop:behaviour_latent_position_matrices}} \label{proof:prop1}

    Write $\boldsymbol{A}_{NM_N} = \tilde{\boldsymbol{X}}^\top \tilde{\boldsymbol{X}}/NM_N$, then entrywise convergence tells us
    \begin{equation}
        \norm{\boldsymbol{A}_{NM_N} - \boldsymbol{Q}_X}_F^2 = \sum_{k,\ell=1}^d(A_{NM_N,k\ell} - Q_{X,k\ell})^2 \to 0,
    \end{equation}
    and so $\norm{\boldsymbol{A}_{NM_N} - \boldsymbol{Q}_X} \leq \norm{\boldsymbol{A}_{NM_N} - \boldsymbol{Q}_X}_F \to 0$. Weyl's inequality then says
    \begin{equation}
        |\lambda_i(\boldsymbol{A}_{NM_N}) - \lambda_i(\boldsymbol{Q}_X)| \leq \norm{\boldsymbol{A}_{NM_N} - \boldsymbol{Q}_X} \to 0,
    \end{equation}
    and hence we have
    \begin{equation}
    \lambda_i(\tilde{\boldsymbol{X}}^\top\tilde{\boldsymbol{X}}/NM_N) \to \lambda_i(\boldsymbol{Q}_X) \quad \Rightarrow \quad \sigma_i(\tilde{\boldsymbol{X}})/N^{1/2}M_N^{1/2} \to \lambda_i(\boldsymbol{Q}_X)^{1/2}.
    \end{equation}
    As $\boldsymbol{Q}_X \succ 0$, we have $\lambda_i(\boldsymbol{Q}_X) \in (0,\infty)$ and so $\sigma_i(\tilde{\boldsymbol{X}}) = \Theta(N^{1/2}M_N^{1/2})$ for each $i \leq d$. Clearly, since $\tilde{\boldsymbol{X}}$ has only $d$ columns, $\sigma_{d+1}(\tilde{\boldsymbol{X}}) = 0$. For the condition number, we can 
    say
    \begin{equation}
        \kappa(\tilde{\boldsymbol{X}}) = \frac{\sigma_1(\tilde{\boldsymbol{X}})}{\sigma_d(\tilde{\boldsymbol{X}})} \to \frac{\lambda_1(\boldsymbol{Q}_X)^{1/2}}{\lambda_d(\tilde{\boldsymbol{Q}}_X)^{1/2}} = \mathcal{O}(1).
    \end{equation}
    The incoherence parameter can be argued by considering the skinny SVD $\tilde{\boldsymbol{X}} = \tilde{\boldsymbol{U}}\tilde{\boldsymbol{\Sigma}}\tilde{\boldsymbol{V}}^\top$ and that this says $\tilde{\boldsymbol{U}} = \tilde{\boldsymbol{X}}\tilde{\boldsymbol{V}}\tilde{\boldsymbol{\Sigma}}^{-1}$. We then have
    \begin{equation}
        \lVert\tilde{\boldsymbol{U}}\rVert_{2,\infty} \leq \lVert \tilde{\boldsymbol{X}}\rVert_{2,\infty}\lVert \boldsymbol{\Sigma}^{-1}\rVert \leq \mathcal{O}(1) \times N^{-1/2}M_N^{-1/2}\lambda_d(\boldsymbol{Q}_X),
    \end{equation}
    in the limit. The result follows. The same argument holds for $\boldsymbol{Y}$. 

\subsection{Proof of Proposition~\ref{prop:poisson_tail_bound}}

For any $\theta \in \mathbb{R}$, Markov's inequality gives
\begin{equation}
	\mathbb{P}(X \geq k) \leq \frac{\mathbb{E}\left\{\exp(\theta X)\right\}}{\exp(\theta k)} = \exp\left(\lambda \exp(\theta) - \lambda - \theta k \right),
\end{equation}
where in the final equality we use the moment generating function of a Poisson random variable. Setting $\theta = 1$ and $k = \beta_\rho \log(n)$, for some $\beta_\rho \in \mathbb{R}$ and $n \in \mathbb{N}$, we see
\begin{equation}
	\mathbb{P}(X \geq \beta_\rho \log(n)) \leq n^{-\beta_\rho}\exp\left(\lambda(e - 1)\right).
\end{equation}
As $\lambda$ is does not change with $n$, if we pick $\beta_\rho = \rho + \lambda(e - 1)$ then some algebra yields
\begin{equation}
	\mathbb{P}(X < \beta_\rho \log(n)) > 1 - n^{-\rho},
\end{equation}
as was required.

\subsection{Proof of Proposition \ref{prop:hat_lambda_bar_lambda}}

Define the set $\mathcal{X}^{N,M_N,L_N} = [NM_N] \times [NL_N]$. Let $\boldsymbol{A}_{ij}$ be the $NM_N \times NL_N$ matrix with entry $(i,j)$ taking the value $A_{ij,ij} = \hat{\Lambda}_{ij} - \bar{\Lambda}_{ij}$ and 0 in all other entries. Clearly $\mathbb{E}\{A_{ij,rs}\} = 0$ for all entries $(r,s)$. In this proof, we make explicit the upgrade from overwhelming probability to almost surely. We can argue that the  events $\{\forall (i,j) \in  \mathcal{X}^{N,M_N,L_N},\ \hat{\Lambda}_{ij} < M_N \log(N)\}$ occur almost surely as $N\to\infty$ through a union bound and by invoking the Borel-Cantelli Lemma. Each $\hat{\lambda}_{\ell ij}^m/M_N \sim \mathrm{Poisson}(\bar{\lambda}_{\ell ij}^m/M_N)$, and by Assumption \ref{ass:bound_and_int}, we have $\bar{\lambda}^m_{\ell ij} = \mathcal{O}(1)$ which implies that there exists a $\lambda_{\text{max}} > 0$ such that $\bar{\lambda}^{m}_{\ell ij} \leq \lambda_{\text{max}}$ eventually as $N \to \infty$. Proposition \ref{prop:poisson_tail_bound} then tells us that
\begin{equation}
	\mathbb{P}\left(\hat{\lambda}_{\ell ij} < \left(\rho M_N + (e-1)\lambda_{\text{max}}\right)\log(N)\right) > 1 - N^{-\rho},
\end{equation}
for any $\rho > 0$.%Define $\beta = \max_{i,j} \beta_\rho^{ij}$, which is a constant dependent on $\rho$ but independent of $N$ by our assumption of a uniform bound. 
We then have
\begin{align}
	\mathbb{P}&\left(\bigcup_{(i,j) \in \mathcal{X}^{N,M_N,L_N}} \left\{\hat{\Lambda}_{ij} \geq \left(\rho M_N + (e-1)\lambda_{\text{max}}\right)\log(N)\right\}\right)\\
	 &\leq \sum_{(i,j) \in \mathcal{X}^{N,M_N,L_N}} \mathbb{P}\left(\hat{\Lambda}_{ij}
	\geq \left(\rho M_N + (e-1)\lambda_{\text{max}}\right)\log(N) \right)\\ 
	&\leq N^{2 - \rho}M_NL_N.
\end{align}
For a large enough choice of $\rho$, we have $\sum_{N \geq 1} N^{2-\rho} < \infty$, and so Borel-Cantelli says 
\begin{equation}
	\mathbb{P}\left(\left\{\exists (i,j) \in \mathcal{X}^{N,M_N,L_N} \ : \ \hat{\Lambda}_{ij} \geq M_N \log(N)\right\} \ \text{infinitely often}\right) = 0.
\end{equation}
It follows that for large enough $N$, the event that every entry of $\hat{\boldsymbol{\Lambda}}$ is bounded by $M \log (N)$ occurs almost surely. Furthermore, by Assumption \ref{ass:bound_and_int}, we have that $\bar{\Lambda}_{ij} = \mathcal{O}(1)$. Using the simple bound
\begin{equation}
	\norm{\boldsymbol{A}_{ij}} = |\hat{\Lambda}_{ij} - \bar{\Lambda}_{ij}| \leq |\hat{\Lambda}_{ij}| + \abs{\bar{\Lambda}_{ij}}, 
\end{equation}
we combine these results to obtain that for large enough $N$, for all $(i,j) \in \mathcal{X}^{N,M_N,L_N}$, the event $\norm{\boldsymbol{A}_{ij}} < M_N \log(N)$ occurs almost surely. We can now apply Bernstein's theorem as stated in Theorem \ref{thm:matrix-Bernstein} to $\boldsymbol{A} = \sum_{(i,j) \in \mathcal{X}^{N,M_N,L_N}} \boldsymbol{A}_{ij}$. To apply Bernstein, we need to first bound $v(\boldsymbol{A})$. To do so, we note that
\begin{equation}
	(\boldsymbol{A}_{ij}\boldsymbol{A}_{ij}^\top)_{rs} = \sum_{k \in [NL]}A_{ij,rk} A_{ij,sk} = 
	\begin{cases}
		\left(\hat{\Lambda}_{ij} - \bar{\Lambda}_{ij}\right)^2, & (r,s) = (i,i), \\
		0, & (r,s) \neq (i,i).
	\end{cases}
\end{equation}
We then have 
\begin{equation}
	\sum_{(i,j) \in \mathcal{X}^{M_N,L_N}} \boldsymbol{A}_{ij}\boldsymbol{A}_{ij}^\top = \mathrm{diag}\left(\sum_{k \in [NL_N]} \left(\hat{\Lambda}_{1k} - \bar{\Lambda}_{1k}\right)^2, \dots, \sum_{k \in [NL_N]} \left(\hat{\Lambda}_{(NM_N)k} - \bar{\Lambda}_{(NM_N)k}\right)^2\right),
\end{equation}
where the notation $B_{(ab)c}$ is to be read as the $(ab,c)$-th entry of $\boldsymbol{B}$. By Popoviciu's inequality, we have $\text{Var}\left\{\hat{\Lambda}_{ij}\right\} \leq M_N^2 \log^2(N)$ for each $(i,j)\in\mathcal{X}^{N,M_N,L_N}$ almost surely. As we have a diagonal matrix, after expectations and invoking Popoviciu's inequality, it follows 
\begin{equation}
	\norm{\,\mathbb{E}\left\{\sum_{(i,j) \in \mathcal{X}^{N,M_N,L_N}} \boldsymbol{A}_{ij}\boldsymbol{A}_{ij}^\top\right\}\,} \leq NL_NM_N^2\log^2(N),
\end{equation}
almost surely. An analogous argument yields
\begin{equation}
	\norm{\,\mathbb{E}\left\{\sum_{(i,j) \in \mathcal{X}^{N,M_N,L_N}} \boldsymbol{A}_{ij}^\top \boldsymbol{A}_{ij}\right\}\,} \leq NM_N^3\log^2(N),
\end{equation}
almost surely. Substituting into Theorem \ref{thm:matrix-Bernstein}, we have that $t \geq 0$:
\begin{equation}
	\mathbb{P}\left(\lVert\hat{\boldsymbol{\Lambda}} - \bar{\boldsymbol{\Lambda}}\rVert \geq t\right) \leq N(M_N + L_N) \exp\left\{-\frac{3t^2}{6\max\{M_N,L_N\}NM_N^2\log^2(N) + 2M_N t\log(N)}\right\},
\end{equation}
almost surely. For sufficiently large $N$, the numerator in the exponential dominates if we select $t$ to be $t = \rho \max\{M_N,L_N\}^{1/2}M_NN^{1/2}\log^{3/2}(N)$. From which it follows that
\begin{equation}
	\lVert\hat{\boldsymbol{\Lambda}} - \bar{\boldsymbol{\Lambda}}\rVert = \mathcal{O}\left(\max\{M_N,L_N\}^{1/2}M_NN^{1/2}\log^{3/2}(N)\right),
\end{equation}
with overwhelming probability, and thus eventually almost surely as $N \to \infty$, as was required.

\subsection{Proof of Proposition \ref{prop:sing_vals_lam_hat}}

    By Corollary 7.3.5 of \citeSM{horn2012}, for any two matrices $\boldsymbol{A}$ and $\boldsymbol{B}$ of the same size, we have
\begin{equation}
	\sigma_i(\boldsymbol{B}) - \norm{\boldsymbol{A} - \boldsymbol{B}} \leq \sigma_i(\boldsymbol{A}) \leq \sigma_i(\boldsymbol{B}) + \norm{\boldsymbol{A} - \boldsymbol{B}}.
\end{equation}
Applying this result with $\boldsymbol{A} = \hat{\boldsymbol{\Lambda}}$ and $\boldsymbol{B} = \bar{\boldsymbol{\Lambda}}$, applying Proposition \ref{prop:behaviour_latent_position_matrices} and noting that $\sigma_{d+1}(\bar{\boldsymbol{\Lambda}}) = 0$ yields the desired rates.

\subsection{Proof of Proposition \ref{prop:U_lambda_hat_lambda_bar_V}}

	We can write 
\begin{equation}
	\left(\bar{\boldsymbol{U}}^\top (\hat{\boldsymbol{\Lambda}} - \bar{\boldsymbol{\Lambda}})\bar{\boldsymbol{V}}\right)_{rs} = \sum_{p=1}^{NM_N}\sum_{q=1}^{NL_N} \bar{U}^\top_{rp}(\hat{\Lambda} - \bar{\Lambda})_{pq}\bar{V}_{qs} = \sum_{p=1}^{NM_N}\sum_{q=1}^{NL_N} \bar{U}_{pr}\bar{V}_{sq}(\hat{\Lambda} - \bar{\Lambda})_{pq},
\end{equation}
which is the sum of independent random variables with expectation 0 and which are almost surely bounded in magnitude by $|\bar{U}_{pr}\bar{V}_{sq}|M_N \log(N)$ by the argument presented in Proposition \ref{prop:hat_lambda_bar_lambda}. Hoeffding's inequality then says
\begin{multline}
	\mathbb{P}\left(\abs{\left(\bar{\boldsymbol{U}}^\top (\hat{\boldsymbol{\Lambda}} - \bar{\boldsymbol{\Lambda}})\bar{\boldsymbol{V}}\right)_{rs}} \geq t\right) \leq \\ 2\exp\left(-\frac{2t^2}{\sum_{p=1}^{NM_N}\sum_{q=1}^{NL_N}\abs{\bar{U}_{pr}\bar{V}_{sq}}^2M_N^2\log^2(N)}\right) \leq 2\exp\left(-\frac{2t^2}{M_N^2\log^2(N)}\right),
\end{multline}
where the final inequality follows since $\bar{\boldsymbol{U}}, \bar{\boldsymbol{V}}$ are unitary. If we pick $t = cM_N\log^{3/2}(N)$, we obtain
\begin{equation}
	\mathbb{P}\left(\abs{\left(\bar{\boldsymbol{U}}^\top (\hat{\boldsymbol{\Lambda}} - \bar{\boldsymbol{\Lambda}})\bar{\boldsymbol{V}}\right)_{rs}} \geq M_N\log^{3/2}(N)\right) \leq 2N^{-2c^2},
\end{equation}
and since the matrix is $d \times d$, a union bound gives 
\begin{equation}
	\lVert\bar{\boldsymbol{U}}^\top(\hat{\boldsymbol{\Lambda}} - \bar{\boldsymbol{\Lambda}})\bar{\boldsymbol{V}}\rVert_F = \mathcal{O}\left(M_N\log^{3/2}(N)\right),
\end{equation}
almost surely.

\subsection{Proof of Proposition \ref{prop:lots_of_bounds}}

We prove the results in order:
\begin{enumerate}
	\item Let $\sigma_1,\dots,\sigma_d$ be the singular values of $\bar{\boldsymbol{U}}^\top\hat{\boldsymbol{U}}$ and $\theta_\ell = \cos^{-1}(\sigma_\ell)$ to be the principal angles. Lemma 2.4 of \citeSM{chen_2021} tells us that the non-zero eigenvalues of $\hat{\boldsymbol{U}}\hat{\boldsymbol{U}}^\top - \bar{\boldsymbol{U}}\bar{\boldsymbol{U}}^\top$ are equal to $\sin(\theta_\ell)$. By invoking a variant of the Davis-Kahan theorem \citepSM[][Theorem 4]{yu_2014}, we find that
	\begin{equation}
		\lVert\hat{\boldsymbol{U}}\hat{\boldsymbol{U}}^\top - \bar{\boldsymbol{U}}\bar{\boldsymbol{U}}^\top\rVert = \max_{\ell \in [d]}\abs{\sin(\theta_\ell)} \leq \frac{2d^{1/2}\lVert\hat{\boldsymbol{\Lambda}} - \bar{\boldsymbol{\Lambda}}\rVert\left(2\sigma_1(\hat{\boldsymbol{\Lambda}}) + \lVert\hat{\boldsymbol{\Lambda}} - \bar{\boldsymbol{\Lambda}}\rVert\right)}{\sigma_d(\bar{\boldsymbol{\Lambda}})^2 - \sigma_{d+1}(\bar{\boldsymbol{\Lambda}})^2},
	\end{equation}
	for sufficiently large $N$. Using that $\sigma_{d+1}(\bar{\boldsymbol{\Lambda}}) = 0$, and $\sigma_d(\bar{\boldsymbol{\Lambda}}) = \Omega(NM_N^{1/2}L_N^{1/2})$ by Proposition \ref{prop:behaviour_latent_position_matrices} and Propositions \ref{prop:hat_lambda_bar_lambda} and \ref{prop:sing_vals_lam_hat}, we obtain
	\begin{equation*}
		\lVert\hat{\boldsymbol{U}}\hat{\boldsymbol{U}}^\top - \bar{\boldsymbol{U}}\bar{\boldsymbol{U}}^\top\rVert = \mathcal{O}\left(\max\{M_N,L_N\}^{1/2}M_N^{1/2}L_N^{-1/2}N^{-1/2}\log^{3/2}(N)\right),
	\end{equation*}
	almost surely. The second result follows analogously.
	\item We use the bound from Part 1 to obtain
	\begin{align}
		\lVert\hat{\boldsymbol{U}} - \bar{\boldsymbol{U}}\bar{\boldsymbol{U}}^\top\hat{\boldsymbol{U}}\rVert_F = \lVert(\hat{\boldsymbol{U}}\hat{\boldsymbol{U}}^\top - \bar{\boldsymbol{U}}\bar{\boldsymbol{U}}^\top)\hat{\boldsymbol{U}}\rVert_F &\leq \lVert\hat{\boldsymbol{U}}\hat{\boldsymbol{U}}^\top - \bar{\boldsymbol{U}}\bar{\boldsymbol{U}}^\top\rVert_F\\ 
		&= \mathcal{O}\left(\max\{M_N,L_N\}^{1/2}M_N^{1/2}L_N^{-1/2}N^{-1/2}\log^{3/2}(N)\right),
	\end{align}
	almost surely. The second result follows analogously.
	\item Observe the following
	\begin{align}
		\bar{\boldsymbol{U}}^\top \hat{\boldsymbol{U}}\hat{\boldsymbol{\Sigma}} - \bar{\boldsymbol{\Sigma}}\bar{\boldsymbol{V}}^\top\hat{\boldsymbol{V}} &= \bar{\boldsymbol{U}}^\top(\hat{\boldsymbol{\Lambda}} - \bar{\boldsymbol{\Lambda}})\hat{\boldsymbol{V}} \\
		&= \bar{\boldsymbol{U}}^\top(\hat{\boldsymbol{\Lambda}} - \bar{\boldsymbol{\Lambda}})(\hat{\boldsymbol{V}} - \bar{\boldsymbol{V}}\bar{\boldsymbol{V}}^\top \hat{\boldsymbol{V}}) + \bar{\boldsymbol{U}}^\top (\hat{\boldsymbol{\Lambda}} - \bar{\boldsymbol{\Lambda}})\bar{\boldsymbol{V}}\bar{\boldsymbol{V}}^\top\hat{\boldsymbol{V}}.
	\end{align}
	Note first that $\lVert\bar{\boldsymbol{U}}^\top\rVert_F = \lVert\bar{\boldsymbol{U}}\rVert_F = d^{1/2} = \mathcal{O}(1)$, and similarly for $\lVert\bar{\boldsymbol{V}}^\top\hat{\boldsymbol{V}}^\top\rVert_F$. Propositions \ref{prop:hat_lambda_bar_lambda} and \ref{prop:U_lambda_hat_lambda_bar_V}, coupled with the result of Part 2. above, then yield
	\begin{align}
		&\lVert\bar{\boldsymbol{U}}^\top \bar{\boldsymbol{U}}\hat{\boldsymbol{\Sigma}} - \bar{\boldsymbol{\Sigma}}\bar{\boldsymbol{V}}^\top\hat{\boldsymbol{V}}\rVert_F \\
		&\leq \lVert\hat{\boldsymbol{U}}^\top\rVert_F\lVert\hat{\boldsymbol{\Lambda}} - \bar{\boldsymbol{\Lambda}}\rVert_F\lVert\hat{\boldsymbol{V}} - \bar{\boldsymbol{V}}\bar{\boldsymbol{V}}^\top\hat{\boldsymbol{V}}\rVert_F  + \lVert\bar{\boldsymbol{U}}^\top (\hat{\boldsymbol{\Lambda}} - \bar{\boldsymbol{\Lambda}})\bar{\boldsymbol{V}}\rVert_F\lVert\bar{\boldsymbol{V}}^\top\hat{\boldsymbol{V}}^\top\rVert_F \\
		&= \mathcal{O}\left(\max\{M_N,L_N\}M_N^{3/2}L_N^{-1/2}\log^3(N)\right) + \mathcal{O}\left(M_N\log^{3/2}(N)\right) \\
		&= \mathcal{O}\left(\max\{M_N,L_N\}M_N^{3/2}L_N^{-1/2}\log^3(N)\right),
	\end{align}
	almost surely. The second result follows analogously.
	\item Some algebra allows us to write
	\begin{align}
		\bar{\boldsymbol{U}}^\top \hat{\boldsymbol{U}} - \bar{\boldsymbol{V}}^\top \hat{\boldsymbol{V}} = \left[\left(\bar{\boldsymbol{U}}^\top \hat{\boldsymbol{U}}\hat{\boldsymbol{\Sigma}} - \bar{\boldsymbol{\Sigma}}\bar{\boldsymbol{V}}^\top\hat{\boldsymbol{V}}\right) + \left(\bar{\boldsymbol{\Sigma}}\bar{\boldsymbol{U}}^\top\hat{\boldsymbol{U}} - \bar{\boldsymbol{V}}^\top\hat{\boldsymbol{V}}\hat{\boldsymbol{\Sigma}}\right)\right]\hat{\boldsymbol{\Sigma}}^{-1} - \bar{\boldsymbol{\Sigma}}(\bar{\boldsymbol{U}}^\top\hat{\boldsymbol{U}} - \bar{\boldsymbol{V}}^\top\hat{\boldsymbol{V}})\hat{\boldsymbol{\Sigma}}^{-1},
	\end{align}
	which yields the identity
	\begin{equation}
		\bar{\boldsymbol{U}}^\top \hat{\boldsymbol{U}} - \bar{\boldsymbol{V}}^\top \hat{\boldsymbol{V}} + \bar{\boldsymbol{\Sigma}}(\bar{\boldsymbol{U}}^\top\hat{\boldsymbol{U}} - \bar{\boldsymbol{V}}^\top\hat{\boldsymbol{V}})\hat{\boldsymbol{\Sigma}}^{-1} = \left[\left(\bar{\boldsymbol{U}}^\top \hat{\boldsymbol{U}}\hat{\boldsymbol{\Sigma}} - \bar{\boldsymbol{\Sigma}}\bar{\boldsymbol{V}}^\top\hat{\boldsymbol{V}}\right) + \left(\bar{\boldsymbol{\Sigma}}\bar{\boldsymbol{U}}^\top\hat{\boldsymbol{U}} - \bar{\boldsymbol{V}}^\top\hat{\boldsymbol{V}}\hat{\boldsymbol{\Sigma}}\right)\right]\hat{\boldsymbol{\Sigma}}^{-1}
		\label{eqn:ij_identity}
	\end{equation}
	Using the definition of $\bar{\boldsymbol{\Sigma}}$ and $\hat{\boldsymbol{\Sigma}}$, we see that the $(i,j)$th entry of the left-hand side of \eqref{eqn:ij_identity} in absolute value is
	\begin{align}
		\abs{\left(\bar{\boldsymbol{U}}^\top \hat{\boldsymbol{U}} - \bar{\boldsymbol{V}}^\top \hat{\boldsymbol{V}} + \bar{\boldsymbol{\Sigma}}(\bar{\boldsymbol{U}}^\top\hat{\boldsymbol{U}} - \bar{\boldsymbol{V}}^\top\hat{\boldsymbol{V}})\hat{\boldsymbol{\Sigma}}^{-1}\right)_{ij}} = \abs{\left(\bar{\boldsymbol{U}}^\top \hat{\boldsymbol{U}} - \bar{\boldsymbol{V}}^\top \hat{\boldsymbol{V}}\right)_{ij}}\left(1 + \frac{\sigma_i(\bar{\boldsymbol{\Lambda}})}{\sigma_j(\hat{\boldsymbol{\Lambda}})}\right).
	\end{align}
	We can bound the absolute value of the right-hand side of \eqref{eqn:ij_identity} by the Frobenius norm of the full matrix to yield
	\begin{align}
		\abs{\left(\bar{\boldsymbol{U}}^\top \hat{\boldsymbol{U}} - \bar{\boldsymbol{V}}^\top \hat{\boldsymbol{V}}\right)_{ij}}\left(1 + \frac{\sigma_i(\bar{\boldsymbol{\Lambda}})}{\sigma_j(\hat{\boldsymbol{\Lambda}})}\right) &\leq \left(\lVert\bar{\boldsymbol{U}}^\top \hat{\boldsymbol{U}}\hat{\boldsymbol{\Sigma}} - \bar{\boldsymbol{\Sigma}}\bar{\boldsymbol{V}}^\top\hat{\boldsymbol{V}}\rVert_F + \lVert\bar{\boldsymbol{\Sigma}}\bar{\boldsymbol{U}}^\top\hat{\boldsymbol{U}} - \bar{\boldsymbol{V}}^\top\hat{\boldsymbol{V}}\hat{\boldsymbol{\Sigma}}\rVert_F\right)\lVert\hat{\boldsymbol{\Sigma}}^{-1}\rVert_F \\
		&= \mathcal{O}\left(\max\{M_N,L_N\}M_NL_N^{-1}N^{-1}\log^3(N)\right)
	\end{align}
	almost surely, where we have used the result of Part 3 above and Proposition \ref{prop:sing_vals_lam_hat}. Noticing that $1 + \sigma_i(\bar{\boldsymbol{\Lambda}}) / \sigma_j(\hat{\boldsymbol{\Lambda}}) > 1$ gives the desired result.
\end{enumerate}

\subsection{Proof of Proposition \ref{prop:def_W}}

From \citeSM{schonemann_1966}, we have that
\begin{equation}
	\boldsymbol{W} = \argmin_{\boldsymbol{W} \in \mathbb{O}(d)}\left[\lVert\bar{\boldsymbol{U}}^\top\hat{\boldsymbol{U}} - \boldsymbol{W}\rVert_F^2 + \lVert\bar{\boldsymbol{V}}^\top\hat{\boldsymbol{V}} - \boldsymbol{W}\rVert_F^2\right]. \label{eqn:schonemann}
\end{equation}
Denote the SVD of $\bar{\boldsymbol{U}}^\top\hat{\boldsymbol{U}}$ by $\boldsymbol{W}_{\boldsymbol{U},1}\boldsymbol{\Sigma}_{\boldsymbol{W}}\boldsymbol{W}_{\boldsymbol{U},2}^\top$ and define the $d \times d$ orthogonal matrix $\boldsymbol{W}_{\boldsymbol{U}} = \boldsymbol{W}_{\boldsymbol{U},1}\boldsymbol{W}_{\boldsymbol{U},2}^\top$. Then, with $\sigma_1,\dots,\sigma_d$ defined as in Part 1 of Proposition \ref{prop:lots_of_bounds}, we have 
\begin{align}
	\lVert\bar{\boldsymbol{U}}^\top\hat{\boldsymbol{U}} - \boldsymbol{W}_{\boldsymbol{U}}\rVert_F &= \lVert\boldsymbol{\Sigma}_{\boldsymbol{U}} - \boldsymbol{I}\rVert_F = \left(\sum_{i=1}^d (1 - \sigma_i)^2\right)^{1/2} \leq \sum_{i=1}^d (1 - \sigma_i) \leq \sum_{i=1}^d (1 - \sigma_i^2) \\
	&= \sum_{i=1}^d \sin^2(\theta_i)\leq d\lVert\hat{\boldsymbol{U}}\hat{\boldsymbol{U}}^\top - \bar{\boldsymbol{U}}\bar{\boldsymbol{U}}^\top\rVert^2\\ 
	&= \mathcal{O}\left(\max\{M_N,L_N\}M_NL_N^{-1}N^{-1}\log^{3}(N)\right),
\end{align}
where we invoke Part 1 of Proposition \ref{prop:lots_of_bounds}. Combining with Part 4 of the same proposition, we find
\begin{align}
	\lVert\bar{\boldsymbol{V}}^\top \hat{\boldsymbol{V}} - \boldsymbol{W}_U\rVert_F &\leq \lVert\bar{\boldsymbol{V}}^\top \hat{\boldsymbol{V}} - \bar{\boldsymbol{U}}^\top \hat{\boldsymbol{U}}\rVert_F + \lVert\bar{\boldsymbol{U}}^\top \hat{\boldsymbol{U}} - \boldsymbol{W}_U\rVert_F \\
	&= \mathcal{O}\left(\max\{M_N,L_N\}M_NL_N^{-1}N^{-1}\log^3(N)\right)
\end{align}
Combining these results gives us
\begin{align}
	\lVert\bar{\boldsymbol{U}}^\top\hat{\boldsymbol{U}} - \boldsymbol{W}\rVert_F^2 + \lVert\bar{\boldsymbol{V}}^\top \hat{\boldsymbol{V}} - \boldsymbol{W}\rVert_F^2 &\leq \lVert\bar{\boldsymbol{V}}^\top \hat{\boldsymbol{V}} - \boldsymbol{W}_U\rVert_F^2 + \lVert\bar{\boldsymbol{U}}^\top\hat{\boldsymbol{U}} - \boldsymbol{W}_U\rVert_F \\
	&= \mathcal{O}\left(\max\{M_N,L_N\}^2M_N^2L_N^{-2}N^{-2}\log^{6}(N)\right),
\end{align}
where we have used that $\boldsymbol{W}$ is the minimiser of \eqref{eqn:schonemann}, which yields the desired bound.

\subsection{Proof of Proposition \ref{prop:W_bounds}}

We address the bounds in order:
\begin{enumerate}
	\item Observe that
	\begin{align}
		\boldsymbol{W}\hat{\boldsymbol{\Sigma}} - \bar{\boldsymbol{\Sigma}}\boldsymbol{W} &= (\boldsymbol{W} - \bar{\boldsymbol{U}}^\top \hat{\boldsymbol{U}})\hat{\boldsymbol{\Sigma}} + \bar{\boldsymbol{U}}^\top \hat{\boldsymbol{U}}\hat{\boldsymbol{\Sigma}} - \bar{\boldsymbol{\Sigma}}\boldsymbol{W} \\
		&= (\boldsymbol{W} - \bar{\boldsymbol{U}}^\top \hat{\boldsymbol{U}})\hat{\boldsymbol{\Sigma}} + (\bar{\boldsymbol{U}}^\top \hat{\boldsymbol{U}}\hat{\boldsymbol{\Sigma}} - \bar{\boldsymbol{\Sigma}}\bar{\boldsymbol{V}}^\top \hat{\boldsymbol{V}}) + \bar{\boldsymbol{\Sigma}}(\bar{\boldsymbol{V}}^\top\hat{\boldsymbol{V}} - \boldsymbol{W}). 
	\end{align}
	Combining Propositions \ref{prop:def_W}, \ref{prop:sing_vals_lam_hat} and \ref{prop:behaviour_latent_position_matrices} says that 
	\begin{equation}
		\lVert(\boldsymbol{W} - \bar{\boldsymbol{U}}^\top \hat{\boldsymbol{U}})\hat{\boldsymbol{\Sigma}}\rVert_F, \lVert\bar{\boldsymbol{\Sigma}}(\bar{\boldsymbol{V}}^\top\hat{\boldsymbol{V}} - \boldsymbol{W})\rVert_F = \mathcal{O}\left(\max\{M_N,L_N\}M_N^{3/2}L_N^{-1/2}\log^{3}(N)\right),
	\end{equation}
	and that 
	\begin{equation}
		\lVert\bar{\boldsymbol{U}}^\top \hat{\boldsymbol{U}}\hat{\boldsymbol{\Sigma}} - \bar{\boldsymbol{\Sigma}}\bar{\boldsymbol{V}}^\top\hat{\boldsymbol{V}}\rVert_F = \mathcal{O}\left(\max\{M_N,L_N\}M_N^{3/2}L_N^{-1/2}\log^3(N)\right),
	\end{equation}
	from which the result follows.
	\item We prove this by bounding the absolute value of the $(i,j)$-th entry. We see
	\begin{align}
		\left(\boldsymbol{W}\hat{\boldsymbol{\Sigma}}^{1/2} - \bar{\boldsymbol{\Sigma}}^{1/2}\boldsymbol{W}\right)_{ij} &= W_{ij}\left[\sigma_i(\hat{\boldsymbol{\Sigma}})^{1/2} - \sigma_j(\bar{\boldsymbol{\Sigma}})^{1/2}\right]\\ &= \frac{W_{ij}\left[\sigma_i(\hat{\boldsymbol{\Sigma}}) - \sigma_j(\bar{\boldsymbol{\Sigma}})\right]}{\sigma_i(\hat{\boldsymbol{\Sigma}})^{1/2} + \sigma_j(\bar{\boldsymbol{\Sigma}})^{1/2}}\\ &= \frac{\left(\boldsymbol{W}\hat{\boldsymbol{\Sigma}} - \bar{\boldsymbol{\Sigma}}\boldsymbol{W}\right)_{ij}}{\sigma_i(\hat{\boldsymbol{\Sigma}})^{1/2} + \sigma_j(\bar{\boldsymbol{\Sigma}})^{1/2}}.
	\end{align}
	Using the result from Part 1, and Propositions \ref{prop:behaviour_latent_position_matrices} and \ref{prop:sing_vals_lam_hat}, we see
	\begin{equation}
		\abs{\left(\boldsymbol{W}\hat{\boldsymbol{\Sigma}}^{1/2} - \bar{\boldsymbol{\Sigma}}^{1/2}\boldsymbol{W}\right)_{ij}} \leq \frac{\abs{\left(\boldsymbol{W}\hat{\boldsymbol{\Sigma}} - \bar{\boldsymbol{\Sigma}}\boldsymbol{W}\right)_{ij}}}{\sigma_d(\bar{\boldsymbol{\Sigma}})^{1/2}} =  \mathcal{O}\left(\max\{M_N,L_N\}M_N^{5/4}L_N^{-3/4}N^{-1/2}\log^3(N)\right).
	\end{equation}
	The result follows from summing over the $d^2$ elements of the matrix.
	\item Following the same approach, we get
	\begin{equation}
		\left(\boldsymbol{W}\hat{\boldsymbol{\Sigma}}^{-1/2} - \bar{\boldsymbol{\Sigma}}^{-1/2}\boldsymbol{W}\right)_{ij} = \frac{W_{ij}\left(\sigma_i(\bar{\boldsymbol{\Lambda}})^{1/2} - \sigma_j(\hat{\boldsymbol{\Lambda}})^{1/2}\right)}{\sigma_{i}(\bar{\boldsymbol{\Lambda}})^{1/2}\sigma_j(\hat{\boldsymbol{\Lambda}})^{1/2}} = \frac{\left(\boldsymbol{W}\hat{\boldsymbol{\Sigma}}^{1/2} - \bar{\boldsymbol{\Sigma}}^{1/2}\boldsymbol{W}\right)_{ij}}{\sigma_i(\bar{\boldsymbol{\Lambda}})^{1/2}\sigma_j(\hat{\boldsymbol{\Lambda}})^{1/2}}.
	\end{equation}
	Taking the Frobenius norm and using the bounds from Part 2 and Propositions \ref{prop:behaviour_latent_position_matrices} and \ref{prop:sing_vals_lam_hat}, we obtain
	\begin{equation}
		\lVert\boldsymbol{W}\hat{\boldsymbol{\Sigma}}^{-1/2} - \bar{\boldsymbol{\Sigma}}^{-1/2}\boldsymbol{W}\rVert_F = \mathcal{O}\left(\max\{M_N,L_N\}M_N^{3/4}L_N^{-5/4}N^{-3/2}\log^3(N)\right).
	\end{equation}
\end{enumerate}

\subsection{Proof of Proposition \ref{prop:KR}}

Define $\boldsymbol{\Pi}_{\tilde{\boldsymbol{X}}} = \left(\tilde{\boldsymbol{X}}^\top \tilde{\boldsymbol{X}}\right)^{1/2}$ and $\boldsymbol{\Pi}_{\boldsymbol{Y}} = \left(\boldsymbol{Y}^\top \boldsymbol{Y}\right)^{1/2}$. Note that
\begin{equation}
	\left(\bar{\boldsymbol{X}}\bar{\boldsymbol{\Sigma}}^{1/2}\right)\left(\bar{\boldsymbol{X}}\bar{\boldsymbol{\Sigma}}^{1/2}\right)^\top = \bar{\boldsymbol{U}}\bar{\boldsymbol{\Sigma}}^2\bar{\boldsymbol{U}}^\top = \bar{\boldsymbol{\Lambda}}\bar{\boldsymbol{\Lambda}}^\top = \tilde{\boldsymbol{X}}\boldsymbol{Y}^\top \boldsymbol{Y}\tilde{\boldsymbol{X}}^\top = \left(\tilde{\boldsymbol{X}}\boldsymbol{\Pi}_{\boldsymbol{Y}}\right)\left(\tilde{\boldsymbol{X}}\boldsymbol{\Pi}_{\boldsymbol{Y}}\right)^\top.
\end{equation}
This implies the existence of some orthogonal matrix $\boldsymbol{Q} \in \mathbb{O}(d)$ such that
\begin{equation}
	\bar{\boldsymbol{X}}\bar{\boldsymbol{\Sigma}}^{1/2} = \tilde{\boldsymbol{X}}\boldsymbol{\Pi}_{\boldsymbol{Y}}\boldsymbol{Q}, 
\end{equation}
then $\breve{\boldsymbol{K}} := \boldsymbol{\Pi}_{\boldsymbol{Y}}\boldsymbol{Q}\bar{\boldsymbol{\Sigma}}^{-1/2}\in \mathrm{GL}(d)$ satisfies the required definition. Similarly, we write
\begin{equation}
	\left(\bar{\boldsymbol{Y}}\bar{\boldsymbol{\Sigma}}^{1/2}\right)\left(\bar{\boldsymbol{Y}}\bar{\boldsymbol{\Sigma}}^{1/2}\right)^\top = \bar{\boldsymbol{V}}\bar{\boldsymbol{\Sigma}}^2\bar{\boldsymbol{V}} = \bar{\boldsymbol{\Lambda}}\bar{\boldsymbol{\Lambda}}^\top = \boldsymbol{Y}\tilde{\boldsymbol{X}}^\top \tilde{\boldsymbol{X}}\boldsymbol{Y}^\top = \left(\boldsymbol{Y}\boldsymbol{\Pi}_{\tilde{\boldsymbol{X}}}\right)\left(\boldsymbol{Y}\boldsymbol{\Pi}_{\tilde{\boldsymbol{X}}}\right)^\top.
\end{equation}
In turn, this implies the existence of $\boldsymbol{Q}^* \in \mathbb{O}(d)$ such that
\begin{equation}
	\bar{\boldsymbol{Y}}\bar{\boldsymbol{\Sigma}}^{1/2} = \boldsymbol{Y}\boldsymbol{\Pi}_{\tilde{\boldsymbol{X}}}\boldsymbol{Q}^*,
\end{equation}
and so $\breve{\boldsymbol{R}} := \boldsymbol{\Pi}_{\tilde{\boldsymbol{X}}}\boldsymbol{Q}^*\bar{\boldsymbol{\Sigma}}^{-1/2} \in \mathrm{GL}(d)$ is the matrix required. Further note
\begin{equation}
	\tilde{\boldsymbol{X}}\breve{\boldsymbol{K}}\breve{\boldsymbol{R}}^\top \boldsymbol{Y}^\top = \bar{\boldsymbol{X}}\bar{\boldsymbol{Y}}^\top = \bar{\boldsymbol{\Lambda}} = \tilde{\boldsymbol{X}}\boldsymbol{Y}^\top.
\end{equation}
Multiplying on the left by $(\tilde{\boldsymbol{X}}^\top \tilde{\boldsymbol{X}})^{-1}\tilde{\boldsymbol{X}}^\top$ and on the right by $\boldsymbol{Y}(\boldsymbol{Y}^\top \boldsymbol{Y})^{-1}$ yields $\breve{\boldsymbol{L}}\breve{\boldsymbol{R}}^\top = \boldsymbol{I}_d$.

\subsection{Proof of Proposition \ref{prop:LOO_incoherence}}

We will show the result for $\hat{\boldsymbol{U}}^{(i)}$, with the other results being shown in an analogous manner. Note that for a matrix $\boldsymbol{A}$ with orthonormal columns, we can write
\begin{equation}
	\lVert \boldsymbol{A}\boldsymbol{A}^\top e_i \rVert^2 = e_i^\top \boldsymbol{A}^\top \boldsymbol{A} \boldsymbol{A} \boldsymbol{A}^\top e_i = e_i^\top \boldsymbol{A} \boldsymbol{A}^\top e_i = (\boldsymbol{A} \boldsymbol{A}^\top)_{ii} = \lVert A_{i,\ast}\rVert^2.
\end{equation}
We can use this identity along with Wedin's Theorem, as stated in Theorem \ref{thm:wedin}, say
\begin{align}
	\lVert \hat{U}^{(i)}_{i,\ast} \rVert = \lVert (\hat{\boldsymbol{U}}^{(i)}\hat{\boldsymbol{U}}^{(i) \top })e_i \rVert &\leq \lVert (\hat{\boldsymbol{U}}^{(i)}\hat{\boldsymbol{U}}^{(i) \top } - \bar{\boldsymbol{U}}\bar{\boldsymbol{U}}^\top)e_i \rVert + \lVert \bar{U}_{i,\ast} \rVert \\
	& \leq \lVert \hat{\boldsymbol{U}}^{(i)}\hat{\boldsymbol{U}}^{(i) \top } - \bar{\boldsymbol{U}}\bar{\boldsymbol{U}}^\top \rVert + \lVert \bar{U}_{i,\ast} \rVert \\
	&\leq \frac{\max\left\{\lVert (\hat{\boldsymbol{\Lambda}}^{(i)} - \bar{\boldsymbol{\Lambda}})^\top \bar{\boldsymbol{U}} \rVert, \lVert (\hat{\boldsymbol{\Lambda}}^{(i)} - \bar{\boldsymbol{\Lambda}}) \bar{\boldsymbol{V}} \rVert  \right\}}{\sigma_d(\bar{\boldsymbol{\Lambda}}) - \sigma_{d+1}(\bar{\boldsymbol{\Lambda}}) - \lVert \hat{\boldsymbol{\Lambda}}^{(i)} - \bar{\boldsymbol{\Lambda}} \rVert} + \lVert \bar{U}_{i,\ast} \rVert.
\end{align}
We look to bound the growth of the first term. Define the row-deletion operator $\boldsymbol{P}^{(i)} := \boldsymbol{I} - e_ie_i^\top$, which is symmetric and idempotent. We can then write
\begin{equation}
	\hat{\boldsymbol{\Lambda}}^{(i)} - \bar{\boldsymbol{\Lambda}} = \boldsymbol{P}^{(i)}(\hat{\boldsymbol{\Lambda}} - \bar{\boldsymbol{\Lambda}}),
\end{equation}
and so we can bound each of the terms in the numerator as
\begin{align}
	\lVert (\hat{\boldsymbol{\Lambda}}^{(i)} - \bar{\boldsymbol{\Lambda}})^\top \bar{\boldsymbol{U}} \rVert &= \lVert (\hat{\boldsymbol{\Lambda}} - \bar{\boldsymbol{\Lambda}})^\top \boldsymbol{P}^{(i)} \bar{\boldsymbol{U}}\rVert \leq \lVert \hat{\boldsymbol{\Lambda}} - \bar{\boldsymbol{\Lambda}}  \rVert,
\end{align}
and
\begin{align}
	\lVert (\hat{\boldsymbol{\Lambda}}^{(i)} - \bar{\boldsymbol{\Lambda}})^\top \bar{\boldsymbol{V}} \rVert &= \lVert \boldsymbol{P}^{(i)}(\hat{\boldsymbol{\Lambda}} - \bar{\boldsymbol{\Lambda}})  \bar{\boldsymbol{V}}\rVert \leq \lVert \hat{\boldsymbol{\Lambda}} - \bar{\boldsymbol{\Lambda}}  \rVert,
\end{align}
both of which are $\mathcal{O}\left(\max\{M_N,L_N\}^{1/2}M_NN^{1/2}\log^{3/2}(N)\right)$ by Proposition \ref{prop:hat_lambda_bar_lambda}. The denominator, by our assumption on the singular values of $\bar{\boldsymbol{\Lambda}}$ and Proposition \ref{prop:hat_lambda_bar_lambda}, is $\mathcal{O}\left(M_N^{1/2}L_N^{1/2}N\right)$, and so it follows 
\begin{equation}
	\lVert \hat{U}_{i,\ast}^{(i)} \rVert \leq \mathcal{O}\left(\max\{M_N,L_N\}^{1/2}M_N^{1/2}L_N^{-1/2}N^{-1/2}\log^{3/2}(N)\right) + \lVert  \bar{U}_{i,\ast} \rVert.
\end{equation}
By Proposition \ref{prop:behaviour_latent_position_matrices}, we know $\mu(\bar{\boldsymbol{\Lambda}}) = \mathcal{O}(1)$, and so we can bound $\lVert\bar{U}_{i,\ast}\rVert = \mathcal{O}(M_N^{-1/2}N^{-1/2})$. The result follows.

\subsection{Proof of Proposition \ref{prop:KR_spec_norm}}

Recall the definitions $\breve{\boldsymbol{K}} = \boldsymbol{\Pi}_{\boldsymbol{Y}}\boldsymbol{Q} \bar{\boldsymbol{\Sigma}}^{-1/2}$ and $\breve{\boldsymbol{R}} = \boldsymbol{\Pi}_{\tilde{\boldsymbol{X}}}\boldsymbol{Q}^* \bar{\boldsymbol{\Sigma}}^{-1/2}$. By Proposition \ref{prop:behaviour_latent_position_matrices}, we have
\begin{equation}
	\lVert\bar{\boldsymbol{\Sigma}}^{1/2}\rVert= \mathcal{O}\left(M_N^{1/4}L_N^{1/4}N^{1/2}\right), \qquad \lVert\bar{\boldsymbol{\Sigma}}^{-1/2}\rVert = \mathcal{O}\left(N^{-1/2}M_N^{-1/4}L_N^{-1/4}\right), 
\end{equation}
Since the entries of $\tilde{\boldsymbol{X}}$ and $\boldsymbol{Y}$ are bounded by Assumption \ref{ass:bound_and_int} and $\boldsymbol{\Pi}_{\tilde{\boldsymbol{X}}}$ and $\boldsymbol{\Pi}_{\boldsymbol{Y}}$ have the same singular values as $\tilde{\boldsymbol{X}}$ and $\boldsymbol{Y}$, respectively, we see 
\begin{equation}
	\lVert\boldsymbol{\Pi}_{\tilde{\boldsymbol{X}}}\rVert = \mathcal{O}\left(N^{1/2}M_N^{1/2}\right), \qquad\lVert\boldsymbol{\Pi}_{\boldsymbol{Y}}\rVert = \mathcal{O}\left(N^{1/2}L_N^{1/2}\right),
\end{equation}
By the rank $d$ assumption of $\tilde{\boldsymbol{X}}$ and $\boldsymbol{Y}$ paired with Proposition \ref{prop:behaviour_latent_position_matrices}, we have  \begin{equation}
	\lVert\boldsymbol{\Pi}_{\tilde{\boldsymbol{X}}}^{-1}\rVert = \mathcal{O}\left(M_N^{-1/2}N^{-1/2}\right), \qquad \lVert\boldsymbol{\Pi}_{\boldsymbol{Y}}^{-1}\rVert = \mathcal{O}\left(L_N^{-1/2}N^{-1/2}\right).
\end{equation}
It follows that
\begin{align}
	&\breve{\norm{\boldsymbol{K}}} \leq \norm{\boldsymbol{\Pi}_{\boldsymbol{Y}}}\norm{\boldsymbol{Q}}\lVert\bar{\boldsymbol{\Sigma}}^{-1/2}\rVert = \mathcal{O}\left(M_N^{-1/4}L_N^{1/4}\right), & & \lVert\breve{\boldsymbol{K}}^{-1}\rVert \leq \lVert\bar{\boldsymbol{\Sigma}}^{1/2}\rVert\lVert\boldsymbol{Q}^{-1}\rVert\lVert\boldsymbol{\Pi}_{\boldsymbol{Y}}^{-1}\rVert = \mathcal{O}\left(M_N^{1/4}L_N^{-1/4}\right)\\
	&\breve{\norm{\boldsymbol{R}}} \leq \lVert\boldsymbol{\Pi}_{\tilde{\boldsymbol{X}}}\rVert\norm{\boldsymbol{Q}^*}\lVert\bar{\boldsymbol{\Sigma}}^{-1/2}\rVert = \mathcal{O}\left(M_N^{1/4}L_N^{-1/4}\right),& & \lVert\breve{\boldsymbol{R}}^{-1}\rVert \leq \lVert\bar{\boldsymbol{\Sigma}}^{1/2}\rVert\lVert\boldsymbol{Q}^{*-1}\rVert\lVert\boldsymbol{\Pi}_{\tilde{\boldsymbol{X}}}^{-1}\rVert = \mathcal{O}\left(M_N^{-1/4}L_N^{1/4}\right).
\end{align} 

\subsection{Proof of Proposition \ref{prop:residuals}}

We will prove the results for $\boldsymbol{R}_{1,\ell}$, for $\ell=1,2,3,4$, with $\boldsymbol{R}_{2,\ell}$ following analogously unless otherwise stated. 
\begin{enumerate}
	\item 
	%Recall that $\bar{\boldsymbol{U}}\bar{\boldsymbol{\Sigma}}^{1/2} =\tilde{\boldsymbol{X}}\breve{\boldsymbol{K}}$, and so using that $\norm{\boldsymbol{A}\boldsymbol{P}}_{2,\infty} \leq \norm{\boldsymbol{A}}_{2,\infty}\norm{\boldsymbol{P}}$ \citepSM[see, for example,][]{cape_2019}, we bound $\norm{\bar{\boldsymbol{U}}}_{2,\infty} \leq \tilde{\norm{\boldsymbol{X}}}_{2,\infty}\breve{\norm{\boldsymbol{K}}}\lVert\bar{\boldsymbol{\Sigma}^{-1/2}}\rVert$. The rows of $\tilde{\boldsymbol{X}}$ are bounded in Euclidean norm by assumption \textcolor{red}{(assumption on finite/bounded row norms of latent positions)}. Combining with the spectral norm bound on $\bar{\boldsymbol{\Sigma}}$ from Proposition \textcolor{red}{(proposition/assumption for the spectral norm bound on $\bar{\boldsymbol{\Lambda}}$)} and the norm bound of $\breve{\boldsymbol{K}}$ from Proposition \ref{prop:KR_spec_norm}, 
	By Proposition \ref{prop:behaviour_latent_position_matrices}, we have $\mathcal{O}(1)$ growth of the incoherence of $\bar{\boldsymbol{\Lambda}}$, and so $\norm{\bar{\boldsymbol{U}}}_{2,\infty} = \mathcal{O}\left(M_N^{-1/2}N^{-1/2}\right)$. It follows that
	\begin{align}
		\norm{\boldsymbol{R}_{1,1}}_{2,\infty} &\leq \lVert\bar{\boldsymbol{U}}\rVert_{2,\infty}\lVert\bar{\boldsymbol{U}}^\top \hat{\boldsymbol{U}}\hat{\boldsymbol{\Sigma}}^{1/2} - \bar{\boldsymbol{\Sigma}}^{1/2}\boldsymbol{W}\rVert \\
		& \leq \lVert\bar{\boldsymbol{U}}\rVert_{2,\infty}\left(\lVert(\bar{\boldsymbol{U}}^\top \hat{\boldsymbol{U}} - \boldsymbol{W})\hat{\boldsymbol{\Sigma}}^{1/2}\rVert_F + \lVert\boldsymbol{W}\hat{\boldsymbol{\Sigma}}^{1/2} - \bar{\boldsymbol{\Sigma}}^{1/2}\boldsymbol{W}\rVert_F\right).
	\end{align}
	Using Propositions \ref{prop:sing_vals_lam_hat}  and \ref{prop:def_W} for the first term and Proposition \ref{prop:W_bounds} for the second, we obtain the desired bound.
	\item Define $\boldsymbol{M}_1 = \bar{\boldsymbol{U}}\bar{\boldsymbol{U}}^\top(\hat{\boldsymbol{\Lambda}} - \bar{\boldsymbol{\Lambda}})(\hat{\boldsymbol{V}} - \bar{\boldsymbol{V}}\boldsymbol{W})\hat{\boldsymbol{\Sigma}}^{-1/2}$ and $\boldsymbol{M}_2 = (\hat{\boldsymbol{\Lambda}} - \bar{\boldsymbol{\Lambda}})(\hat{\boldsymbol{V}} - \bar{\boldsymbol{V}}\boldsymbol{W})\hat{\boldsymbol{\Sigma}}^{-1/2}$ so that $\boldsymbol{R}_{1,2} = \boldsymbol{M}_2 - \boldsymbol{M}_1$ and we can bound $\norm{\boldsymbol{R}_{1,2}}_{2,\infty} \leq \norm{\boldsymbol{M}_1}_{2,\infty} + \norm{\boldsymbol{M}_2}_{2,\infty}$. 
	
	\hspace{1cm}\textbf{Bound on $\boldsymbol{M}_1$:}
	
	For the first norm, we split as:
	\begin{align}
		\norm{\boldsymbol{M}_1}_{2,\infty} &\leq \lVert\bar{\boldsymbol{U}}\rVert_{2,\infty} \lVert\hat{\boldsymbol{\Lambda}} - \bar{\boldsymbol{\Lambda}}\rVert\lVert\hat{\boldsymbol{V}} - \bar{\boldsymbol{V}}\boldsymbol{W}\rVert\lVert\hat{\boldsymbol{\Sigma}}^{-1/2}\rVert \\
		&\leq \lVert\hat{\boldsymbol{V}} - \bar{\boldsymbol{V}}\boldsymbol{W}\rVert\mathcal{O}\left(\max\{M_N,L_N\}^{1/2}M_N^{1/4}L_N^{-1/4}N^{-1/2}\log^{3/2}(N)\right),
	\end{align}
	where we have used the bound on $\lVert\bar{\boldsymbol{U}}\rVert_{2,\infty}$ and Propositions \ref{prop:hat_lambda_bar_lambda} and \ref{prop:sing_vals_lam_hat}. We then observe that by Propositions \ref{prop:lots_of_bounds} and \ref{prop:def_W} we have
	\begin{align}
		\lVert\hat{\boldsymbol{V}} - \bar{\boldsymbol{V}}\boldsymbol{W}\rVert &\leq \lVert\hat{\boldsymbol{V}} - \bar{\boldsymbol{V}}\bar{\boldsymbol{V}}^\top\hat{\boldsymbol{V}}\rVert + \lVert\bar{\boldsymbol{V}}(\bar{\boldsymbol{V}}^\top \hat{\boldsymbol{V}} - \boldsymbol{W})\rVert \\
		&= \mathcal{O}\left(\max\{M_N,L_N\}^{1/2}M_N^{1/2}L_N^{-1/2}N^{-1/2}\log^{3/2}(N)\right) \\
		&+ \mathcal{O}\left(\max\{M_N,L_N\}M_NL_N^{-1}N^{-1}\log^{3}(N)\right) \\
		&= \mathcal{O}\left(\max\{M_N,L_N\}^{1/2}M_N^{1/2}L_N^{-1/2}N^{-1/2}\log^{3/2}(N)\right)
	\end{align}
	Combining gives
	\begin{equation}
		\norm{\boldsymbol{M}_1}_{2,\infty} = \mathcal{O}\left(\max\{M_N,L_N\}M_N^{3/4}L_N^{-3/4}N^{-1}\log^3(N)\right), \label{eqn:M_1_bound}
	\end{equation}
	almost surely.
	
	\hspace{1cm}\textbf{Bound on $\boldsymbol{M}_2$:}
	
	Note that we can write
	\begin{equation}
		\boldsymbol{M}_2 = \underbrace{(\hat{\boldsymbol{\Lambda}} - \bar{\boldsymbol{\Lambda}})(\boldsymbol{I} - \bar{\boldsymbol{V}}\bar{\boldsymbol{V}}^\top)\hat{\boldsymbol{V}}\hat{\boldsymbol{\Sigma}}^{-1/2}}_{\boldsymbol{T}_1} + \underbrace{(\hat{\boldsymbol{\Lambda}} - \bar{\boldsymbol{\Lambda}})\bar{\boldsymbol{V}}(\bar{\boldsymbol{V}}^\top \hat{\boldsymbol{V}} - \boldsymbol{W})\hat{\boldsymbol{\Sigma}}^{-1/2}}_{\boldsymbol{T}_2}.
	\end{equation}
	We bound $\lVert\boldsymbol{T}_2\rVert_{2,\infty}$ using Propositions \ref{prop:hat_lambda_bar_lambda}, \ref{prop:sing_vals_lam_hat} and \ref{prop:def_W} and that $\lVert\boldsymbol{T}_2\rVert_{2,\infty} \leq \lVert\boldsymbol{T}_2\rVert$:
	\begin{align}
		\lVert\boldsymbol{T}_2\rVert_{2,\infty} &\leq \lVert\hat{\boldsymbol{\Lambda}} - \bar{\boldsymbol{\Lambda}}\rVert\lVert\bar{\boldsymbol{V}}\rVert\lVert(\bar{\boldsymbol{V}}^\top \hat{\boldsymbol{V}} - \boldsymbol{W})\rVert\lVert\hat{\boldsymbol{\Sigma}}^{-1/2}\rVert\\
		&= \mathcal{O}\left(\max\{M_N,L_N\}^{3/2}M_N^{7/4}L_N^{-5/4}N^{-1}\log^{9/2}(N)\right)\label{eqn:T_2_bound}.
	\end{align}
	For $\boldsymbol{T}_1$, first note that we can write
	\begin{equation}
		\lVert\boldsymbol{T}_1\rVert_{2,\infty} \leq \lVert\underbrace{(\hat{\boldsymbol{\Lambda}} - \bar{\boldsymbol{\Lambda}})(\boldsymbol{I} - \bar{\boldsymbol{V}}\bar{\boldsymbol{V}}^\top)\hat{\boldsymbol{V}}\hat{\boldsymbol{V}}^\top}_{\boldsymbol{T}_3}\rVert_{2,\infty}\lVert\underbrace{\hat{\boldsymbol{V}}\hat{\boldsymbol{\Sigma}}^{-1/2}}_{\boldsymbol{T}_4}\rVert. \label{eqn:T_1_decomp} 
	\end{equation}
	We have 
	\begin{equation}
		\lVert\boldsymbol{T}_4\rVert \leq \lVert\hat{\boldsymbol{V}}\rVert\lVert\hat{\boldsymbol{\Sigma}}^{-1/2}\rVert = \mathcal{O}\left(M_N^{-1/4}L_N^{-1/4}N^{-1/2}\right), \label{eqn:T_4_bound}
	\end{equation} 
	and so it remains only to bound the norm of $\boldsymbol{T}_3$. Define $\bar{\boldsymbol{P}} = \bar{\boldsymbol{V}}\bar{\boldsymbol{V}}^\top$, $\hat{\boldsymbol{P}} = \hat{\boldsymbol{V}}\hat{\boldsymbol{V}}^\top$ and $\boldsymbol{E} = \hat{\boldsymbol{\Lambda}} - \bar{\boldsymbol{\Lambda}}$, so that $\boldsymbol{T}_3 = \boldsymbol{E}(\boldsymbol{I}-\bar{\boldsymbol{P}})\hat{\boldsymbol{P}}$. We look to obtain a bound on 
	\begin{equation}
		\lVert\boldsymbol{T}_3\rVert_{2,\infty} = \max_{i\in [NM]}\lVert e_i^\top \boldsymbol{T}_3\rVert,
	\end{equation}
	where $e_i$ is the $i$-th standard basis vector. There is a dependence between $\hat{\boldsymbol{P}}$ and $\boldsymbol{E}$ that prevents the application of standard concentration inequalities. To circumvent this, we proceed using a leave-one-out argument. For each $i \in [NM_N]$, define a matrix %$\boldsymbol{E}^{(i)} = \hat{\boldsymbol{\Lambda}}^{(i)} - \bar{\boldsymbol{\Lambda}}$, 
	$\hat{\boldsymbol{\Lambda}}^{(i)}$ as
	\begin{equation}
		\hat{\Lambda}^{(i)}_{jk} = \begin{cases}
			\hat{\Lambda}_{jk}, & j \neq i, \\
			\bar{\Lambda}_{jk} & j = i
		\end{cases}.
	\end{equation}
	Then $\hat{\boldsymbol{\Lambda}}^{(i)}$ is $\hat{\boldsymbol{\Lambda}}$ with its $i$-th row replaced with its expectation. That is to say, $\hat{\boldsymbol{\Lambda}}^{(i)}$ is $\hat{\boldsymbol{\Lambda}}$ but with the randomness removed from the $i$-th row. In this way, we can write $\hat{\boldsymbol{\Lambda}} = \hat{\boldsymbol{\Lambda}}^{(i)} + \boldsymbol{E}^{(i)}$, where 
	\begin{equation}
		E^{(i)}_{pq} = 
		\begin{cases}
			\hat{\Lambda}_{pq} - \bar{\Lambda}_{pq}, & p \neq i, \\
			0, & p = i.
		\end{cases}
	\end{equation}%Note this says that the $i$-th row of $\boldsymbol{E}^{(i)}$ is zero. 
	We compute the singular value decomposition of this de-noised matrix $\hat{\boldsymbol{\Lambda}}^{(i)}$ as
	\begin{equation}
		\hat{\boldsymbol{\Lambda}}^{(i)} = \hat{\boldsymbol{U}}^{(i)}\hat{\boldsymbol{\Sigma}}^{(i)}\hat{\boldsymbol{V}}^{(i) \top} + \hat{\boldsymbol{U}}^{(i)}_\perp\hat{\boldsymbol{\Sigma}}^{(i)}_\perp\hat{ \boldsymbol{V}}^{(i)\top}_\perp,
	\end{equation}
	and define $\hat{\boldsymbol{P}}^{(i)} = \hat{\boldsymbol{V}}^{(i)}\hat{\boldsymbol{V}}^{(i)\top}$. It is important to note that this construction ensures that $\hat{\boldsymbol{\Lambda}}^{(i)}$, and therefore $\hat{\boldsymbol{P}}^{(i)}$, is independent of the $i$-th row of $\boldsymbol{E}$. 
	
	Recall that we are trying to bound each $\lVert e_i^\top \boldsymbol{T}_3\rVert$. With these new definitions, for a given $i \in [NM_N]$, we consider a decomposition of the row norm as
	\begin{equation}
		\lVert e_i^\top \boldsymbol{T}_3\rVert \leq \lVert\underbrace{ e_i^\top \boldsymbol{E} (\boldsymbol{I} - \bar{\boldsymbol{P}})\hat{\boldsymbol{P}}^{(i)}}_{T_5}\rVert + \lVert\underbrace{ e_i^\top \boldsymbol{E}(\boldsymbol{I} - \bar{\boldsymbol{P}})(\hat{\boldsymbol{P}} - \hat{\boldsymbol{P}}^{(i)})}_{T_6}\rVert. \label{eqn:row_bound_T_3}
	\end{equation}
	We start by bounding $\lVert T_6\rVert$. To do so, we note
	\begin{equation}
		\lVert T_6 \rVert \leq \lVert e_i^\top \boldsymbol{E}\rVert \lVert\boldsymbol{I} - \bar{\boldsymbol{P}} \rVert\lVert \hat{\boldsymbol{P}} - \hat{\boldsymbol{P}}^{(i)}\rVert = \lVert e_i^\top \boldsymbol{E}\rVert\lVert \hat{\boldsymbol{P}} - \hat{\boldsymbol{P}}^{(i)}\rVert. \label{eqn:T_6_decomp}
	\end{equation} 
	Wedin's Theorem, as stated in Theorem \ref{thm:wedin}, tells us
	\begin{align}
		\lVert \hat{\boldsymbol{P}} - \hat{\boldsymbol{P}}^{(i)}\rVert = \lVert \hat{\boldsymbol{V}}\hat{ \boldsymbol{V}}^\top - \hat{\boldsymbol{V}}^{(i)}\hat{\boldsymbol{V}}^{(i)\top}\rVert &\leq \frac{2^{1/2}\max \left\{\lVert\boldsymbol{E}^{(i)\top}\hat{\boldsymbol{U}}^{(i)}\rVert, \lVert\boldsymbol{E}^{(i)} \hat{\boldsymbol{V}}^{(i)}\rVert\right\}}{\sigma_d(\hat{\boldsymbol{\Lambda}}^{(i)}) - \sigma_{d+1}(\hat{\boldsymbol{\Lambda}}^{(i)}) - \lVert \boldsymbol{E}^{(i)}\rVert}  \\
		&= \frac{2^{1/2}\max\left\{\lVert(\hat{\Lambda}_{i,\ast} - \bar{\Lambda}_{i,\ast})^\top e_i^\top\hat{\boldsymbol{U}}^{(i)}\rVert,\lVert(\hat{\Lambda}_{i,\ast} - \bar{\Lambda}_{i,\ast})\hat{\boldsymbol{V}}^{(i)}\rVert\right\}}{\sigma_d(\hat{\boldsymbol{\Lambda}}^{(i)}) - \sigma_{d+1}(\hat{\boldsymbol{\Lambda}}^{(i)}) - \lVert\hat{\Lambda}_{i,\ast} - \bar{\Lambda}_{i,\ast}\rVert}. \label{eqn:P_hat_minus_P_hat_i}
	\end{align}
	We first consider the numerator of \eqref{eqn:P_hat_minus_P_hat_i} and bound each of the terms inside of the max operator. 
	%\begin{equation}
	%\max\left\{\lVert\hat{\Lambda}_{i,\ast} - \bar{\Lambda}_{i,\ast}\rVert \lVert U^{(i)}_{i,\ast}\rVert,\lVert(\hat{\Lambda}_{i,\ast} - \bar{\Lambda}_{i,\ast})\hat{\boldsymbol{V}}^{(i)}\rVert\right\}
	%\end{equation}
	For the second of the numerator terms, we note that the $k$th element of the vector $(\hat{\Lambda}_{i,\ast} - \bar{\Lambda}_{i,\ast})\hat{\boldsymbol{V}}^{(i)}$ takes the form $\sum_{j=1}^{NL_N} (\hat{\Lambda}_{i,\ast} - \bar{\Lambda}_{i,\ast})_j\hat{V}^{(i)}_{jk}$. By construction, each element of $\hat{\boldsymbol{V}}^{(i)}$ is independent of the vector $(\hat{\Lambda}_{i,\ast} - \bar{\Lambda}_{i,\ast})$, and so we can invoke a standard subexponential Bernstein inequality, as stated in Theorem \ref{thm:subexp_bern}, using that a centred Poisson random variable is subexponential. By the definition of an SVD, the $k$th column of $\hat{\boldsymbol{V}}^{(i)}$ satisfies $\lVert \hat{V}^{(i)}_{\ast, k}\rVert = 1$ and $\lVert\hat{V}^{(i)}_{\ast, k}\rVert_\infty \leq 1$, almost surely. Under Assumption \ref{ass:bound_and_int}, the subexponential norm of each vector entry is bounded. That is $\lVert(\hat{\Lambda}_{i,j} - \bar{\Lambda}_{i,j})/M_N\rVert_{\psi_1} = \mathcal{O}(1)$ for each $j \in [NL_N]$. We can thus apply Theorem \ref{thm:subexp_bern} to $\sum_{j=1}^{NL_N} (\hat{\Lambda}_{i,j} - \bar{\Lambda}_{i,j})\hat{V}^{(i)}_{jk}$ to get that for any $t \geq 0$, 
	\begin{equation}
		\mathbb{P}\left(\abs{\sum_{j=1}^{NL_N} (\hat{\Lambda}_{ij} - \bar{\Lambda}_{ij})\hat{V}^{(i)}_{jk}} \geq Mt\right) \leq \exp\left(-C^{(i)}_{1,k} \min\left\{\frac{t^2}{C^{(i)}_{2,k}}, \frac{t}{C^{(i)}_{3,k}}\right\}\right),
	\end{equation}
	almost surely, for some constants $C_{1,k}^{(i)}, C_{2,k}^{(i)}, C_{3,k}^{(i)} > 0$. For large $t$, the quadratic term dominates and so the bound reduces to
	\begin{equation}
		\mathbb{P}\left(\abs{\sum_{j=1}^{NL_N} (\hat{\Lambda}_{ij} - \bar{\Lambda}_{ij})\hat{V}^{(i)}_{jk}} \geq Mt\right) \leq \exp\left(-C_{4,k}^{(i)}t\right).
	\end{equation}
	Choosing $t = C_{5,k}^{(i)}\log(N)$, taking a union bound over the $d$ (a constant) elements of the vector $(\hat{\Lambda}_{i,\ast} - \bar{\Lambda}_{i,\ast})\hat{V}^{(i)}$ and invoking that $\lVert(\hat{\Lambda}_{i,\ast} - \bar{\Lambda}_{i,\ast})\hat{\boldsymbol{V}}^{(i)}\rVert \leq d^{1/2}\lVert(\hat{\Lambda}_{i,\ast} - \bar{\Lambda}_{i,\ast})\hat{\boldsymbol{V}}^{(i)}\rVert_\infty$ shows
	\begin{equation}
		\lVert(\hat{\Lambda}_{i,\ast} - \bar{\Lambda}_{i,\ast})\hat{\boldsymbol{V}}^{(i)}\rVert = \mathcal{O}\left(M_N\log(N)\right), \label{eqn:right_numerator_bound}
	\end{equation}
	almost surely. 
	
	For the first term, the we cannot apply the same technique as we do not reduce to the norm of a length $d$ vector of weighted Poisson random variables. This raises issues with controlling the spectral norm. We note the following decomposition of the operator norm in the case of a rank-one matrix: for $a \in \mathbb{R}^m$ and $b\in \mathbb{R}^n$, we have $\lVert ab^\top \rVert = \lVert a \rVert \lVert b \rVert$. We can apply this to write the first term of the numerator, up to the multiplicative constant, as $\lVert\hat{\Lambda}_{i,\ast} - \bar{\Lambda}_{i,\ast}\rVert\lVert \hat{U}^{(i)}_{i,\ast}\rVert$. The second term is bounded by Proposition \ref{prop:LOO_incoherence}. We can then use that the two-to-infinity norm is upper bounded by the spectral norm, and so Proposition \ref{prop:hat_lambda_bar_lambda} says that $\lVert\hat{\Lambda}_{i,\ast} - \bar{\Lambda}_{i,\ast}\rVert = \mathcal{O}\left(\max\{M_N,L_N\}^{1/2}M_NN^{1/2}\log^{3/2}(N)\right)$ almost surely. It follows that
	\begin{equation}
		\lVert(\hat{\Lambda}_{i,\ast} - \bar{\Lambda}_{i,\ast})e_i^\top\boldsymbol{U}^{(i)}\rVert = \mathcal{O}\left(\max\{M_N,L_N\}M_N^{3/2}L_N^{-1/2}\log^{3}(N)\right). \label{eqn:left_numerator_bound}
	\end{equation}
	Comparing \eqref{eqn:left_numerator_bound} and \eqref{eqn:right_numerator_bound}, we see that the numerator of \eqref{eqn:P_hat_minus_P_hat_i} is
	\begin{equation}
		2^{1/2}\max\left\{\lVert(\hat{\Lambda}_{i,\ast} - \bar{\Lambda}_{i,\ast})e_i^\top\boldsymbol{U}^{(i)}\rVert,\lVert(\hat{\Lambda}_{i,\ast} - \bar{\Lambda}_{i,\ast})\hat{\boldsymbol{V}}^{(i)}\rVert\right\} = \mathcal{O}\left(\max\{M_N,L_N\}M_N^{3/2}L_N^{-1/2}\log^{3}(N)\}\right)\label{eqn:numerator_bound},
	\end{equation}
	almost surely.
	
	Next we bound the denominator of \eqref{eqn:P_hat_minus_P_hat_i}. To do so, we again recall Corollary 7.3.5 of \citeSM{horn2012}, which implies that for any two matrices $\boldsymbol{A}$ and $\boldsymbol{B}$ of the same size, we have
	\begin{equation}
		\sigma_i(\boldsymbol{B}) - \norm{\boldsymbol{A} - \boldsymbol{B}} \leq \sigma_i(\boldsymbol{A}) \leq \sigma_i(\boldsymbol{B}) + \norm{\boldsymbol{A} - \boldsymbol{B}}.
	\end{equation}
	This tells us,
	\begin{equation}
		\sigma_d(\hat{\boldsymbol{\Lambda}}) - \lVert\hat{\boldsymbol{\Lambda}}^{(i)} - \hat{\boldsymbol{\Lambda}}\rVert \leq \sigma_d(\hat{\boldsymbol{\Lambda}}^{(i)}) \leq \sigma_d(\hat{\boldsymbol{\Lambda}}) + \lVert\hat{\boldsymbol{\Lambda}}^{(i)} - \hat{\boldsymbol{\Lambda}}\rVert.
	\end{equation}
	Propositions \ref{prop:hat_lambda_bar_lambda} and \ref{prop:sing_vals_lam_hat} the combine to tells us
	\begin{multline}
		\Theta\bigl(M_N^{1/2}L_N^{1/2}N\bigr)
		- \mathcal{O}\!\bigl(\max\{M_N,L_N\}^{1/2}M_NN^{1/2}\log^{3/2}(N)\bigr) \\[0.3em]
		\le \sigma_d\bigl(\hat{\boldsymbol{\Lambda}}^{(i)}\bigr)
		\le \Theta\bigl(M_N^{1/2}L_N^{1/2}N\bigr)
		+ \mathcal{O}\!\bigl(\max\{M_N,L_N\}^{1/2}M_NN^{1/2}\log^{3/2}(N)\bigr),
	\end{multline}
	almost surely, and so $\sigma_d(\hat{\boldsymbol{\Lambda}}^{(i)}) = \Omega\left(M_N^{1/2}L_N^{1/2}N\right)$. Similarly, we see
	\begin{multline}
		\mathcal{O}\!\bigl(\max\{M_N,L_N\}^{1/2}M_NN^{1/2}\log^{3/2}(N)\bigr)
		- 
		\mathcal{O}\!\bigl(\max\{M_N,L_N\}^{1/2}M_NN^{1/2}\log^{3/2}(N)\bigr)
		\\[0.3em]
		\le
		\sigma_{d+1}\bigl(\hat{\boldsymbol{\Lambda}}^{(i)}\bigr)
		\le
		\mathcal{O}\!\bigl(\max\{M_N,L_N\}^{1/2}M_NN^{1/2}\log^{3/2}(N)\bigr)
		+
		\mathcal{O}\!\bigl(\max\{M_N,L_N\}^{1/2}M_NN^{1/2}\log^{3/2}(N)\bigr),
	\end{multline}
	and so $\sigma_{d+1}(\hat{\boldsymbol{\Lambda}}^{(i)}) = \mathcal{O}\!\bigl(\max\{M_N,L_N\}^{1/2}M_NN^{1/2}\log^{3/2}(N)\bigr)$. It follows that
	\begin{equation}
		\left(\sigma_d(\hat{\boldsymbol{\Lambda}}^{(i)}) - \sigma_{d+1}(\hat{\boldsymbol{\Lambda}}^{(i)}) - \lVert\hat{\Lambda}_{i,\ast} - \bar{\Lambda}_{i,\ast}\rVert\right)^{-1} = \mathcal{O}\left(M_N^{-1/2}L_N^{-1/2}N^{-1}\right), \label{eqn:denominator_bound}
	\end{equation}
	almost surely. Combining We can then combine \eqref{eqn:numerator_bound} and \eqref{eqn:denominator_bound}, we find
	\begin{equation}
		\lVert \hat{\boldsymbol{P}} - \hat{\boldsymbol{P}}^{(i)} \rVert = \mathcal{O}\left(\max\{M_N,L_N\}^{1/2}M_NL_N^{-1}N^{-1}\log^{3}(N)\right), \label{eqn:P_hat_P_hat_i}
	\end{equation}
	almost surely. Finally, upon noting that the two-to-infinity norm is bounded by the spectral norm, we combine \eqref{eqn:P_hat_P_hat_i} with Proposition \ref{prop:hat_lambda_bar_lambda}, we can bound $\norm{T_6}$ as decomposed in \eqref{eqn:T_6_decomp}:
	\begin{equation}
		\norm{T_6} = \mathcal{O}\left(\max\{M_N,L_N\}M_N^{2}L_N^{-1}N^{-1/2}\log^{9/2}(N)\right), \label{eqn:T_6_bound}
	\end{equation}
	almost surely.
	
	To bound $\norm{T_5}$, we note
	\begin{equation}
		\norm{T_5} \leq \lVert e_i^\top \boldsymbol{E}(\boldsymbol{I} - \bar{\boldsymbol{P}})\hat{\boldsymbol{V}}^{(i)}\rVert = \lVert(\hat{\Lambda}_{i,\ast} - \bar{\Lambda}_{i,\ast})(\boldsymbol{I} - \bar{\boldsymbol{P}})\hat{\boldsymbol{V}}^{(i)}\rVert.
	\end{equation}
	By construction, the random matrix $\hat{\boldsymbol{V}}^{(i)}$ is independent of $(\hat{\Lambda}_{i,\ast} - \bar{\Lambda}_{i,\ast})$ and so we will look to apply concentration inequalities We have
	\begin{align}
		\lVert(\boldsymbol{I} - \bar{\boldsymbol{P}})\hat{\boldsymbol{V}}^{(i)}\rVert &\leq \lVert(\boldsymbol{I} - \bar{\boldsymbol{P}})\hat{\boldsymbol{V}}\rVert + \min_{\boldsymbol{W} \in \mathbb{O}(d)}\lVert(\boldsymbol{I} - \bar{\boldsymbol{P}})\rVert\lVert\hat{\boldsymbol{V}}^{(i)} - \hat{\boldsymbol{V}}\boldsymbol{W}\rVert\\ 
		&= \lVert(\boldsymbol{I} - \bar{\boldsymbol{P}})\hat{\boldsymbol{V}}\rVert + \min_{\boldsymbol{W} \in \mathbb{O}(d)}\lVert\hat{\boldsymbol{V}}^{(i)} - \hat{\boldsymbol{V}}\boldsymbol{W}\rVert,
	\end{align}
	which is valid as the triangle inequality will hold for any $\boldsymbol{W}$, and we use its assumed orthogonality to remove it from the first of the two norms. By Proposition \ref{prop:lots_of_bounds}, the first term satisfies
	\begin{equation}
		\lVert(\boldsymbol{I} - \bar{\boldsymbol{P}})\hat{\boldsymbol{V}}\rVert = \lVert\hat{\boldsymbol{V}} - \bar{\boldsymbol{V}}\bar{\boldsymbol{V}}^\top\hat{\boldsymbol{V}}\rVert =  \mathcal{O}\left(\max\{M_N,L_N\}^{1/2}M_N^{1/2}L_N^{-1/2}N^{-1/2}\log^{3/2}(N)\right), \label{eqn:I_minus_P_hat_first_term}
	\end{equation}
	almost surely. We use \eqref{eqn:P_hat_P_hat_i} to bound the second term by noting that we can write
	\begin{align}
		\min_{\boldsymbol{W} \in \mathbb{O}(d)}\lVert\hat{\boldsymbol{V}}^{(i)} - \hat{\boldsymbol{V}}\boldsymbol{W}\rVert \leq\lVert \hat{\boldsymbol{P}} - \hat{\boldsymbol{P}}^{(i)}\rVert = \mathcal{O}\left(\max\{M_N,L_N\}^{1/2}M_NL_N^{-1}N^{-1}\log^{3}(N)\right). \label{eqn:I_minus_P_hat_second_term}
	\end{align}
	Combining \eqref{eqn:I_minus_P_hat_first_term} and \eqref{eqn:I_minus_P_hat_second_term}, we find 
	\begin{equation}
		\lVert(\boldsymbol{I} - \bar{\boldsymbol{P}})\hat{\boldsymbol{V}}^{(i)}\rVert = \mathcal{O}\left(\max\{M_N,L_N\}^{1/2}M_N^{1/2}L_N^{-1/2}N^{-1/2}\log^{3/2}(N)\right), \label{eqn:I_minus_P_hat}
	\end{equation}
	almost surely. In particular, as this matrix is of maximal rank $d$, we have 
	\begin{equation}
		\lVert(\boldsymbol{I} - \bar{\boldsymbol{P}})\hat{\boldsymbol{V}}^{(i)}\rVert_{2,\infty} \leq \lVert(\boldsymbol{I} - \bar{\boldsymbol{P}})\hat{\boldsymbol{V}}^{(i)}\rVert_F \leq d^{1/2}\lVert(\boldsymbol{I} - \bar{\boldsymbol{P}})\hat{\boldsymbol{V}}^{(i)}\rVert,
	\end{equation}
	and so we obtain almost sure bounds on the Frobenius and two-to-infinity norms. 
	
	 We know $\lVert [(\boldsymbol{I} - \bar{\boldsymbol{P}})\hat{\boldsymbol{V}}^{(i)}]_{\ast,k}\rVert, \lVert [(\boldsymbol{I} - \bar{\boldsymbol{P}})\hat{\boldsymbol{V}}^{(i)}]_{\ast,k}\rVert_\infty$ decay at least as fast as $\lVert(\boldsymbol{I} - \bar{\boldsymbol{P}})\hat{\boldsymbol{V}}^{(i)}\rVert_F$. So, after recalling that the subexponential norm of $(\hat{\Lambda}_{i,k} - \bar{\Lambda}_{i,k})/M_N$ is $\mathcal{O}(1)$, we again apply Theorem \ref{thm:subexp_bern} as before to say 
	\begin{align}
		\mathbb{P}\left(\abs{\sum_{j=1}^{NL_N} (\hat{\Lambda}_{i,j} - \bar{\Lambda}_{i,j})Q^{(i)}_{jk}} \geq Mt\right) \leq \exp\Bigg(-D_{1,k}^{(i)} \min\Bigg\{&\frac{t^2L_NN}{D_{2,k}^{(i)}\max\{M_N,L_N\}M_N\log^{3}(N)}, \\ &
		\frac{tL_N^{1/2}N^{1/2}}{D_{3,k}^{(i)}\max\{M_N,L_N\}^{1/2}M_N^{1/2}\log^{3/2}(N)}\Bigg\}\Bigg),
	\end{align}
	for constants $D_{1,k}^{(i)}, D_{2,k}^{(i)}, D_{3,k}^{(i)} > 0$. The linear regime dominates with a choice of 
	\begin{equation}
		t = D_{5,k}^{(i)}\max\{M_N,L_N\}^{1/2}M_N^{1/2}L_N^{-1/2}N^{-1/2}\log^{5/2}(N). 
	\end{equation}
	Taking a union bound over the $d$ elements of the vector as before, it follows that
	\begin{equation}
		\norm{T_5} = \mathcal{O}\left(\max\{M_N,L_N\}^{1/2}M_N^{3/2}L_N^{-1/2}N^{-1/2}\log^{5/2}(N)\right), \label{eqn:T_5_bound}
	\end{equation}
	almost surely. Combining \eqref{eqn:T_6_bound} with \eqref{eqn:T_5_bound}, we finally obtain the desired bound on $\lVert e_i^\top \boldsymbol{T}_3\rVert$ of:
	\begin{equation}
		\lVert e_i^\top \boldsymbol{T}_3\rVert = \mathcal{O}\left(\max\{M_N,L_N\}^{1/2}M_N^{3/2}L_N^{-1/2}N^{-1/2}\log^{5/2}(N)\right), \label{eqn:T_3_bound}
	\end{equation} 
	almost surely independently of $i$. We couple \eqref{eqn:T_3_bound} with \eqref{eqn:T_4_bound} using the upper bound of \eqref{eqn:T_1_decomp} to bound $\lVert \boldsymbol{T}_1 \rVert_{2,\infty}$ as
	\begin{equation}
		\norm{\boldsymbol{T}_1}_{2,\infty} = \mathcal{O}\left(\max\{M_N,L_N\}^{1/2}M_N^{5/4}L_N^{-3/4}N^{-1}\log^{5/2}(N)\right), \label{eqn:T_1_bound}
	\end{equation}
	almost surely. Finally, we combine our bounds on $\boldsymbol{T}_1$ and $\boldsymbol{T}_2$ to bound the two-to-infinity norm of $\boldsymbol{M}_2$ as
	\begin{equation}
		\lVert\boldsymbol{M}_2\rVert_{2,\infty} = \mathcal{O}\left(\max\{M_N,L_N\}^{3/2}M_N^{7/4}L_N^{-5/4}N^{-1}\log^{9/2}(N)\right), \label{eqn:M_2_bound}
	\end{equation}
	almost surely. Using \eqref{eqn:M_2_bound} with the bound on $\lVert\boldsymbol{M}_2\rVert_{2,\infty}$ in \eqref{eqn:M_1_bound}, we obtain the stated result. The result for $\boldsymbol{R}_{2,2}$ follows analogously.
	\item We see that
	\begin{align}
		\lVert\boldsymbol{R}_{1,3}\rVert_{2,\infty} &\leq \lVert\bar{\boldsymbol{U}}\rVert_{2,\infty}\lVert\bar{\boldsymbol{U}}^\top(\hat{\boldsymbol{\Lambda}} - \bar{\boldsymbol{\Lambda}})\bar{\boldsymbol{V}}\boldsymbol{W}\hat{\boldsymbol{\Sigma}}^{-1/2}\rVert \\
		&\leq \lVert\bar{\boldsymbol{U}}\rVert_{2,\infty} \lVert\bar{\boldsymbol{U}}^\top(\hat{\boldsymbol{\Lambda}} - \bar{\boldsymbol{\Lambda}})\bar{\boldsymbol{V}}\rVert_F\lVert W\hat{\boldsymbol{\Sigma}}^{-1/2}\rVert_F.
	\end{align}
	We know $\lVert\bar{\boldsymbol{U}}\rVert_{2,\infty} = \mathcal{O}\left(M_N^{-1/2}N^{-1/2}\right)$, and by Proposition \ref{prop:sing_vals_lam_hat} we have the bound
	$\lVert \boldsymbol{W}\hat{\boldsymbol{\Sigma}}^{-1/2}\rVert_F = \mathcal{O}\left(N^{-1/2}M_N^{-1/4}L_N^{-1/4}\right)$. Combining these with Proposition \ref{prop:U_lambda_hat_lambda_bar_V} gives the desired result on $\lVert\boldsymbol{R}_{1,3}\rVert_{2,\infty}$. The bound on $\lVert\boldsymbol{R}_{2,3}\rVert_{2,\infty}$ follows analogously.
	\item We get
	\begin{equation}
		\lVert\boldsymbol{R}_{1,4}\rVert_{2,\infty} \leq \lVert\hat{\boldsymbol{\Lambda}} - \bar{\boldsymbol{\Lambda}}\rVert\lVert\bar{\boldsymbol{V}}\rVert_F\lVert \boldsymbol{W}\hat{\boldsymbol{\Sigma}}^{-1/2} - \bar{\boldsymbol{\Sigma}}^{-1/2}\boldsymbol{W}\rVert_F.
	\end{equation}
	We can invoke Propositions \ref{prop:hat_lambda_bar_lambda} and \ref{prop:W_bounds} to write
	\begin{align}
		\lVert\boldsymbol{R}_{1,4}\rVert_{2,\infty} &= \mathcal{O}\left(\max\{M_N,L_N\}^{1/2}M_NN^{1/2}\log^{3/2}(N)\max\{M_N,L_N\}N^{-3/2}M_N^{3/4}L_N^{-5/4}\log^3(N)\right) \\
		&= \mathcal{O}\left(\max\{M_N,L_N\}^{3/2}M_N^{7/4}L_N^{-5/4}N^{-1}\log^{9/2}(N)\right).
	\end{align}
\end{enumerate}

\subsection{Proof of Proposition \ref{prop:bias_decay_no_rotation_lipschitz}}

For some $t_m \in B_m$, we can write the following:
\begin{equation}
	\lVert\tilde{\boldsymbol{X}}^{(m,\ast)} - \boldsymbol{X}(t)\rVert_{2,\infty} = M_N\int_{B_m}\lVert\boldsymbol{X}(s) - \boldsymbol{X}(t_m)\rVert_{2,\infty} \dd s + \lVert\boldsymbol{X}(t_m) - \boldsymbol{X}(t)\rVert_{2,\infty}.
\end{equation}
By Assumption \ref{ass:no_rotation_lipschitz_continuity}, we can then say
\begin{align}
	\lVert\tilde{\boldsymbol{X}}^{(m,\ast)} - \boldsymbol{X}(t)\rVert_{2,\infty} \leq M_NK\int_{B_m}|t_m - s|\dd s + K|t_m - t| \leq M_NK \int_{B_m}M_N^{-1}\dd s + KM_N^{-1} = \mathcal{O}(KM_N^{-1}).
\end{align}

\subsection{Proof of Proposition \ref{prop:bias_decay_subspace_lipschitz}}

Define $\boldsymbol{W}_m$ to be
\begin{equation*}
	\boldsymbol{W}_m \in \argmin_{\boldsymbol{Q} \in \mathbb{O}(d)} \lVert \boldsymbol{P}(t_m)\tilde{\boldsymbol{X}}^{(m,\ast)} \boldsymbol{Q} - \boldsymbol{X}(t_m)\rVert_{2,\infty},
\end{equation*}
and decompose our norm into the following two terms:
\begin{equation}
	\lVert \tilde{\boldsymbol{X}}^{(m,\ast)} \boldsymbol{W}_m - \boldsymbol{X}(t) \rVert_{2,\infty} \leq \lVert (\boldsymbol{I} - \boldsymbol{P}(t))\tilde{\boldsymbol{X}}^{(m,\ast)}\boldsymbol{W}_m\rVert_{2,\infty} + \lVert \boldsymbol{P}(t) \tilde{\boldsymbol{X}}^{(m,\ast)} \boldsymbol{W}_m - \boldsymbol{X}(t)\rVert_{2,\infty}.
\end{equation}
Note that $\boldsymbol{W}_m$ is guaranteed to exists as $\mathbb{O}(d)$ is compact and the mapping is continuous as it is a composition of a linear map and the continuous two-to-infinity norm. For the first of these terms, we use that the two-to-infinity norm is invariant upon right multiplication by an orthogonal matrix and that $\boldsymbol{X}(u) = \boldsymbol{P}(u)$ to write
\begin{equation}
	(\boldsymbol{I} - \boldsymbol{P}(t))\tilde{\boldsymbol{X}}^{(m,\ast)} = M_N\int_{B_m} (\boldsymbol{I} - \boldsymbol{P}(t))\boldsymbol{P}(u)\boldsymbol{X}(u)\dd u.
\end{equation}
We then use the following equality:
\begin{equation}
	\lVert (\boldsymbol{I} - \boldsymbol{P}(t))\boldsymbol{P}(u)\rVert_{2,\infty} = \lVert \boldsymbol{P}(u)^2 - \boldsymbol{P}(t)\boldsymbol{P}(u)\rVert_{2,\infty} = \lVert \boldsymbol{P}(u) - \boldsymbol{P}(t)\rVert_{2,\infty},
\end{equation}
to bound our expression as
\begin{align}
	\lVert (\boldsymbol{I} - \boldsymbol{P}(t))\tilde{\boldsymbol{X}}^{(m,\ast)}\boldsymbol{W}_m\rVert_{2,\infty} &\leq M_N\int_{B_m}\lVert (\boldsymbol{I} - \boldsymbol{P}(t))\boldsymbol{P}(u)\rVert_{2,\infty} \lVert \boldsymbol{X}(u)\rVert_{2,\infty} \dd u, \\
	&\leq M_N \int_{B_m}\lVert \boldsymbol{P}(u) - \boldsymbol{P}(t)\rVert_{2,\infty} \lVert \boldsymbol{X}(u)\rVert_{2,\infty} \dd u, \\
	& \leq M_N \int_{B_m}\lVert \boldsymbol{P}(u) - \boldsymbol{P}(t)\rVert_{2,\infty}  \dd u \times \mathcal{O}\left(1\right), \\
	& \leq M_NK_1\int_{B_m} |t - u|\dd u \times \mathcal{O}\left(1\right), \\
	& \leq K_1/(2M_N) \times\mathcal{O}\left(1\right) \\
    &= \mathcal{O}\left(K_1 M_N^{-1}\right).
\end{align}
where we invoked Assumptions \ref{ass:bound_and_int} and \ref{ass:subspace_lipschitz_continuity}. The second term we decompose as follow:
\begin{align}
    \lVert \boldsymbol{P}(t) \tilde{\boldsymbol{X}}^{(m,\ast)} \boldsymbol{W}_m - \boldsymbol{X}(t)\rVert_{2,\infty} &\leq \lVert (\boldsymbol{P}(t) - \boldsymbol{P}(t_m))\tilde{\boldsymbol{X}}^{(m,\ast)} \rVert_{2,\infty} + \lVert \boldsymbol{P}(t_m)\tilde{\boldsymbol{X}}^{(m,\ast)} \boldsymbol{W}_m - \boldsymbol{X}(t_m)\rVert_{2,\infty}\\ 
    &+ \lVert\boldsymbol{X}(t_m) - \boldsymbol{X}(t) \rVert_{2,\infty}.
\end{align}
The first term is clearly $\mathcal{O}\left(K_1M_N^{-1}\right)$. The third term is is bounded as
\begin{align}
    \lVert\boldsymbol{X}(t_m) - \boldsymbol{X}(t) \rVert_{2,\infty} &\leq \lVert\boldsymbol{P}(t_m)(\boldsymbol{X}(t_m) - \boldsymbol{X}(t))\rVert_{2,\infty} + \lVert (\boldsymbol{I} - \boldsymbol{P}(t_m))\boldsymbol{X}(t) \rVert_{2,\infty},\\
    %&\leq \lVert\boldsymbol{P}(t_m)(\boldsymbol{X}(t_m) - \boldsymbol{X}(t))\rVert_{2,\infty} + \lVert (\boldsymbol{I} - \boldsymbol{P}(t_m))\boldsymbol{P}(t)\rVert_{2,\infty} \times \mathcal{O}(1),\\
   % &\leq \lVert\boldsymbol{P}(t_m)(\boldsymbol{X}(t_m) - \boldsymbol{X}(t))\rVert_{2,\infty} + \lVert\boldsymbol{P}(t_m) - \boldsymbol{P}(t)\rVert_{2,\infty} \times \mathcal{O}(1),\\
    & \leq K_2|t_m - t| + K_1|t_m - t| \times \mathcal{O}(1), \\
    &= \mathcal{O}\left((K_1 + K_2)M_N^{-1}\right).
\end{align}
Finally, the second term is bounded by invoking the definition of $\boldsymbol{W}_m$ to say
\begin{align}
    \lVert \boldsymbol{P}(t_m)\tilde{\boldsymbol{X}}^{(m,\ast)} \boldsymbol{W}_m - \boldsymbol{X}(t_m)\rVert_{2,\infty} \leq \lVert \boldsymbol{P}(t_m)\tilde{\boldsymbol{X}}^{(m,\ast)} - \boldsymbol{X}(t_m)\rVert_{2,\infty} = \mathcal{O}\left(K_1M^{-1}\right).
\end{align}
Combining, we see
\begin{equation}
    \lVert \boldsymbol{P}(t) \tilde{\boldsymbol{X}}^{(m,\ast)} \boldsymbol{W}_m - \boldsymbol{X}(t)\rVert_{2,\infty} = \mathcal{O}\left((K_1 + K_2)M_N^{-1}\right),
\end{equation}
and so
\begin{equation}
    \lVert \tilde{\boldsymbol{X}}^{(m,\ast)} \boldsymbol{W}_m - \boldsymbol{X}(t) \rVert_{2,\infty} = \mathcal{O}\left((K_1 + K_2)M_N^{-1}\right).
\end{equation}
This result is clearly uniform over $t$ and $m$ and the result follows.
% \end{appendices}